\documentclass[modern,dvipsnames,trackchanges]{aastex62}

\usepackage{natbib}
\usepackage{amsmath,amssymb}
\usepackage{textcomp} % for \degree
\usepackage{gensymb} % for \degree
\usepackage[normalem]{ulem} % for strikeout \sout

\graphicspath{{./}{figures/}{movies/}}
\shorttitle{MHD simulations of AR 11283 }
\shortauthors{Prasad et al.}
\begin{document}
\title{Magnetohydrodynamic Simulation of Magnetic Null-point 
Reconnections and Coronal dimmings during the X2.1 flare in NOAA AR 11283}

%===== Author information =======
\correspondingauthor{Avijeet Prasad, Karin Dissauer}
\email{ap0162@uah.edu, karin.dissauer@uni-graz.at}
\author[0000-0003-0819-464X]{Avijeet Prasad}
\affiliation{Center for Space Plasma \& Aeronomic Research,
The University of Alabama in Huntsville,
Huntsville, Alabama 35899, USA}

\author[0000-0001-5661-9759]{Karin Dissauer}
\affiliation{Institute of Physics,
University of Graz,
8010 Graz, Universit\"{a}tsplatz 5, Austria}
\affiliation{NorthWest Research Associates Inc, 3380 Mitchell Ln, Boulder, Colorado 80301, USA}

\author[0000-0002-7570-2301]{Qiang Hu}
\affiliation{Center for Space Plasma \& Aeronomic Research,
The University of Alabama in Huntsville,
Huntsville, Alabama 35899, USA}
\affiliation{Department of Space Science, The University of Alabama in Huntsville, Huntsville, AL 35899, USA}
\author[0000-0003-4522-5070]{R. Bhattacharyya}
\affiliation{Udaipur Solar Observatory, Physical Research Laboratory,
Dewali, Bari Road, Udaipur 313001, India}

\author[0000-0003-2073-002X]{Astrid M. Veronig}
\affiliation{Institute of Physics, University of Graz,
8010 Graz, Universit\"{a}tsplatz 5, Austria}
\affiliation{Kanzelh\"ohe Observatory for Solar and Environmental Research, University of Graz,  9521 Treffen, Austria}

\author{Sanjay Kumar}
\affiliation{Post Graduate Department of Physics, 
Patna University, Patna 800005, India}

\author[0000-0001-5042-2170]{Bhuwan Joshi}
\affiliation{Udaipur Solar Observatory, Physical Research Laboratory,
Dewali, Bari Road, Udaipur 313001, India}
%=========== ABSTRACT =====================================
\begin{abstract}
The magnetohydrodynamics of active region NOAA 11283 is simulated using an initial non-force-free magnetic field extrapolated from its photospheric vector magnetogram. We focus on the magnetic reconnections at a magnetic null point that participated in the X2.1 flare on 2011 September 6 around 22:21 UT (SOL2011-09-06T22:21X2.1) followed by the appearance of circular flare ribbons and coronal dimmings. The initial magnetic field from extrapolation displays a three-dimensional (3D) null topology overlying a sheared arcade. Prior to the flare, magnetic loops rise due to the initial Lorentz force, and reconnect at the 3D null, leading to expansion and loss of confined plasma that produce the observed pre-flare coronal dimmings. Further, the simulated dynamics documents the transfer of twist from the arcade to the overlying loops through  reconnections, developing a flux rope. The non-parallel field lines comprising the rope and lower-lying arcades form an X-type geometry. Importantly, the simultaneous reconnections at the 3D null and the X-type geometry can explain the observed circular and parallel flare ribbons. Reconnections at the 3D null transform closed inner spine field lines into open field lines of the outer spine. The footpoints of these open field lines correspond to a ring-shaped coronal dimming region, tracing the dome.
Further, the flux rope bifurcates because of these reconnections which also results in the generation of open magnetic field lines. The plasma loss along the open field lines can potentially explain the observed coronal dimming.

%Additionally the reconnections bifurcate the rope, which can lead to the loss of its confined plasma and provide a viable explanation for the distinct coronal dimmings observed during the impulsive phase of the flare. 
\end{abstract}
%======== KEYWORDS ======================

%============ SECTION 1 ===================
\section{Introduction} \label{sec:intro}
The solar corona can be treated as a magnetized plasma having a large electrical conductivity with its evolution being governed by the magnetohydrodynamic (MHD) equations {\citep{priest2014book}}. The magnetic Reynolds number $R_M$ ($vL/\eta$, in usual notations) for the corona is of the order of $10^{10}$ {\citep{aschwanden2004book}}. Under such conditions, the Alfv\'{e}n's  theorem of flux freezing is valid, which  ensures that the plasma-parcels remain tied to the magnetic field lines (MFLs) during their evolution {\citep{alfven1942nat}}. Eruptive events occurring in the corona like solar flares and coronal mass ejections (CMEs) are thought to be signatures of magnetic reconnection: a process involving topological rearrangement of MFLs with conversion of  magnetic energy into heat and kinetic energy of mass motion \citep{shibata&magara2011lrsp}. Notably, the requirement to onset magnetic reconnections is small $R_M$ which corresponds to small $L$, the length scale over which the magnetic field varies.  The smallness of  $L$ can either be pre-existing in a magnetic topology, manifested as magnetic nulls and quasi-separatrix layers (QSLs), or can develop autonomously during the evolution of the magnetofluid. Such autonomous developments (owing to discontinuities in magnetic field) are expected from Parker's magnetostatic theorem \citep{parker1972apj,parker1988apj,parker1994book}.
%which states that for a perfect electrically conducting plasma, the conditions of flux-freezing and the equilibrium cannot be satisfied simultaneously by a magnetic field which is continuous everywhere.
The reduction of $L$ and the consequent spontaneous magnetic reconnections during a quasi-static evolution of the plasma under a near-precise maintenance of the flux-freezing have been identified in contemporary MHD simulations performed with initial analytically constructed magnetic fields {\citep{kumard+2015phpl,kumar&bhattacharyya2016phpl,kumar+2016apj}}. 

Typically, the coronal magnetic field is extrapolated from the photospheric magnetic field observations because of a lack of accurate direct magnetic field measurements in the corona. In recent years, the  nonlinear-force-free-fields (NLFFFs) has been widely used for these extrapolations by the solar community \citep[e.g.][]{wiegelmann2008jgra,wiegelmann&sakurai2012lrsp,duan+2017apj}. Recent MHD simulations based on NLFFF extrapolations were successful in simulating  the coronal dynamics leading to eruptions \citep{jiang+2013apjl,kliem+2013apj,amari+2014nat, inoue+2014apj,inoue+2015apj,savcheva+2015apj,savcheva+2016apj,inoue2016peps}. 
However, the use of NLFFF extrapolations has a serious limitation, that in the solar photosphere, where the vector magnetograms are taken, the plasma beta is of the order of unity \citep{gary2001soph}, so that the Lorentz force is non-negligible. Generally, to mitigate this problem within the framework of NLFFF, a technique called `preprocessing' is often performed on the photospheric data which minimizes the Lorentz force in the vector magnetograms and provides a boundary condition suitable for NLFFF extrapolations \citep{wiegelmann+2006soph,jiang&feng2014soph}.

A novel alternative to NLFFF is the extrapolation using non-force-free-fields (NFFFs), which are described by the double-curl Beltrami equation for the magnetic field ${\bf{B}}$, derived from a variational principle of the minimum energy dissipation rate \citep{bhattacharyya+2007soph}. The equation was first solved analytically to obtain MFLs resembling coronal loops \citep{bhattacharyya+2007soph,kumar&bhattacharyya2011phpl}.
In a previous study, a semi-analytical construction based on maximizing correlations of non-axisymmetric NFFFs with photospheric vector magnetograms of NOAA AR 11283 
successfully mimicked the event of filament bifurcation by tracking the MHD evolution of a pre-existing flux rope \citep{prasad&bhattacharyya2016phpl,prasad+2017apj}.
However, missing from the simulations were the small-scale magnetic features and their influence on the MFL dynamics--which cannot be captured
by analytical/semi-analytical models. 
A numerical NFFF extrapolation model developed by \citet{hu&dasgupta2008soph,hu+2008apj,gary2009soph,hu+2010jastp} was used to initialize the MHD evolution of NOAA AR 12192 using the HMI vector magnetogram taken approximately 30 minutes prior to a confined X3.1 flare \citep{prasad+2018apj}.
%Particularly, the focus was on the magnetic reconnections occurring close to a magnetic null point that resulted in the appearance of circular chromospheric flare ribbons on 2014 October 24 around 21:21 UT. They found slipping reconnections at these quasi-separatrix layers, which were co-located with the post-flare circular brightening observed at chromospheric heights. Moreover, the initial field and its simulated evolution were found to be devoid of any flux rope, which were congruent with the confined nature of the flare.
Another NFFF initiated simulation, for the case of a blowout jet event was recently studied in \citet{nayak+2019apj}. 

In this study we continue our numerical studies of flaring active regions (ARs) which are initiated by NFFF
% To include these magnetic features and determine their role in overall magnetofluid evolution, here we numerically simulate evolution of AR 11283 initiated with the NFFF extrapolation model developed by \citet{hu&dasgupta2008soph,hu+2008apj,gary2009soph}. 
%Discussions onsen Karin's papers on dimming and 11283.
with a novel focus on coronal dimmings, which are temporary regions of strongly reduced coronal emission in EUV and soft X-rays that form in the wake of CMEs \citep[e.g.][]{hudson+1996apj,sterling&hudson1997apj,zarro+1999apj,thompson+2000georl}. In general, their formation is interpreted as density depletion due to the expansion and expulsion of plasma during the early CME evolution  \citep[e.g.][]{hudson+1996apj,harrison&lyons2000aa,veronig+2019apj}. Recently, distinct statistical relationships between decisive dimming parameters and CME and flare quantities were derived \citep{dissauer+2018b_apj,dissauer+2019apj}. Using a newly developed detection algorithm, so far not resolved fine structure within the dimming region could be identified for the first time \citep{dissauer+2018a_apj}. Both aspects verify the importance of coronal dimmings in the early diagnostics of solar eruptions. 
%The possible relation between dimming and flux rope eruption was also discussed in \citet{cheng&qiu2016apj,wang&liu2012apj}.
In order to exploit this potential further, in this paper, we analyze the X2.1 flare/CME event on 2011 September 6 by combining extreme-ultraviolet (EUV) observations of coronal dimmings and the associated flare with MHD simulations using an intital NFFF extrapolated field. 
%We follow the evolution of AR 11283, starting at 22:00 UT and study the X2.1 flare occurring at 22:21 UT. 
% ------------ Summary of on AR 11283 simulation results from earlier papers ---------------------
%The X2.1 flare of AR 11283 was also accompanied by a halo CME. 

Several aspects of this event have been studied already in literature, including both observational and modeling efforts \citep[e.g.][]{petrie:2012apj,feng+2013apj,yang:2014apj,roman0+2015aa,janvier+2016,dissauer+2016apj,jiang+2013apjl,jiang+2014apj, jiang+2016nat, jiang+2018apj,vanninathan+2018apj}. For example, \citet{feng+2013apj} estimated the magnetic energy partition between the flare and CME. They concluded that, within the uncertainties, both the flare and the CME might have consumed free energy of around $6.5 \times 10^{31}$ erg.
\cite{janvier+2016} studied
the morphology and time evolution of photosperic traces of
current density and flare ribbons, and compared it with the
topological features found by NLFFF modeling.  They identified a spine-fan
configuration of the overlying field lines, due to the presence of a parasitic positive polarity, embedding a flux rope.
In a series of papers on this AR, \citet{jiang+2013apjl,jiang+2014apj,jiang+2016nat,jiang+2018apj} also have explored the dynamical evolution of this region through different MHD simulations. In their latest study, \citet{jiang+2018apj} discuss the complex sigmoid eruption in the active region characterized by a multipolar configuration embedding a null-point topology and a sigmoidal magnetic flux rope. Based on EUV observations and MHD simulations, they suggest that a three-stage magnetic reconnection scenario might explain this complex flare event.
%Importantly, a non-zero Lorentz force, instead of prescribed flows \citep{amari+2003apj,aulanier+2010apj}, is envisaged here to initiate dynamics. 

In the present paper we aim to use the dynamics, location, and intensity distribution of coronal dimmings together with additional observational information of the associated flare and CME (e.g. signatures of flare ribbons, a hot sigmoid, the flux rope eruption etc.) as guidance for the non-force-free magnetic field modeling and MHD simulations of the X2.1 flare/CME/dimming event on 2011 September 6 in order to understand this complex eruption in more detail.
%Here, the pre- and post-eruptive magnetic field configurations at the observed coronal dimming locations are investigated, with respect to the coronal dimming fine structure. The dynamics of the dimming evolution is studied in the form of timing maps, which code when each pixel of the dimming region is detected for the first time. These timing maps are used to investigate the initiation of the eruption and to identify which flux systems might be involved. \kd{still needs to be adapted to fill this gap...}
%Coronal dimming locations are also compared with squashing factor Q maps, indicating locations in favor of magnetic reconnection, in order to check whether preferential locations for the formation of coronal dimmings exist prior to the eruption. Most importantly, the comprehensive MHD evolution reported here provides a potential explanation of the coronal dimmings observed after the event.
The paper is organized as follows: Section~\ref{sec:data} summarizes the data used in this study and Section~\ref{sec:observations} gives a detailed observational overview of the event and the associated coronal dimming evolution. In Section~\ref{sec:nfff}, we present the details of the initial non-force-free extrapolated field. The MHD model is discussed in Section~\ref{sec:mhd} along with the results of the simulation and their comparison to the observations.  Section~\ref{sec:summary} summarizes our most important findings.

\section{Data and Pre-processing}\label{sec:data}
We use high-cadence (12 s) data from six different ultraviolet and extreme-ultraviolet wavelengths of the Atmospheric Imaging Assembly (AIA; \citealt{lemen+2012soph}) on board the Solar Dynamics Observatory (SDO; \citealt{pesnell+2012soph}) covering a temperature range of \mbox{about $10^4$ to $10^7$ K}. The 171 and \mbox{211 \AA}~channels represent plasma at quiet Sun temperatures (0.6--2 MK), while 335 and \mbox{94 \AA}~are sensitive to hot plasma of active regions and flares (up to 6~MK). The temperature response function of the \mbox{304 \AA~}filter peaks at \mbox{$\approx$ 50,000} K~and plasma at this temperature is likely to origin from the transition region and chromosphere. The ultraviolet 1600 \AA~filter is sensitive to plasma at $\approx$ 10,000~K and resolves structures of the upper photosphere and transition region.

In order to generate suitable extrapolated coronal magnetic field, the photospheric vector magnetogram of AR 11283 is obtained from the Helioseismic Magnetic Imager \citep[HMI;][]{schou+2012soph} on board SDO at 22:00~UT on 2011 September 6. The magnetogram is taken from the `hmi.b\_720s' data series that provides full-disk vector magnetograms of the Sun with a temporal cadence of 12 minutes and a spatial resolution of $0''.5$. The field of view was chosen from the full-disk magnetogram to ensure that all coronal dimming regions 
%observed in Figure~\ref{f2:dimming_evolution}
are located within the computational domain.
In order to obtain the magnetic field on a Cartesian grid, the magnetogram  is initially remapped onto a Lambert cylindrical equal-area (CEA) projection and then transformed into heliographic coordinates \citep{gary&hagyard1990soph}. This results in a field of view of 1024$\times$800 pixels centered at 226.00\degree~and 17.00\degree~Carrington longitude and latitude, respectively. To adequately compare simulation results with observations, SDO/AIA filtergrams are also CEA projected, remapped to the same spatial resolution as the magnetic field data and the same field of view is used.

\section{Event Evolution and Observations}\label{sec:observations}
\label{sec:2}
\begin{figure}
    \centering
    \includegraphics[width=0.98\linewidth]{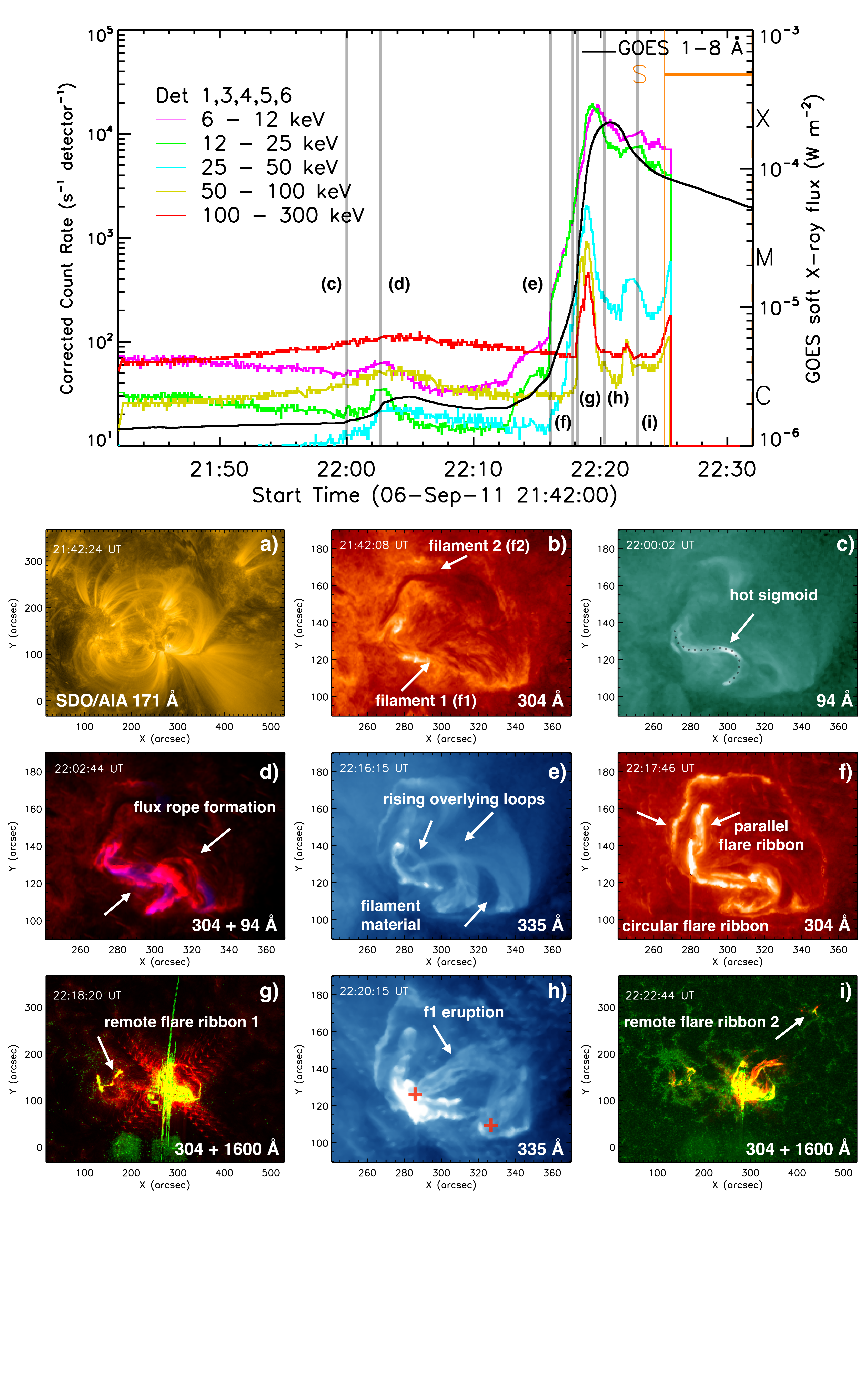}
    \caption{Observational overview of the September 6, 2011 flare/CME/dimming event. Top: Time evolution of the GOES soft X-ray flux together with RHESSI hard X-ray lightcurves from 6--300~keV. Bottom: Panels (a)-(i) are SDO/AIA images showing the formation of the flux rope, the 
    initiation of the flare followed by the evolution of the main circular and remote flare ribbons, as well as the flux rope eruption. In panel(h), the 
    footpoints of the erupting flux rope are identified and marked as red crosses.}
    \label{f1:event_overview}
\end{figure}
\begin{figure}
    \centering
    \includegraphics[width=1.0\linewidth]{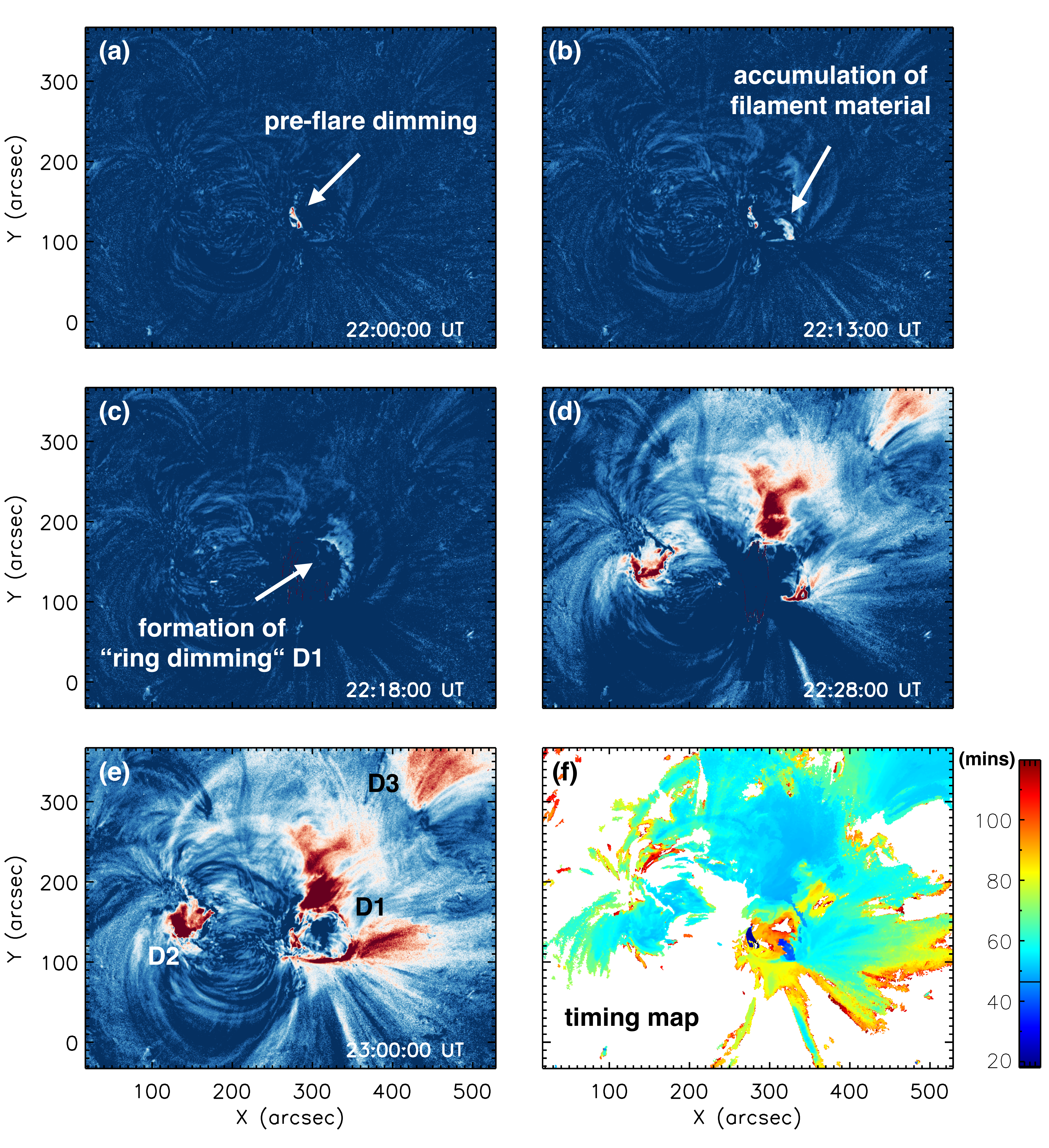}
    %dimming_overview_20110906_detail.pdf
    \caption{Time evolution of the coronal dimming regions associated with the flare/CME event on 2011 September 6. Panels~(a-d) show the formation of small-scale pre-flare coronal dimmings before the flare onset, the accumulation of filament material at the right footpoint of the flux rope, the early formation of the ``ring-shaped" dimming region, and the expansion of the overall dimming to the east and to the north-west of the main flare site. The final extent of the coronal dimming region at the end of its main development phase is shown in Panel~(e), where dimming regions of interest are marked. D1 marks the early-on formed ring-shaped dimming, D2 indicates the peculiar circular shaped dimming region to the east and D3 denotes the remote dimming region to the north-west of the main eruption site. Panel~(f) shows the overall time evolution of the dimming region in the form of a timing map, where each dimming pixel is color-coded based on the time of its first detection in minutes after 21:30~UT.}
    \label{f2:dimming_evolution}
\end{figure}

AR 11283 was a very flare-productive region which produced many flares and CMEs during its disk passage \citep{roman0+2015aa}.
The selected X2.1 flare in AR 11283 occurred on 2011 September 6 close to the disk center at heliographic position N14\degree W18\degree~(SOL2011-09-06T22:21X2.1). It was associated with a fast halo CME with a speed of $v=990$ km/s (as derived from lateral view by the STEREO-A coronagraphs), a fast EUV wave, a type II burst, and pronounced coronal dimmings \citep{dissauer+2016apj,vanninathan+2018apj}. The impulsive phase of the flare started at 22:12~UT and reached its peak around 22:21~UT as evident from the top panel of Figure~\ref{f1:event_overview} which shows the GOES soft X-ray flux in the 1--8 \AA~ band~together with the RHESSI hard X-ray emission in several energy bands from 6--300~keV. 
We use observations of SDO/AIA to outline important observational features that occurred during this event. The bottom panels of Figure~\ref{f1:event_overview} present an observational overview of the event including the formation of the flux rope (associated with filament 1), the initiation of the flare, the evolution of the main circular and remote flare ribbons, and finally the flux rope eruption. The formation and time evolution of the associated coronal dimming regions in SDO/AIA 211 \AA~are shown in Figure~\ref{f2:dimming_evolution}.

Figure~\ref{f1:event_overview}(a--c) show filtergrams in SDO/AIA 171, 304, and 94 \AA~illustrating the pre-flare conditions in the corona and the chromosphere. Figure~\ref{f1:event_overview}(a) outlines the connectivities of the overlying loops, joining the different polarities involved later in the eruption. Two filaments, filament~1 (f1) and filament~2 (f2) could be identified prior to the flare (indicated by the white arrows in panel b). The filament f1, which is also the first one that erupts, is located along the main polarity inversion line (PIL) that is involved in the X2.1 flare/eruption and extends in the east-west direction. Filament f2 is located north of the initial flare site and, like f1, oriented in the east-west direction but slightly bent towards south. 

Around 22:00 UT, the signature of a hot sigmoid is observed, which is co-spatial with filament 1 (see Figure~\ref{f1:event_overview}(c)). During this interval, we also note a small enhancement in the soft X-ray light curves (Figure ~\ref{f1:event_overview}(top panel)).
From the composite image of SDO/AIA 304 (red channel) and 94 \AA~(blue channel) in Figure~\ref{f1:event_overview}(d), the hot sigmoid appears to be growing with time and localized brightenings are observed, leading to the growth of the original flux rope. 
%This occurs co-temporal with a hard X-ray enhancement in the 12-25 keV, 25-50~keV, and 50-100~keV channels. The source of the hard X-ray enhancement most likely originates close to the solar limb, and is not co-spatial with the localized brightenings along the hot sigmoid. However, due to the dynamic range of RHESSI it is still possible that the hard X-ray signatures originate from this region, as the brightenings are also observed in SDO/AIA 1600~\AA.
Until 22:11~UT, ongoing activity can be observed in the AIA EUV imagery, which results in the accumulation of filament material (of f1) at the right leg of the flux rope (Figure~\ref{f1:event_overview}(e)). The corresponding H$\alpha$ observations from Big Bear Observatory at 22:11:54~UT confirm the existence of filament material at this location. 

Notably, close to the onset of the flare at 22:16 UT, the rise of flare loops in SDO/AIA 335 \AA~filtergrams is observed (see Figure~\ref{f1:event_overview}(e)). These loops could be either part of the overlying arcade above the flux rope or its outer envelope. This is followed by the simultaneous formation of a circular flare ribbon, surrounding the main flare site and the standard parallel flare ribbons forming as part of the circular ribbon (see Figure~\ref{f1:event_overview}(f)). Relevantly, the appearance of circular ribbons is generally considered to be caused from the magnetic configuration of a three-dimensional (3D) magnetic null point \citep{masson+2009apj,wang&liu2012apj,hernandez-perez+2017apj,devi+2020SoPh,liu+2020ApJ}.

During the main phase of the flare, two strong and impulsive hard X-ray bursts are observed by RHESSI, indicative of efficient acceleration of high-energy electrons in the flare (top panel, Figure~\ref{f1:event_overview}). The first one occurs between 22:18 -- 22:20~UT and produced detectable hard X-ray emission up to energies $>$800 keV. It is co-temporal with the appearance of the circular flare ribbon at the main flaring site and a remote flare ribbon to the east, shown in Figure~\ref{f1:event_overview}(g) as a composite image of SDO/AIA 304 (red channel) and 1600 \AA ~(green channel). 
Close to the peak time of the flare, f1 erupts, as shown by the SDO/AIA 335 \AA~image in Figure~\ref{f1:event_overview}(h). Both footpoints of the erupting flux rope are identifiable and are marked as red crosses. Interestingly, during the decay phase of the flare, a second hard X-ray burst is observed (up to about 300 keV) during 22:21--22:24~UT, which is concurrent with the activity at the main circular flare ribbon and a new remote flare ribbon to the north-west of the main flaring site (see composite image of SDO/AIA 304 and 1600 \AA~in Figure~\ref{f1:event_overview}(i)). We note that filament~2 also began to erupt around 22:37~UT (not shown), which marks the end of the activity during this event. 

Figure~\ref{f2:dimming_evolution} shows SDO/AIA 211 \AA ~logarithmic base-ratio images illustrating the evolution of coronal dimming regions. To increase the visibility and to resolve the fine structure of dimmings, regions of increased intensity are set to 1.0, small to moderate intensity decreases are shown from lightblue to white and strong intensity decreases appear in red \citep[][]{dissauer+2018a_apj}. Panel~(a) shows the signature of small-scale, bipolar pre-flare dimmings \citep[][]{qiu+2017apj, zhang+2017aa} close to the footpoints of the sigmoid associated with f1 about 12 minutes prior to the onset of the flare and the associated eruption. Until the start of the flare, cool filament material is accumulated at the right footpoint of the flux rope (see also Figure~\ref{f1:event_overview}(d--e)), observed as a dark region in panel~(b) marked by the white arrow. This region does not result from plasma evacuation but from the cool filament material that is darkening in the 211 \AA~passband. 
%\ap{What processes might be involved in the accumulation of the filament material? Can this be linked with what we observe in the simulations?}

\begin{figure}[ht!]
\plotone{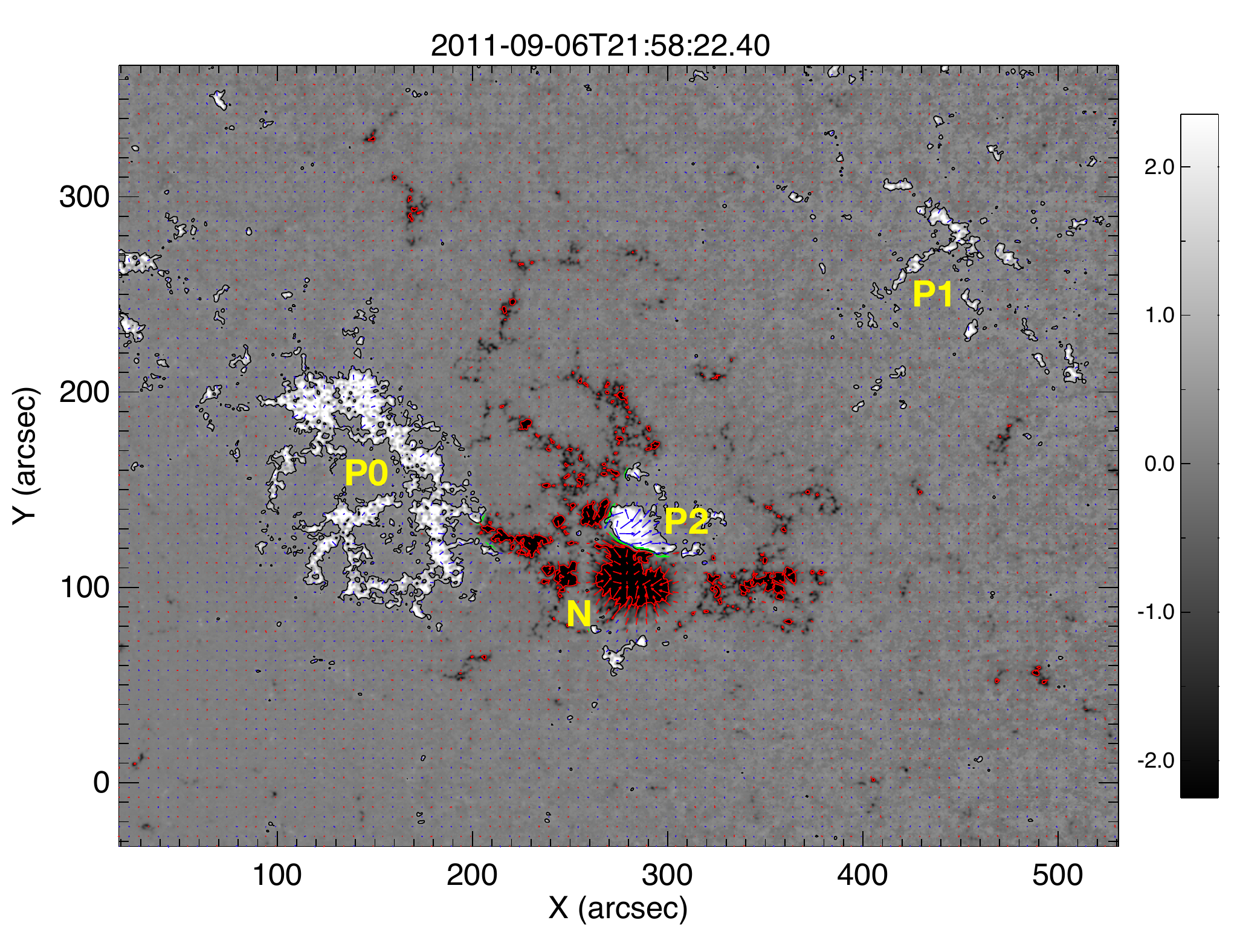}
%HMI_ar11283fd_T2200.pdf
\caption{HMI vector magnetogram of AR 11283 at 22:00 UT  on September 6, 2011. The red and blue arrows depict the strength and direction of the transverse magnetic field and the colorbar on the right shows the vertical field strength in kG. The Carrington longitude and latitude of the field of view of the center are 226.00\degree~and 17.00\degree, respectively.}
\label{f3:hmi_mag}
\end{figure}

Moreover, at the beginning of the flare, a weak semi-circular coronal dimming region close to the main flare site is formed (see Figure \ref{f2:dimming_evolution}(c)). Over the course of the event, this region will develop into the ring-shaped dimming region D1. Panels~(d--e) show the formation and evolution of the coronal dimming during the impulsive phase of the flare, expanding to a remote location to north-west of the main flare site and to east. At the end of the major development phase of the dimming, three main coronal dimming regions are identified (see Figure~\ref{f2:dimming_evolution}~(e)). Region 1 (marked by D1) is characterized by a ring-shaped dimming region that is associated with the main flare site. Region 2 (denoted by D2) is a strongly decreased circular shaped dimming region located mainly in the neighbouring positive polarity to the east of the flare location (marked by P0 in Figure~\ref{f3:hmi_mag}). 
Dimming region 3 (indicated by D3) is located further away from the main flare site to the north-west in a positive polarity (marked by P1 in Figure~\ref{f3:hmi_mag}). 
 Panel~(f) shows the overall time evolution of the dimming region in the form of a timing map, where each dimming pixel is color-coded based on the time of its first detection in minutes after 21:30 UT. This representation of the coronal dimming allows us to identify which parts of the lower corona are affected during the eruption and at which time they are activated.

In this paper, our focus is to understand the processes of the flare initiation and coronal dimming formation. Therefore, the presented simulation is initiated at 22:00~UT when the pre-flare activity starts (see Figure~\ref{f1:event_overview}). Figure~\ref{f3:hmi_mag} shows the magnetogram of the active region at 22:00~UT where the positive and the negative polarities of the longitudinal component of the magnetic field are depicted in white and black, respectively, and the gray represents the background. The transverse positive and negative fields are shown by blue and red arrows, respectively, while the PIL is represented by green lines.
Following \citet{jiang+2018apj}, we mark the main positive polarities as  P0, P1, P2 and the central negative polarity as N. Here P0 and P1 are more dispersed than P2, which is an emerging region close to N. The flare and eruption took place near the PIL between N and P2, where the field is most sheared and non-potential. As a parasitic polarity of N, P2 is surrounded by the negative flux which is supportive of the existence of a magnetic null. To obtain the 3D coronal magnetic field consistent with this photospheric boundary, we use the non-force-free extrapolation, which is described in the next section.

\section{Non-force-free extrapolation of magnetic field}\label{sec:nfff}
\label{sec:3}

The extrapolated coronal magnetic field of AR 11283 at 22:00 UT corresponding to the photospheric boundary shown in Figure \ref{f3:hmi_mag} is obtained numerically by using the non-force free extrapolation technique developed by \citet{hu&dasgupta2008soph,hu+2008apj,hu+2010jastp}. In this approach the magnetic field ${\bf{B}}$ is constructed as
\begin{equation}
\mathbf{B} = \mathbf{B_1}+\mathbf{B_2}+\mathbf{B_3}; \quad \nabla\times\mathbf{B_i}=\alpha_i\mathbf{B_i}
\label{e:b123}
\end{equation}
with $i=1,2,3$. Here, each sub-field ${\bf{B}}_i$ corresponds to a linear-force-free field (LFFF) with corresponding constants $\alpha_i$. Further, without loss of generality, we choose $\alpha_1\ne\alpha_3$ and $\alpha_2 = 0$, making $\mathbf{B_2}$ a potential field. Subsequently, an optimal pair $\alpha=\{\alpha_1 , \alpha_3\}$ is obtained by an iterative method which finds the pair that minimizes the average deviation between the observed ($\mathbf{B}_t$) and the calculated ($\mathbf{b}_t$) transverse field on the photospheric boundary. This is estimated by the following metric \citep{prasad+2018apj}:
\begin{equation}
E_n =\left(\sum_{i=1}^M |\mathbf{B}_{t,i}-\mathbf{b}_{t,i}|\times|\mathbf{B}_{t,i}|\right)/\left(\sum_{i=1}^M |\mathbf{B}_{t,i}|^2\right)
\label{en}
\end{equation}
where $M=N^2$, represents the total number of grid points on the transverse plane. To minimize the contribution from the weaker fields, here the grid points are weighted with respect to the strength of the observed transverse field \citep[see][for further details]{hu&dasgupta2008soph,hu+2010jastp}.

The extrapolated field ${\bf{B}}$ is a solution of an auxiliary higher-curl equation
\begin{equation}
\nabla\times\nabla\times\nabla\times\mathbf{B}+a_1\nabla\times\nabla\times\mathbf{B}+b_1\nabla\times\mathbf{B}=0,\label{e:bnff2}
\end{equation}
where $a_1$ and $b1$ are constants. Equation \eqref{e:bnff2} contains a second order derivative $(\nabla\times\nabla\times\mathbf{B})_z=-\nabla^2 B_z$ at $z=0$, necessitating the requirement 
of vector magnetograms at two or more layers for evaluating the ${\bf{B}}$. In order to work with the available single layer vector magnetograms, an algorithm was devised by \citet{hu+2010jastp}, which involved additional iterations to successively correct the potential subfield $\mathbf{B_2}$. Starting with an initial guess, $\mathbf{B_2}=0$, the system is reduced to second-order which allows for the determination of boundary conditions for $\mathbf{B_1}$ and $\mathbf{B_3}$ using the process as described above. If the resulting minimum $E_n$ value is not satisfactory, then a corrector potential field to $\mathbf{B_2}$ is derived from the difference transverse field, i.e., $\mathbf{B}_t-\mathbf{b}_t$, and added to the previous $\mathbf{B_2}$, in anticipation of an improved match between the transverse fields, as measured by $E_n$. The algorithm relies on the implementation of fast calculations of the LFFFs including the potential field.

\begin{figure}[ht!]
%\epsscale{0.7}
\gridline{
        \fig{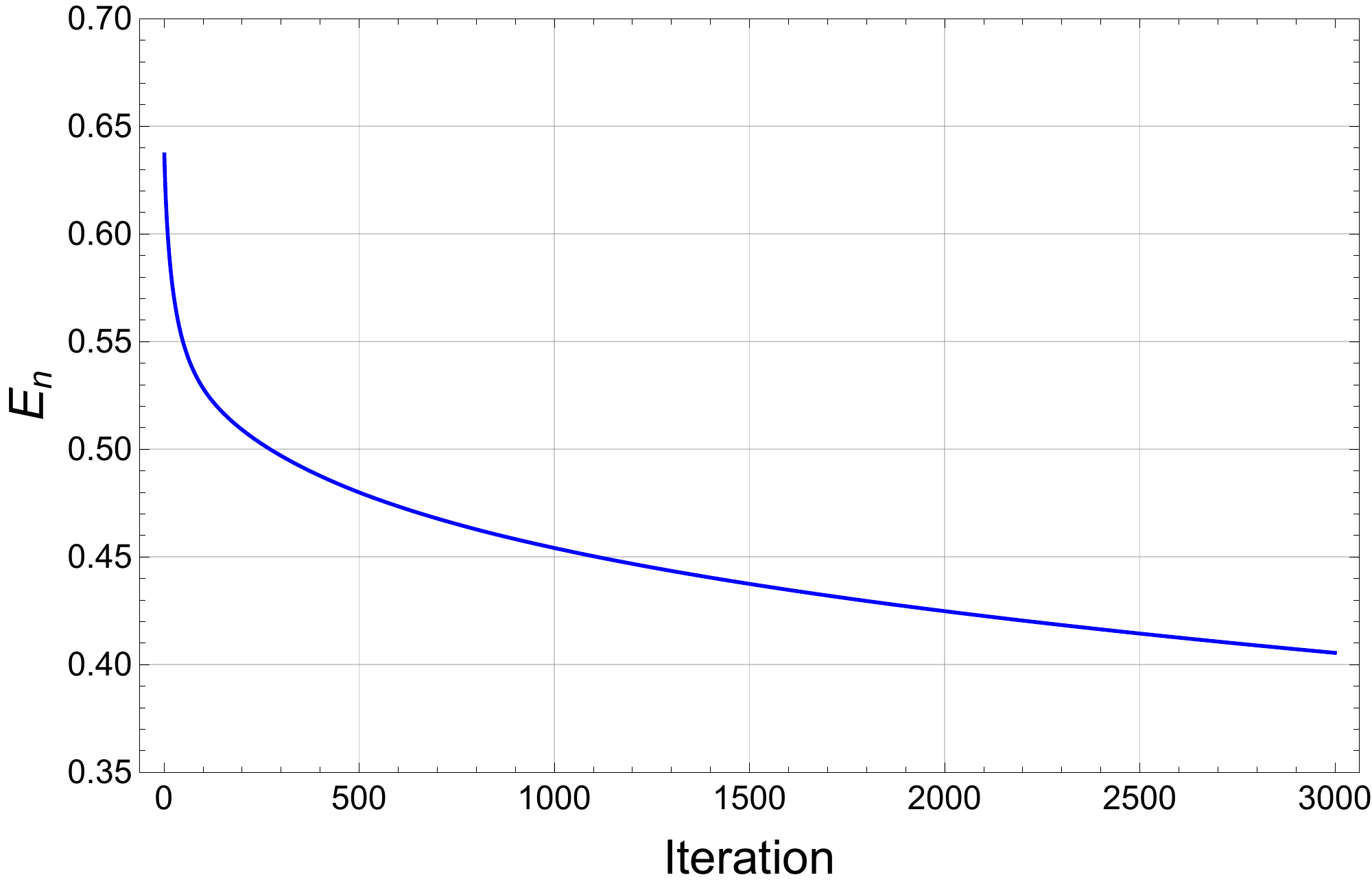}{0.45\textwidth}{(a)}
        %{enplot_minerr.pdf}
        \fig{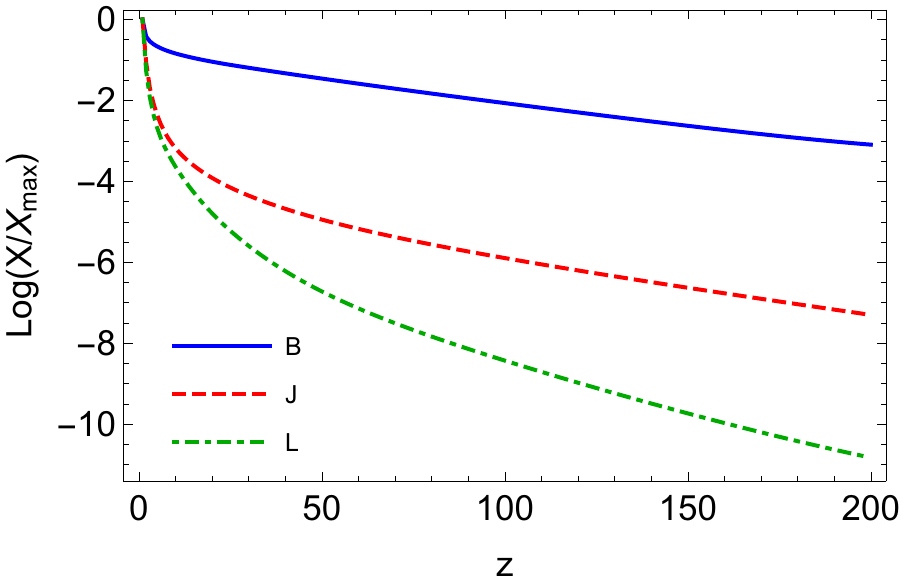}{0.45\textwidth}{(b)}
        %{jblz_bnfff.pdf}
        }
%\plottwo{fig04a.pdf}{fig04b.pdf}
\caption{Panel~(a) shows the variation of $E_n$ as a function of iteration number during the NFFF extrapolation. Panel~(b) depicts the logarithmic variation of strength for the horizontally-averaged magnetic field ($X=B$), the current density ($X=J$) and the Lorentz force density ($X=L$) with pixel height $z$. All the values are normalized with respect to their maximum values as we are mostly interested in the rate of decay with height.}
\label{f4:err_jbl}
\end{figure}

%\subsection{Initial extrapolated NFFF for AR 11283}
The vector magnetogram shown in Figure \ref{f3:hmi_mag} corresponds to an original  cutout of dimension  $1024\times 800$ pixels. To reduce the computational  cost, the original field is re-scaled and extrapolated over a $256\times 200 \times 200$ pixels grid volume in the $x$, $y$ and $z$ directions. The variation of minimum error in the transverse field $E_n$ with iteration number is shown in the left panel of Figure \ref{f4:err_jbl}. Here, we find that the curve reaches a saturation value of 0.4 after 3000 iterations. We stop the iterations at this point to save the computational cost. Noticeably, the final value of $E_n$ is higher compared to those obtained in earlier works \citep{prasad+2018apj,mitra+2018apj,nayak+2019apj}, but this is expected as we have chosen a much larger field of view here which results in more contribution from the weaker fields. The variations of horizontally-averaged strength for the magnetic field, current density and Lorentz force density with pixel height $z$ are shown in the right panel of Figure \ref{f4:err_jbl}. As expected, the horizontally-averaged value of the Lorentz force density falls off fastest with height, followed by that of the current density and the field strength. Notably, the Lorentz force density is non-zero near the photosphere and almost vanishes at coronal heights (cf. Figure \ref{f4:err_jbl} (b)). As a result, in our model, the corona is considered to be reasonably force-free while the photosphere supports the Lorentz force \citep{yalim+2020ApJ,liu+2020ApJ}.

\begin{figure}[ht!]
\gridline{
        \fig{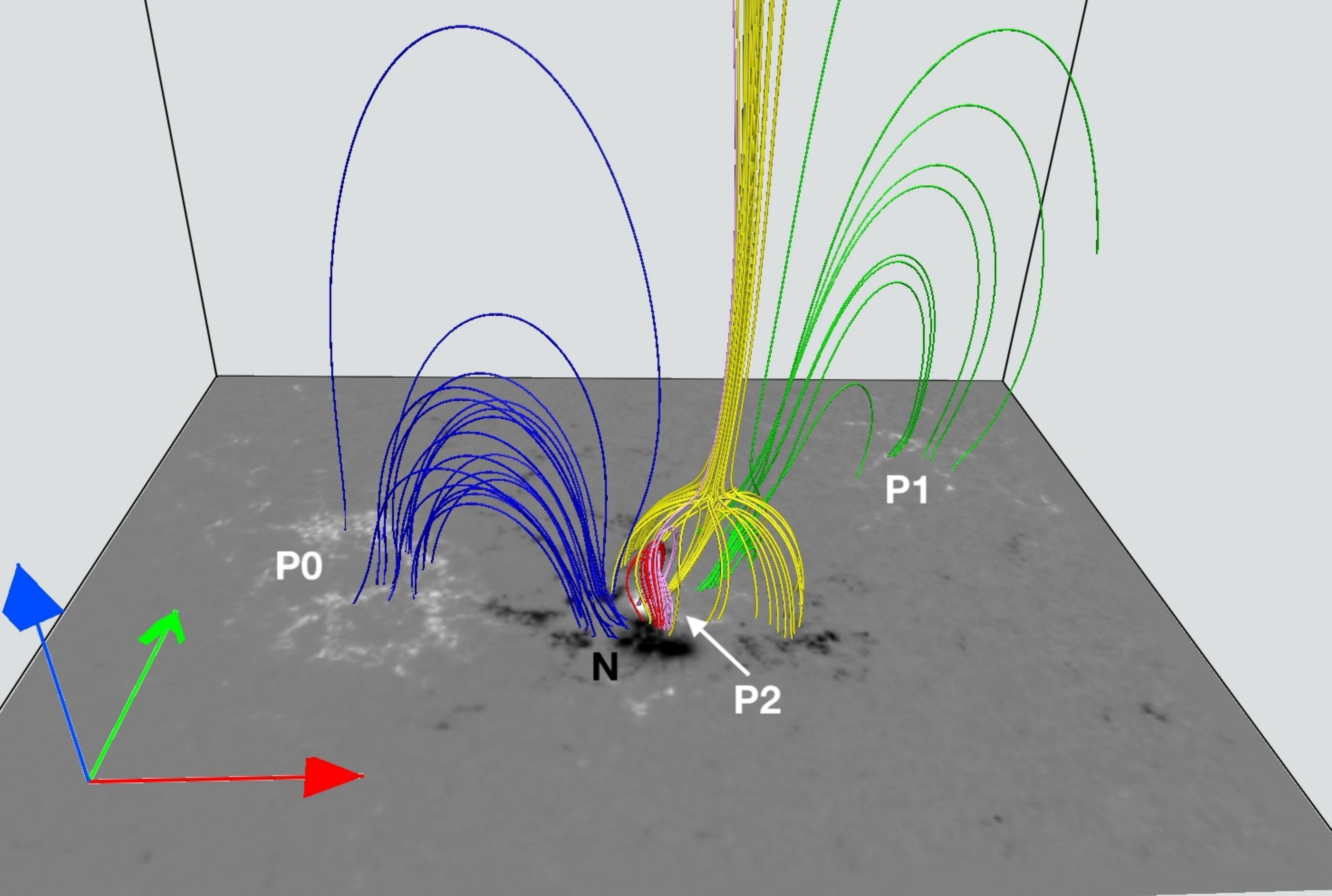}{0.48\textwidth}{(a)}
        %initial_field_side.pdf
        \fig{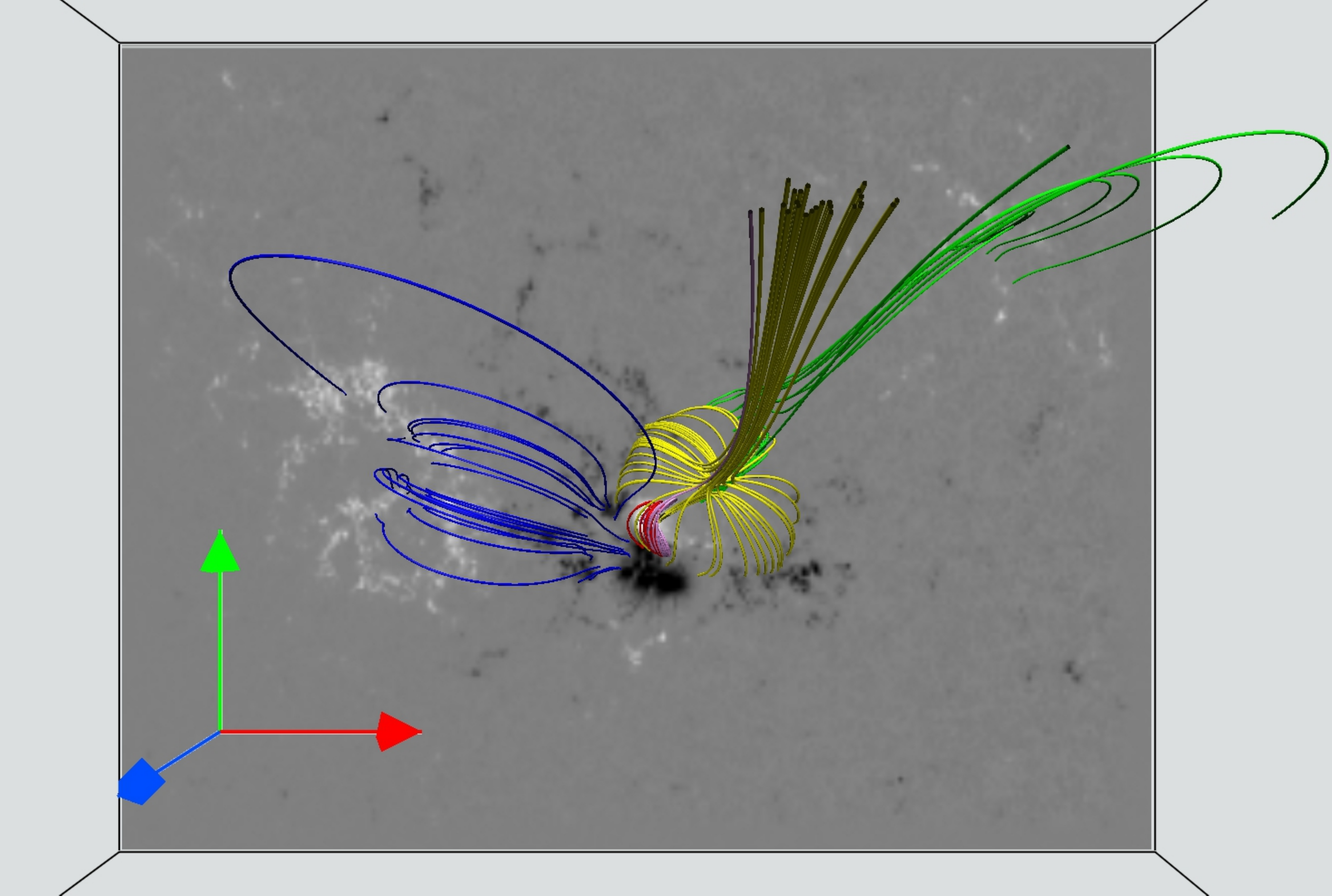}{0.48\textwidth}{(b)}
        %initial_field_top.pdf
        }
\gridline{
        \fig{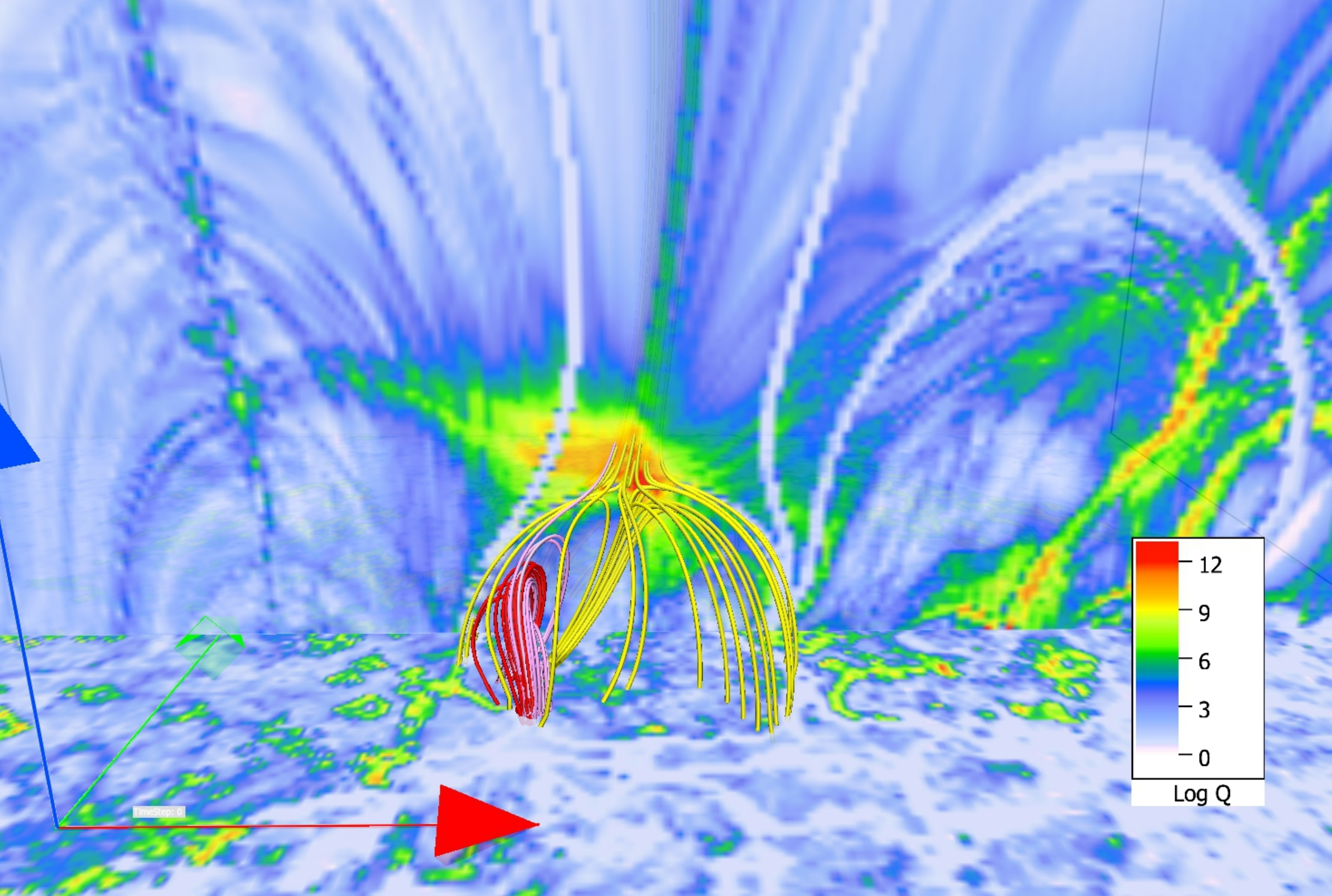}{0.48\textwidth}{(c)}
        %logQ.pdf
        \fig{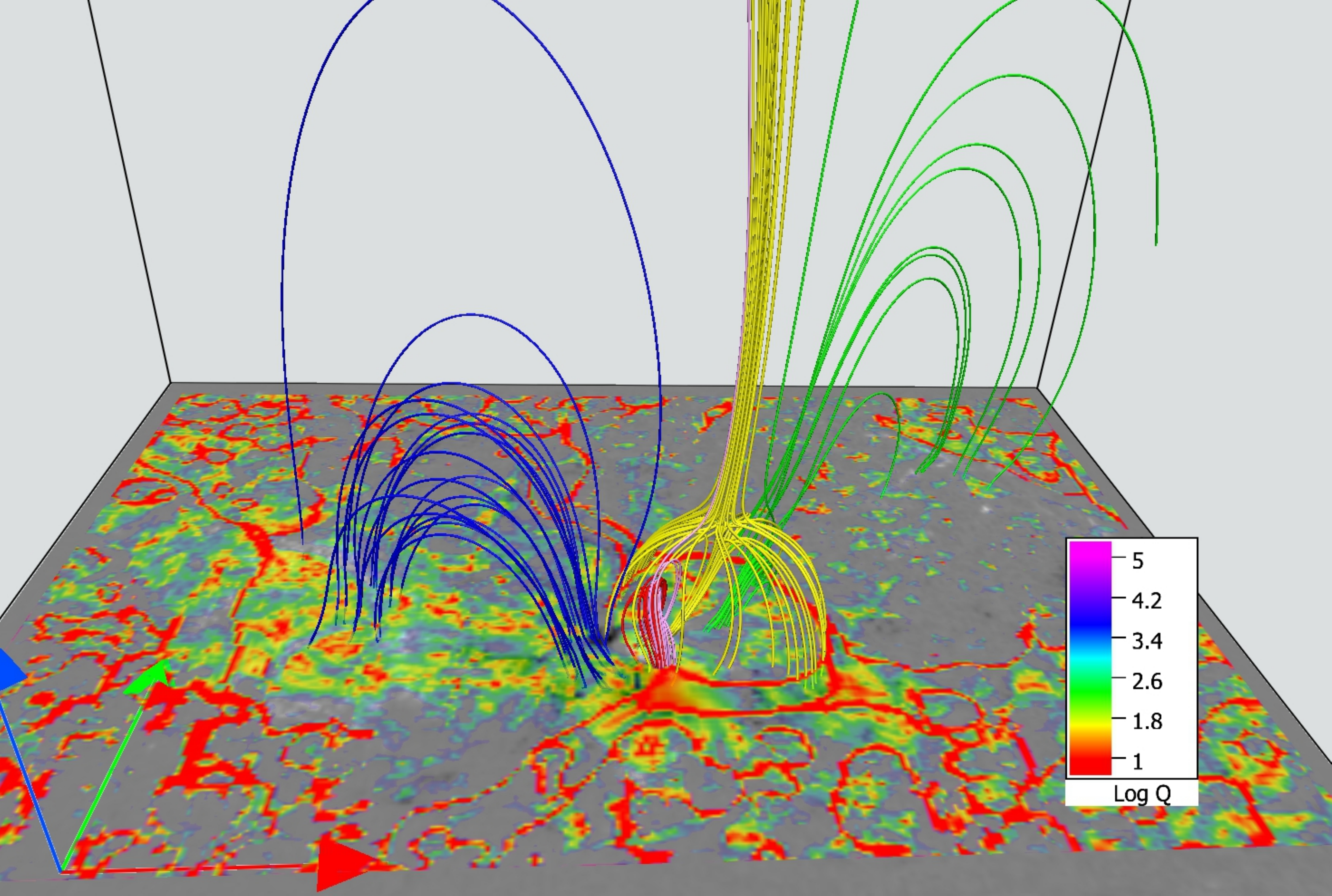}{0.48\textwidth}{(d)}
        %qsl_t0.jpg
        }
\caption{Side (a) and top (b) view of the extrapolated magnetic field highlighting the different connectivity with the magnetogram as the bottom boundary. The field lines in red and purple depict sheared field lines over the PIL. The yellow MFLs correspond to the topology of a 3D null point, while the MFLs in blue and green represent the remote connetivities P0--N and P1--N as earlier marked in Figure \ref{f3:hmi_mag}. Panel (c) depicts the values log $Q$ in the $y-z$ plane passing through the 3D null. Panel (d) overlays the values of log $Q$ between 1 and 5 which helps us to identify different regions of connectivity on the bottom boundary. The red, green and blue arrows represent the $x$, $y$ and $z$ directions respectively.}
\label{f5:initial_field_overview}
\end{figure}
The MFL topology of the extrapolated field is shown in Figure \ref{f5:initial_field_overview} where the field lines are plotted in red, purple, yellow, green, and blue. In this and all the subsequent figures, the arrows in red, green, and blue denote the $x$, $y$, and $z$ axes, respectively.
The yellow MFLs resemble the topology of a 3D magnetic null \citep{lau&finn1990apj}, where  the  MFLs constituting the dome  intersect the bottom boundary to generate footpoints that are distributed in a circular pattern. The MFLs corresponding to the spine-axis of the null extend through the upper boundary and do not close in the domain.
A similar complex magnetic field topology is also suggested in \citet{janvier+2016} based on NLFFF modeling where the spine of the 3D null closes in the polarity P0. In contrast, \citet{jiang+2018apj}, using NLFFF extrapolations, find a different morphology with the 3D null spine axis connecting to the polarity P1. In our case, the field lines marked green which originate very close to the dome have a similar connectivity (P1 to N), while the field lines marked blue show the connectivity between P0 and N. The strongly sheared field lines connecting the main polarities of P2 and N are shown in red and overlying loops building their outer envelope are shown in purple. The difference in the field-line connectivity can be attributed to the key differences in the methods used to generate the extrapolated fields. For instance, \citet{jiang+2018apj} utilize the full vector magnetogram to obtain the coronal magnetic field, while the NLFFF model used in \citet{janvier+2016} is based on the flux rope insertion method which only requires a line-of-sight magnetogram. With different models, these studies were able to provide significant insight into the various aspects of a complex flaring event (as mentioned in Section {\ref{sec:intro}}). Therefore, it becomes imperative to analyze the results with different extrapolation models to obtain an in-depth understanding of the complex flaring processes.   

The field lines pertaining to the 3D null point are shown in greater detail in Figure \ref{f5:initial_field_overview}(c), where the values of the squashing factor are shown in the $y-z$ plane passing through the 3D null  point. The location of the null can be easily identified from the high values of the squashing factor ($Q$) \citep{liu+2016ApJ} shown here in logarithmic scale. The height of the null point is found to be approximately 25 Mm from the photosphere. Figure  \ref{f5:initial_field_overview}(d) overlays the values of $\log~Q$ (shown in the range 1  $\leq \log~Q \leq $  5) along with $B_z$ at the bottom boundary and the field lines shown in Figure~\ref{f5:initial_field_overview}(a) to highlight the different regions of connectivities of the MFLs.

\begin{figure}[ht!]
\gridline{
        \fig{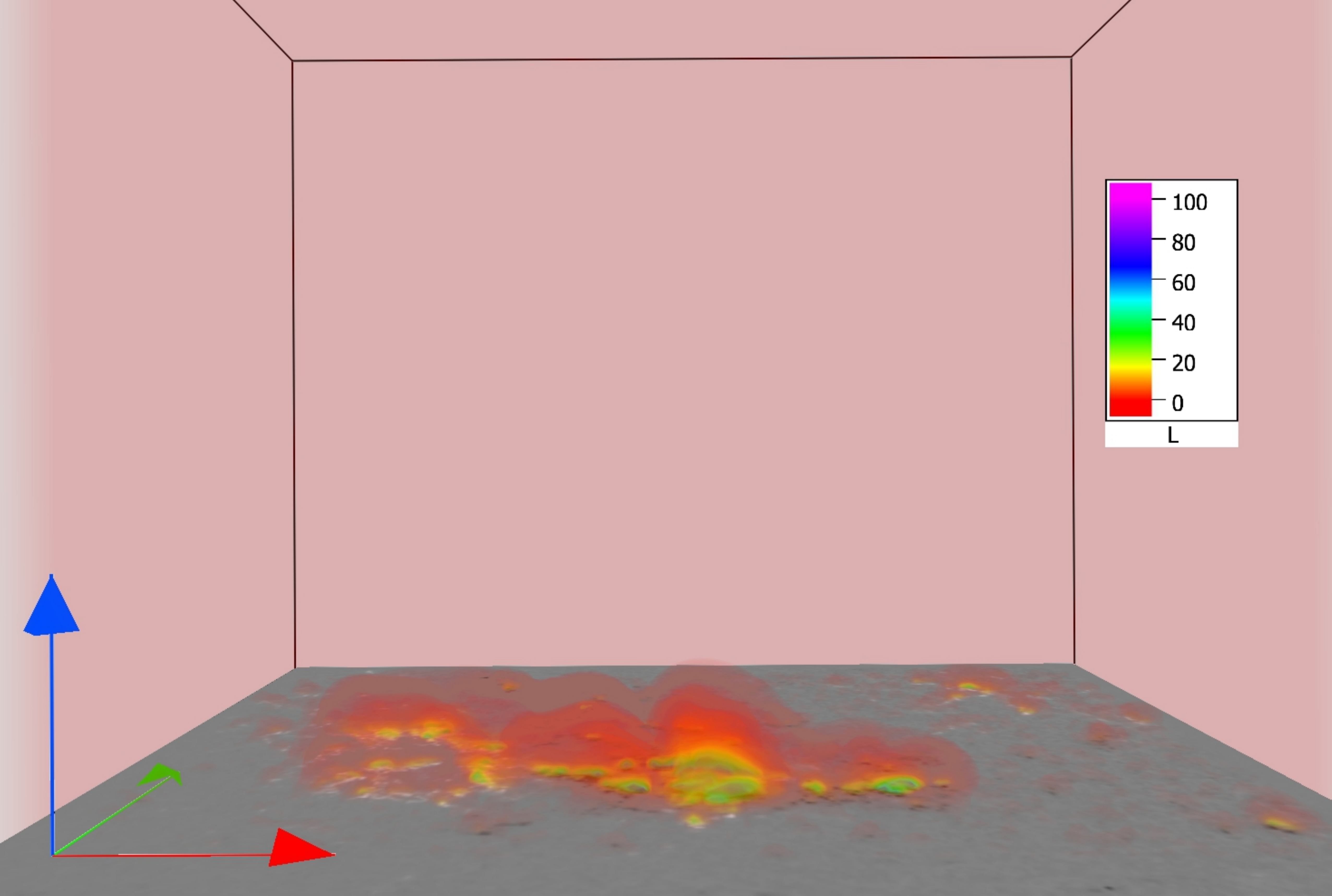}{0.48\textwidth}{(a)}
        %{lorentz_side.pdf}
        \fig{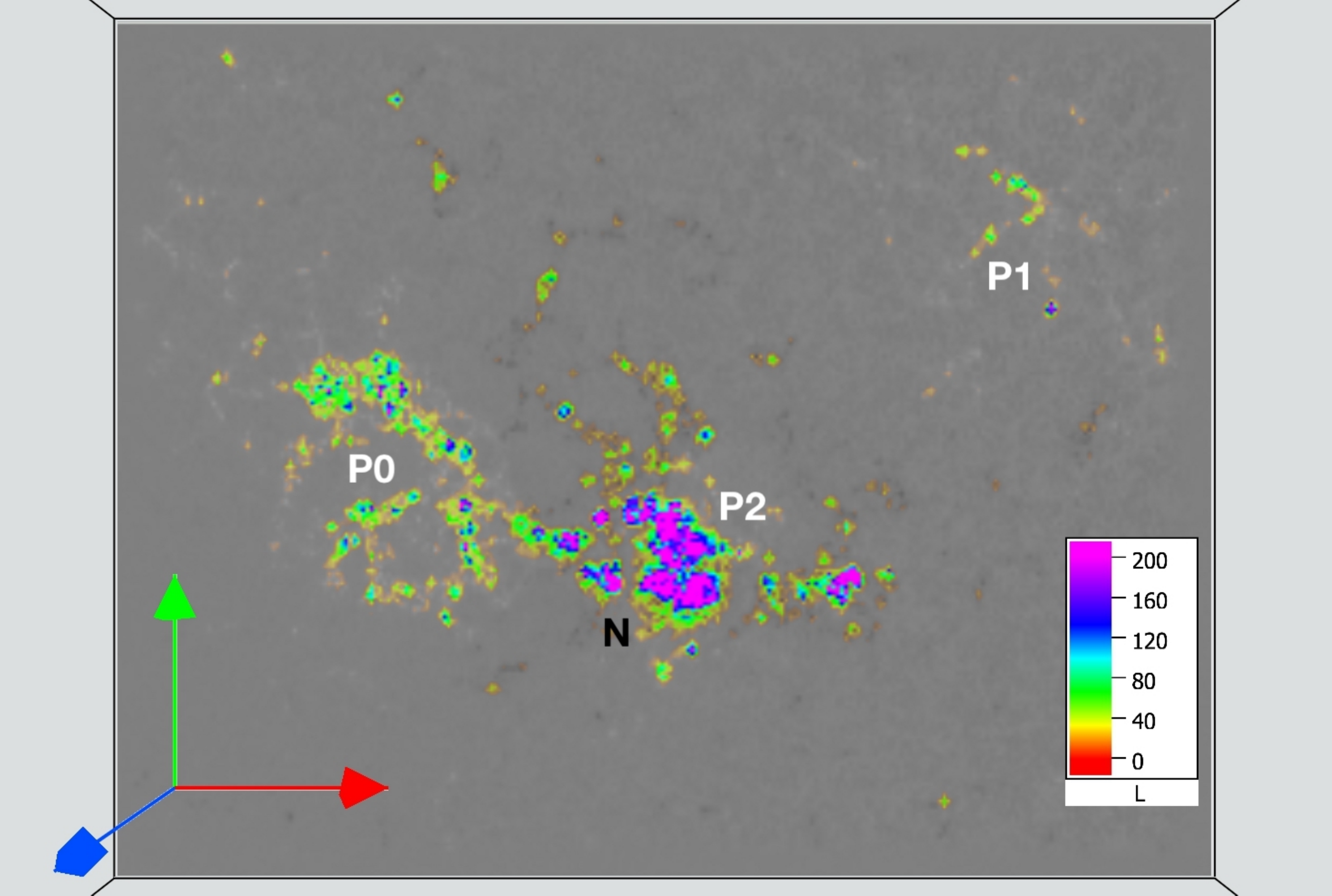}{0.48\textwidth}{(b)}
       %{lorentz_top.pdf}
        }
\caption{Side (a) and top (b) view of the distribution of the magnitude of the Lorentz force density in the computational domain for the initial extrapolated field. The figure clearly depicts the high values of the Lorentz force density near the central region and its exponential decrease in strength with height. Thus the Lorentz force is critical in driving the flows near the bottom boundary during the MHD evolution.}
\label{f6:lorentz_force}
\end{figure}

In Figure \ref{f6:lorentz_force}, the direct volume renderings of the Lorentz force density are illustrated from side and top views. Noticeably,  the regions of large Lorentz force density overlap with those of high values of $B_z$. The figure (along with the right panel of Figure \ref{f4:err_jbl}) also reveals a sharp decay of the Lorentz force density with height, making the magnetic field force-free in the asymptotic limit as previously indicated in Figure \ref{f4:err_jbl}(b).
Figure \ref{f6:lorentz_force}(b) clearly identifies the presence of strong Lorentz force  between polarities P2 and N and the corresponding PIL. Importantly, the Lorentz force plays a central role in driving the simulated evolution that is favorable to initiate the flare.

\begin{figure}[ht!]
\gridline{
        \fig{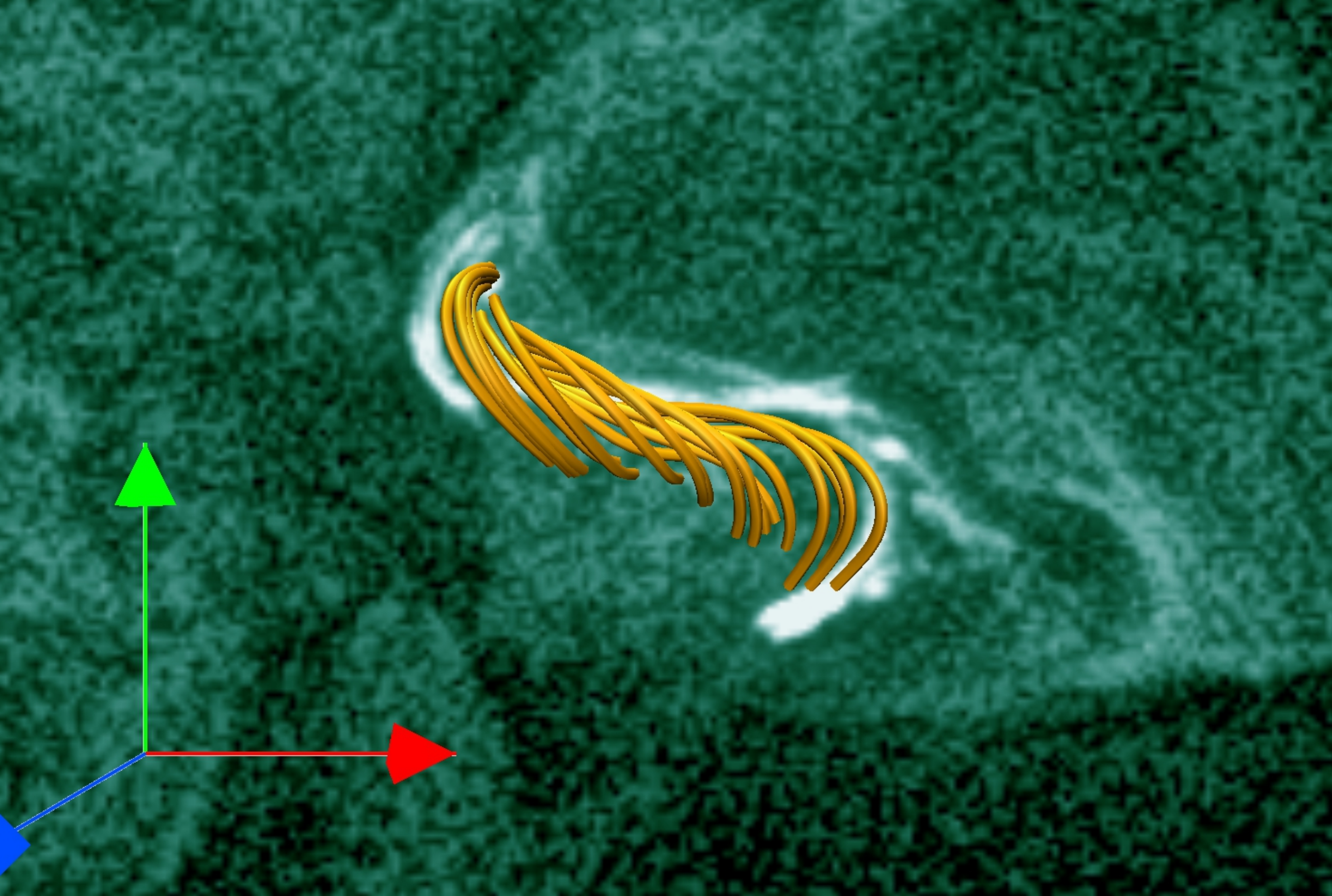}{0.48\textwidth}{(a)}
        %{sigmoid.pdf}
        \fig{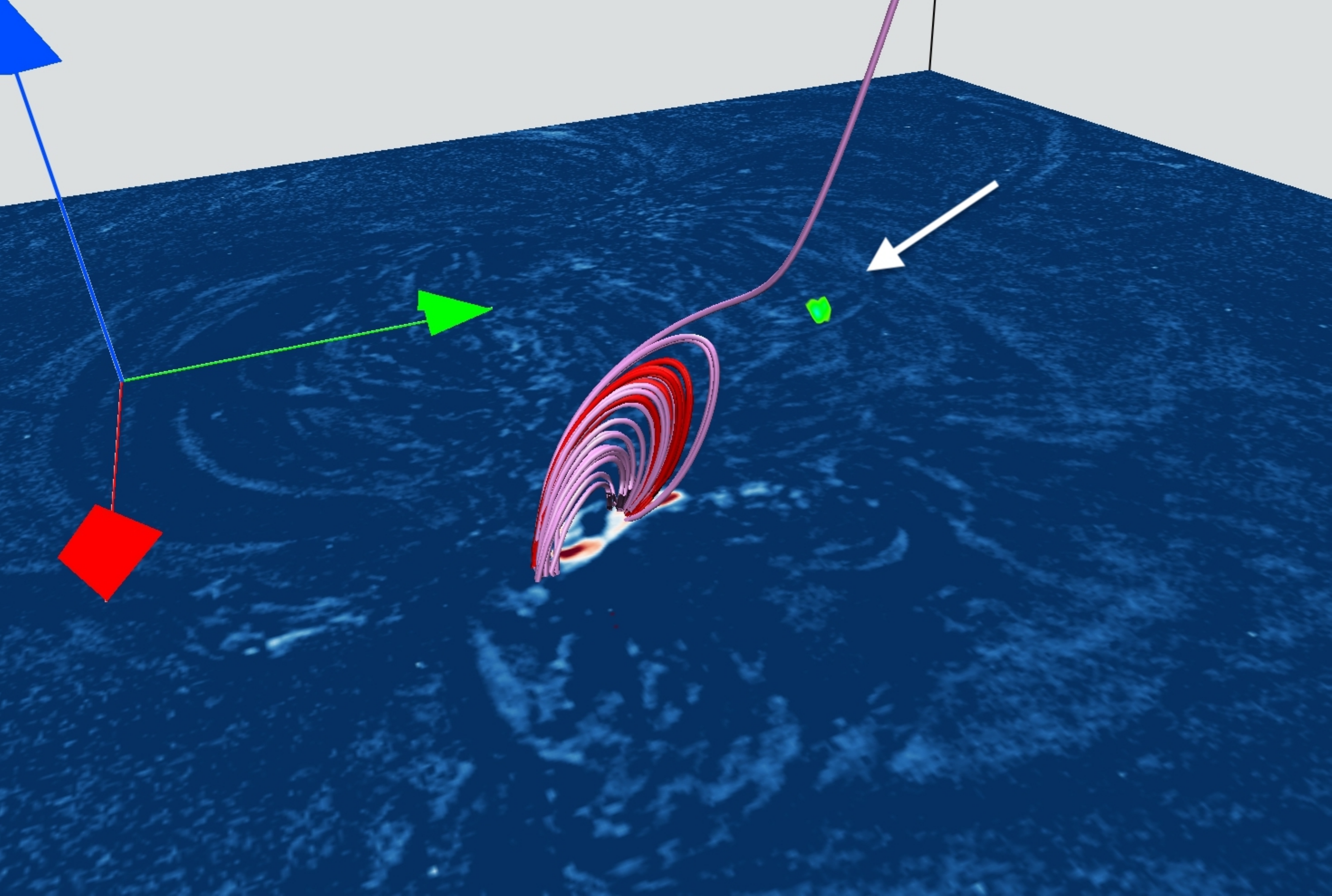}{0.48\textwidth}{(b)}
        %{compare_fig2a.jpg}
        }
\caption{ Panel (a) shows the hot sigmoid in SDO/AIA 94 \AA~ (Figure \ref{f1:event_overview}(c)) together with the highly sheared orange field lines from the extrapolation. Panel (b) shows the small-scale, bipolar pre-flare dimming (Figure \ref{f2:dimming_evolution}(a)) in good correspondence with the outer envelope (purple) of the flux rope (red). Panel (b) is further overlaid with the 3D null depicted by a green spot (also marked by a white arrow).  }
\label{f7:pre_flare_sigmoid}
\end{figure}

To relate the extrapolated field with the observational features, in Figure \ref{f7:pre_flare_sigmoid}(a), we plot field lines in orange which correspond to the sigmoidal brightenings as seen in Figure \ref{f1:event_overview}(c) at $t=0$ corresponding to 22:00 UT. We note a good correspondence between the brightenings observed in SDO/AIA 94 \AA~channel and the field lines shown in orange. The highly sheared nature of the field lines indicates the presence of strong field-aligned currents. The Joule heating of plasma due to dissipation of these currents may explain the EUV and X-ray emissions which lead to the appearance of the sigmoid \citep{jiang+2013apjl}.
Figure~\ref{f7:pre_flare_sigmoid}~(b) shows the extrapolated field that originates at the pre-flare dimming locations (cf.~Figure~\ref{f2:dimming_evolution}~(a)). As will be discussed in more detail in Section~\ref{sec:mhd}, the red field lines correspond to the flux rope, while the purple field lines indicate its outer envelope.

%

%------
\section{MHD Simulation for AR 11283 and X2.1 flare}\label{sec:mhd}
\subsection{Governing MHD equations and EULAG-MHD numerical model}

The presented dynamical evolution of the coronal plasma is governed by the incompressible Navier-Stokes MHD equations under the assumption of thermal homogeneity and perfect electrical 
conductivity  \citep{bhattacharyya+2010phpl,kumar+2014phpl,kumar+2015phpl}. The relevant MHD equations in dimensionless form are: 

\begin{subequations}
\begin{align}
\label{stokes}
&  \frac{\partial{\bf{v}}}{\partial t} 
+ \left({\bf{v}}\cdot\nabla \right) {\bf{ v}} =-\nabla p
+\left(\nabla\times{\bf{B}}\right) \times{\bf{B}}+\frac{\tau_a}{\tau_\nu}\nabla^2{\bf{v}},\\  
\label{incompress1}
&  \nabla\cdot{\bf{v}}=0, \\
\label{induction}
&  \frac{\partial{\bf{B}}}{\partial t}=\nabla\times({\bf{v}}\times{\bf{B}}), \\
\label{solenoid}
 &\nabla\cdot{\bf{B}}=0, 
\end{align}  
\label{e:mhd}
\end{subequations}
written in usual notations. The various variables in Equations \eqref{e:mhd} are normalized as follows
%---------
\begin{equation}
\label{norm}
{\bf{B}}\longrightarrow \frac{{\bf{B}}}{B_0},\quad{\bf{v}}\longrightarrow \frac{\bf{v}}{v_a},\quad
 L \longrightarrow \frac{L}{L_0},\quad t \longrightarrow \frac{t}{\tau_a},\quad
 p  \longrightarrow \frac{p}{\rho {v_a}^2}. 
\end{equation}

%-------
The constants $B_0$ and $L_0$ are fixed using the average magnetic field strength and length-scale of the vector magnetogram respectively. Here, $v_a \equiv B_0/\sqrt{4\pi\rho_0}$ is the Alfv\'{e}n speed and $\rho_0$ is the constant mass density. The constants $\tau_a$ and $\tau_\nu$ represent the Alfv\'{e}nic transit time ($\tau_a=L_0/v_a$) and viscous dissipation time scale ($\tau_\nu= L_0^2/\nu$), respectively, with $\nu$ being the kinematic viscosity. Notably, the simplified choice of incompressibility (Equation \ref{incompress1}) leads to the volume preserving flow --- an assumption routinely used in other works \citep{dahlburg+1991apj,aulanier+2005aa}. While compressibility plays an important role in the thermodynamics of
coronal loops \citep{ruderman&roberts2002apj}, in this work, our focus is on the changes in magnetic topology idealized with a thermally homogeneous magnetofluid. Utilizing the discretized incompressibility constraint, the pressure perturbation, denoted by $p$,  satisfies an elliptic boundary value problem on the discrete integral form of the momentum equation (Equation \ref{stokes}); cf.\citep[][and the references therein]{bhattacharyya+2010phpl}.

The MHD Equations (\ref{stokes})-(\ref{solenoid}) are solved  utilizing the well established magnetohydrodynamic numerical model EULAG-MHD \citep{smolarkiewicz&charbonneau2013jcoph}. The model is an extension of the hydrodynamic model EULAG predominantly used in atmospheric and climate research \citep{prusa2008cf}. Here we discuss only important features of the EULAG-MHD and refer the readers to \citet{smolarkiewicz&charbonneau2013jcoph} and references therein for detailed discussions. The model is based on the spatio-temporally second-order accurate non-oscillatory forward-in-time multidimensional positive definite advection transport algorithm, MPDATA \citep{smolarkiewicz2006ijnmf}.  
Importantly, MPDATA has the proven dissipative property which, intermittently and adaptively, regularizes the under-resolved scales by simulating magnetic reconnections 
and mimicking the action of explicit subgrid-scale turbulence models \citep{margolin+2006jtb} in the spirit of
Implicit Large Eddy Simulations (ILES) \citep{grinstein2007book}. Such ILESs conducted with the model have already been successfully utilized to simulate reconnections to understand their role in the coronal dynamics \citep{prasad+2017apj,prasad+2018apj,nayak+2019apj}. In this work, the simulation continues to rely on the effectiveness of ILES in regularizing the onset of magnetic reconnections.
%-------------------

\subsection{Numerical setup}
\label{s:results}
The simulation is performed in a computational domain having $256\times200\times 200$ grid points which resolve a physical domain spanning $[0,1.28]\times [0,1]\times [0,1]$ units, respectively, in $x$, $y$, and $z$, where an unit length is roughly equivalent to 290  Mm. A motionless state $(\mathbf{v}=0)$ with the NFFF extrapolated magnetic field (Figure \ref{f5:initial_field_overview}(a)) is selected as an initial state for the simulation. Moreover, the magnetofluid is idealized to be thermally homogeneous and having perfect electrical conductivity. The mass density is set to $\rho_0=1$ and kinematic viscosity to $\nu = 0.0002$, in scaled units. The dynamics results from the initial Lorentz force which pushes the magnetofluid.
To ensure the net magnetic flux to be zero in the computational domain, all components of volume ${\bf{B}}$ except for $B_z$, are continued to the boundaries \citep{prasad+2018apj}.
At the bottom boundary, $B_z$ is kept constant (line-tied boundary condition). For the simulation, we  set the dimensionless constant $\tau_a / \tau_\nu \approx 3.5 \times 10^{-4}$, which is roughly two orders of magnitude larger than its coronal value. The higher value of $\tau_a / \tau_\nu$ speeds up the relaxation because of a more efficient viscous dissipation without affecting the magnetic topology.  The spatial unit step $\Delta x = 0.005$ and time step (normalized by the Alfv\'{e}n  transit time $\tau_a \sim 20s$) $\Delta t = 2\times10^{-3}$ are selected to satisfy the Courant-Friedrichs-Lewy (CFL) stability condition \citep{courant1967jrd}.
% there were 4 runs of nt =1000 and ntout = 20 => 200 frames, but we are using only first 100 frames.
The results presented here pertain to a run for 2000 $\Delta t$ which roughly corresponds to an observation time of 2 hours. For the sake of convenience in comparison with observations, we present the time in units of $20 \tau_a$ (which is close to a minute) in the discussions of the figures in the subsequent sections. Notably, the $R_M$ throughout  the simulation is set to infinity except during magnetic reconnections facilitated by the MPDATA driven dissipation. 

\subsection{Pre-flare stage and sigmoid to flux rope transition}
%In the event under study, small-scale, bipolar pre-flare coronal dimmings are observed $\approx$ 30 minutes before the flare onset (see Figure~\ref{f2:dimming_evolution}(a)). The MHD simulation revealed that they correspond to the outer envelope of the flux rope. It rises during the pre-flare phase leading to stretched field lines on the one hand and partly also reconnects with the overlying 3D null resulting in open field lines. 

\begin{figure}[ht!]
\gridline{
        %twist-transfer0000, 0010,0020,0030
        \fig{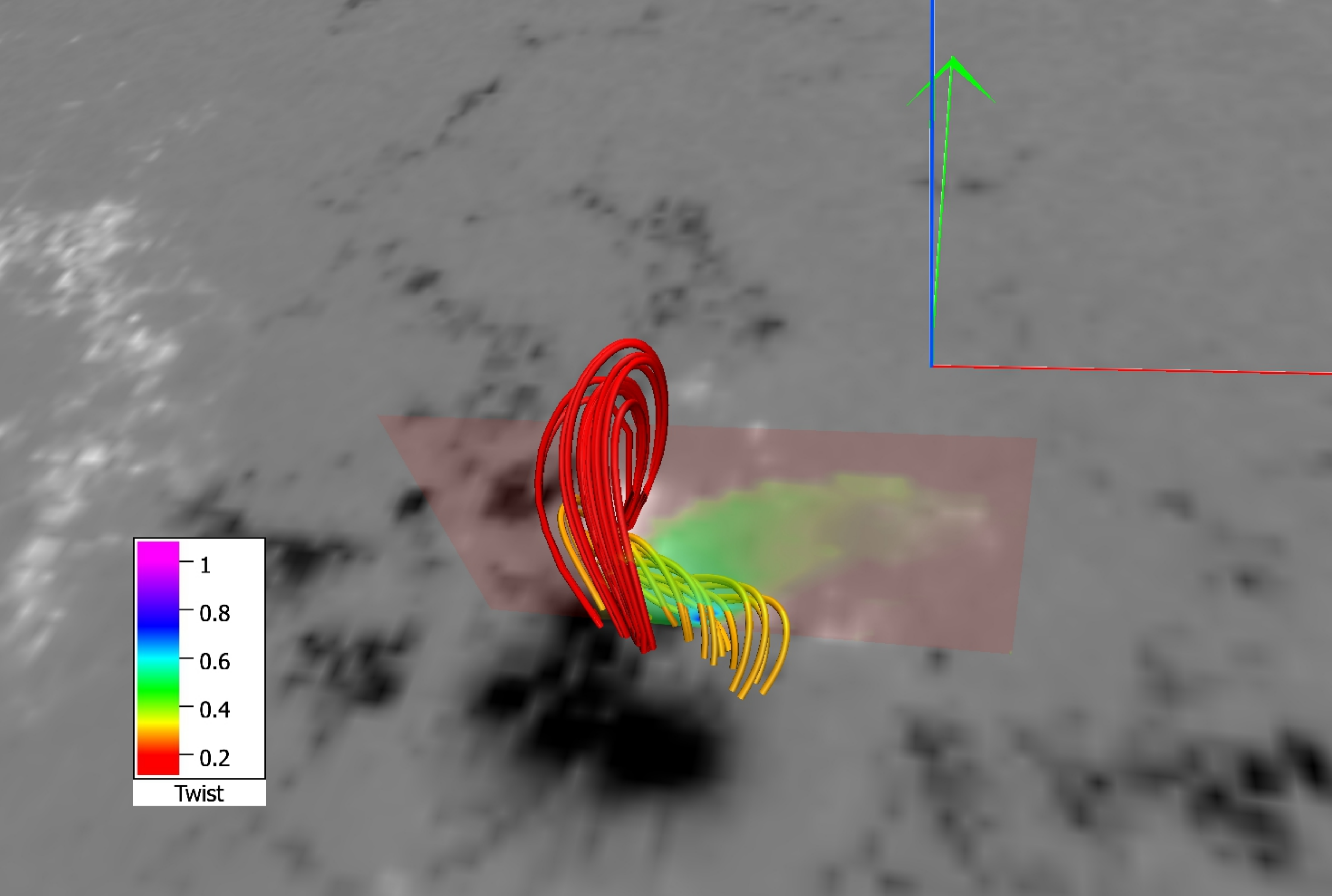}{0.48\textwidth}{(a. t=0)}
        \fig{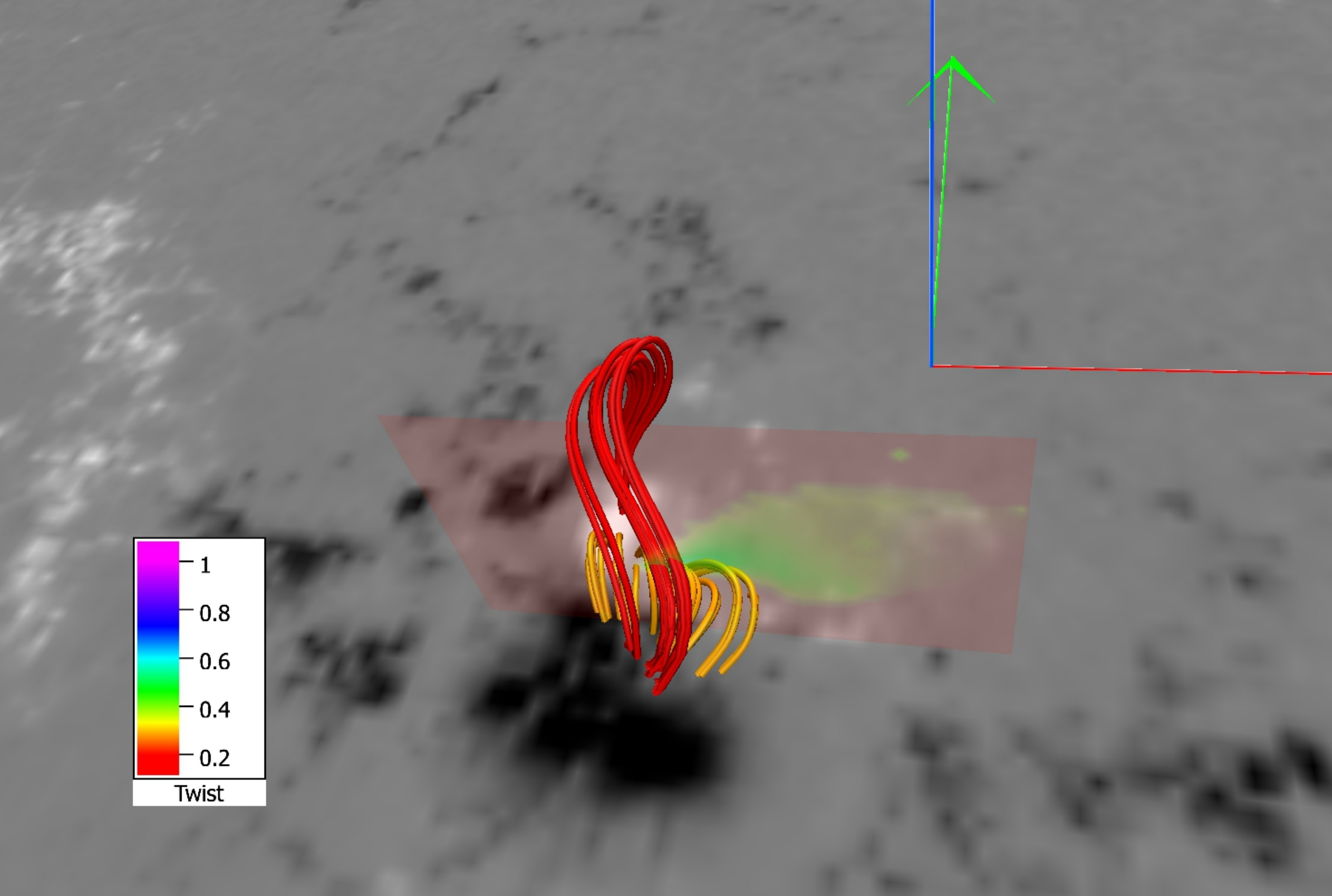}{0.48\textwidth}{(b. t=10)}
        }
\gridline{
        \fig{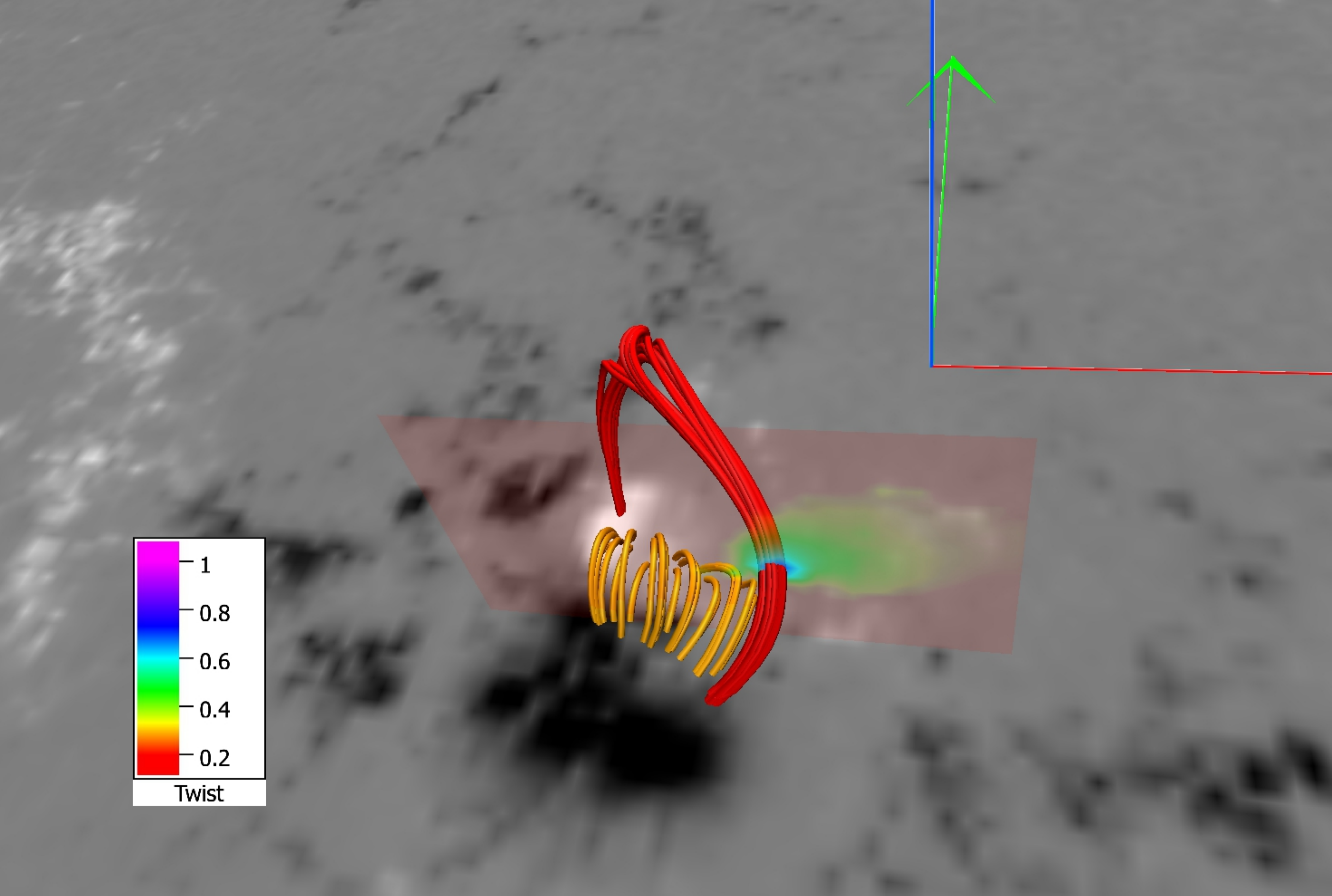}{0.48\textwidth}{(c. t=20)}
        \fig{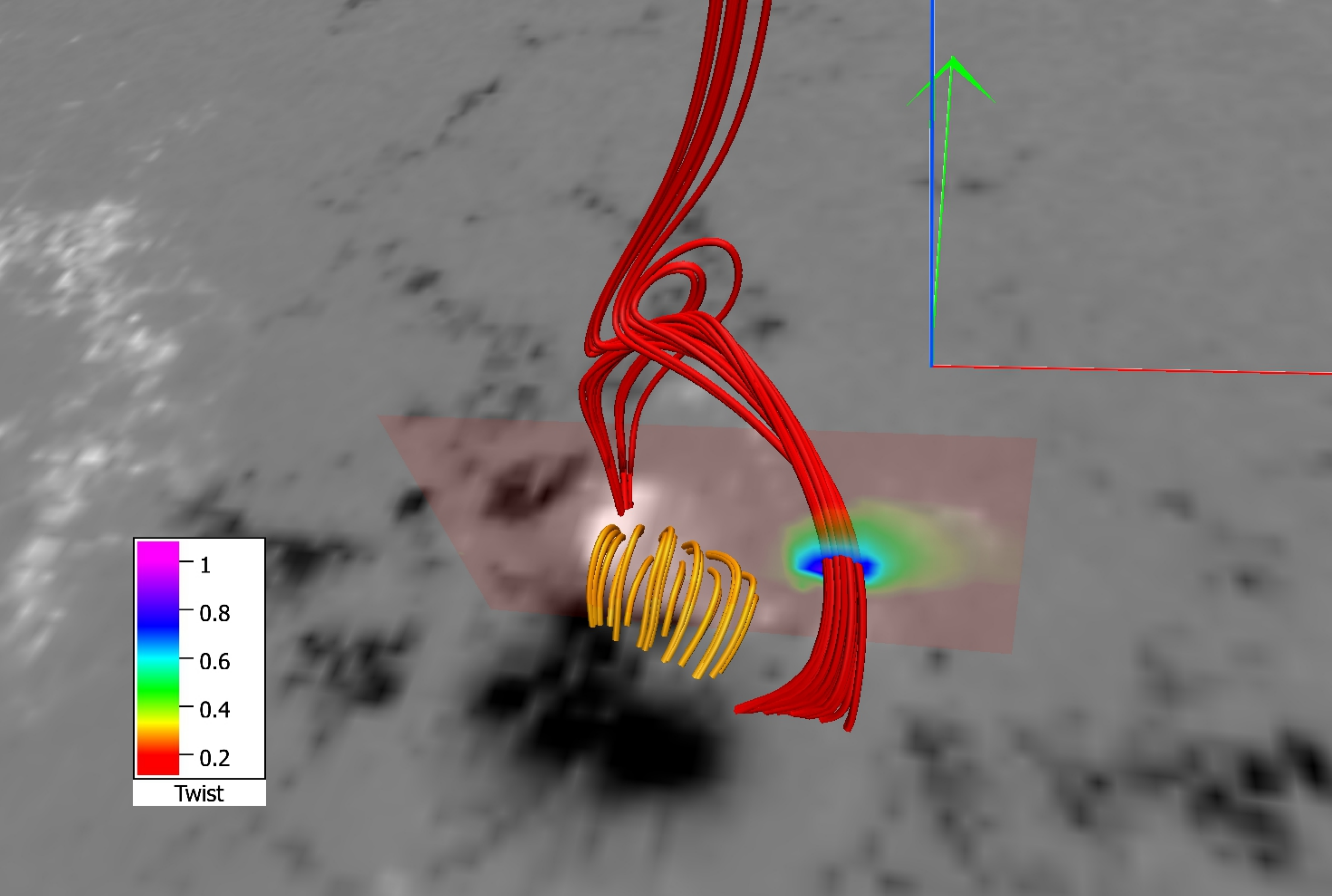}{0.48\textwidth}{(d. t=30)}
        }
\gridline{
        \fig{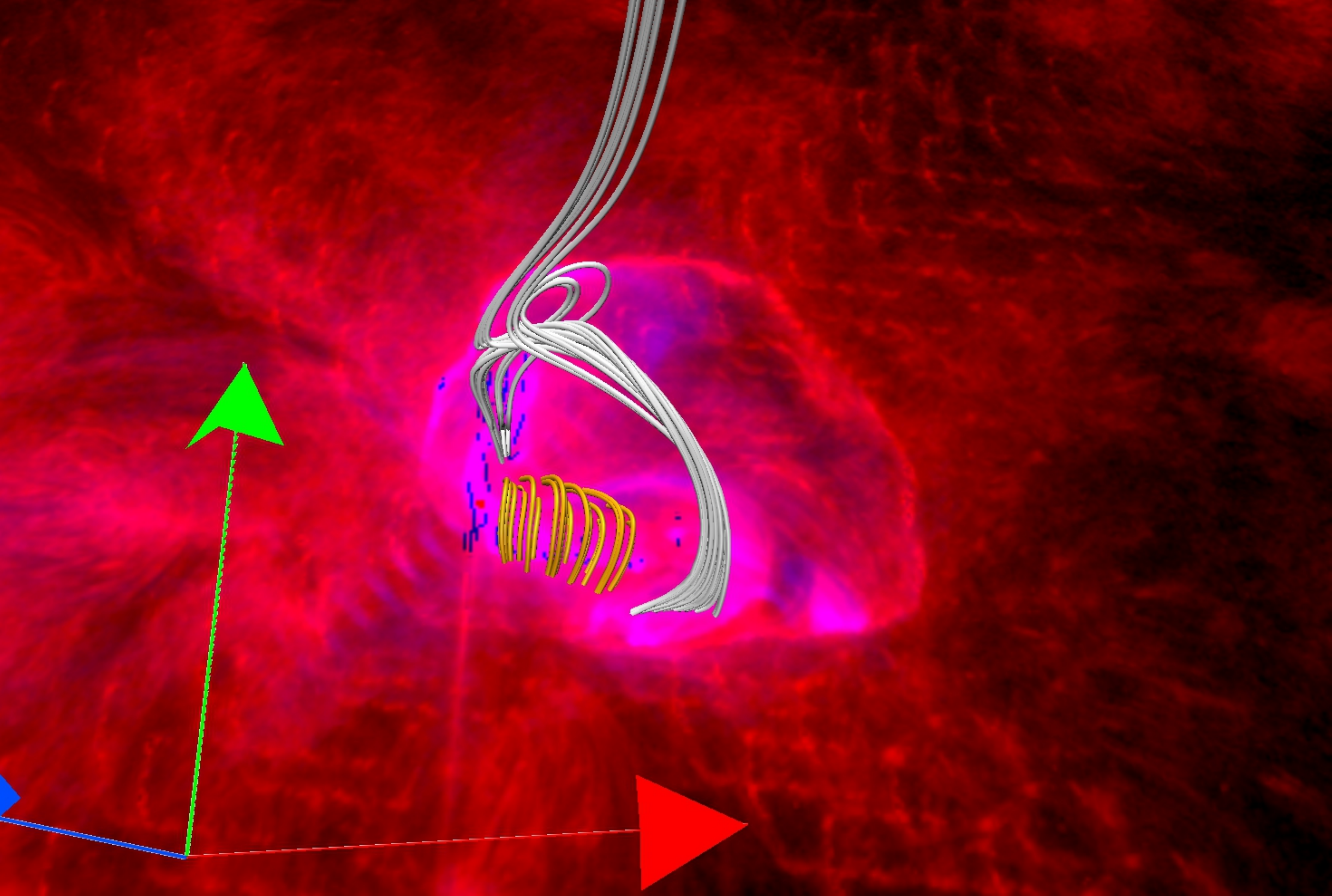}{0.48\textwidth}{(e. t=30)}
        %{20110906_304_94_precursor_late_t30.jpg
        \fig{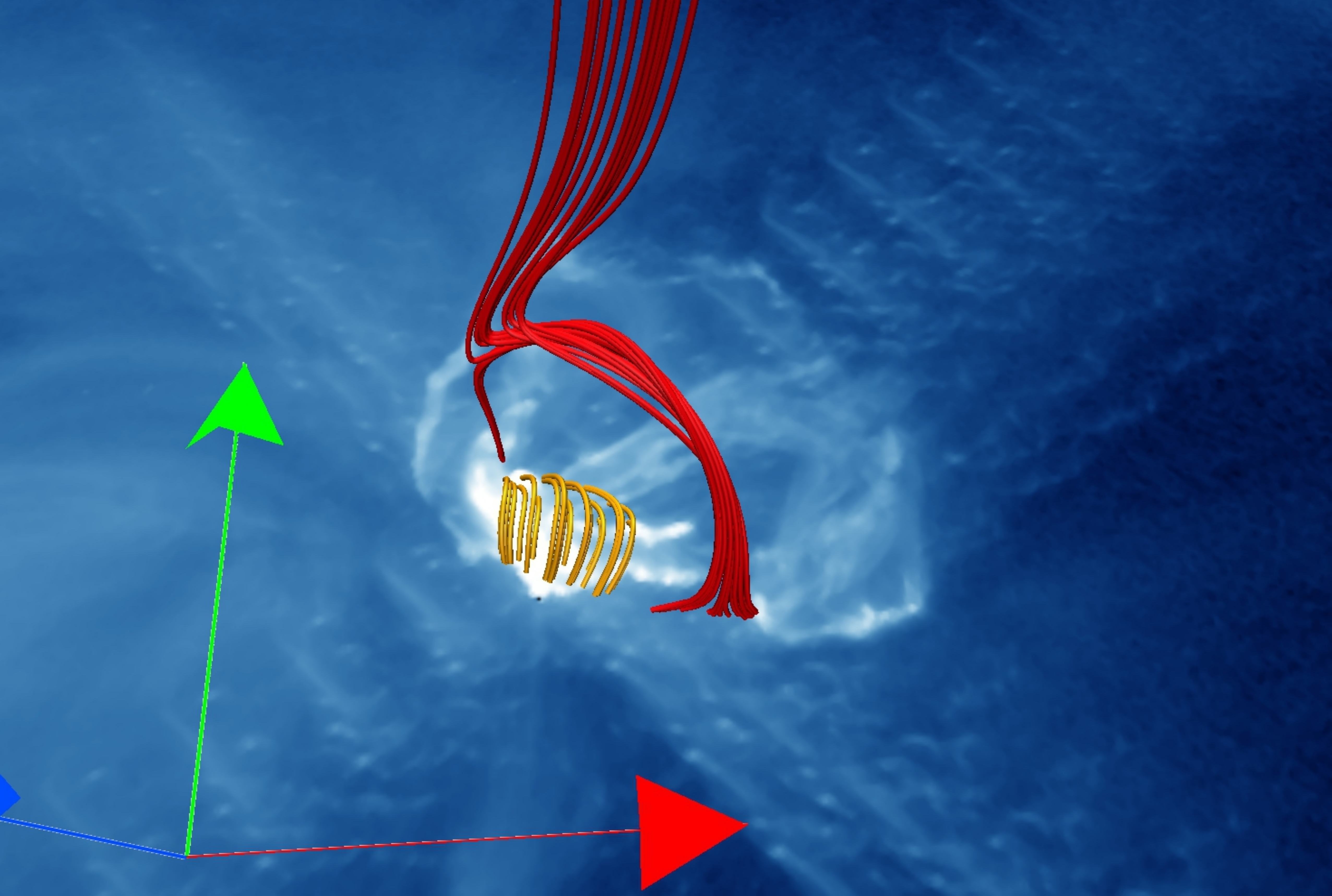}{0.48\textwidth}{(f. t=33)}
        }
        %20110906_335_fr1_t40.jpg
\caption{Panels (a-d) depict the transfer of twist from the underlying sigmoid (Figure \ref{f7:pre_flare_sigmoid}(a)) to the overlying flux rope through small-scale reconnections under the flux rope. 
The panels are overlaid with a vertical cross section of the magnetic twist number.
The orange MFLs can be observed to be almost potential by $t=30$, while the red MFLs are seen to become more twisted.  Panel~(d) also shows the bifurcation of the flux rope due to reconnections. In Panel~(e) the MFLs are overlayed with an SDO/AIA 304 and 94 \AA~composite image shortly after the flare onset ($\sim$~22:17~UT) and panel~(f) uses Figure~\ref{f1:event_overview}(h) as the bottom boundary. In particular, these panels clearly show the correspondence between the reconnection site and the localized brightening in 94 \AA~as well as the match between the footpoints of the erupting flux rope in 335 \AA~with that inferred from the simulations.
(An animation of this figure is available.)}
% \href{https://youtu.be/iIwLv0vB8bo}{Click here for the movie}.) 
\label{f8:mhd_sig2rope}
\end{figure}

To understand the dynamics of the pre-flare stage of this event, we first focus on the formation of the pre-flare dimming and the evolution of initially highly sheared MFLs, representing the sigmoidal brightening situated over the PIL (Figure \ref{f7:pre_flare_sigmoid}(a)).
Magnetic field lines that originate at the pre-flare dimming location (Figure~\ref{f2:dimming_evolution}(a)) are shown in Figure~\ref{f7:pre_flare_sigmoid}(b) in purple color, which also represent the outer envelope of a flux rope (shown in red here, and in Figure \ref{f8:mhd_sig2rope}) which forms during the evolution. 
Overall, the simulation reveals
two mechanisms causing the formation of the pre-flare dimming region. On one hand, the outermost magnetic field lines of the flux rope reconnect at the site of the 3D null point, which leads to the opening of closed field lines and results in dimming at their corresponding footpoints. 
This is discussed in more detail later in relation to Figure \ref{f10:mhd_reconnections}.
On the other hand, the outer envelope of the flux rope rises due to the initial Lorentz force and the observed dimming signature is a result of the stretching and expansion of these field lines.
In general, pre-flare coronal dimmings are observed $\sim$30--90  minutes  before  the  flare  onset \citep{qiu+2017apj,zhang+2017aa}.
There is a growing consensus that overlying fields are stretched due to the gradual and slow rise of the flux rope, prior to the formation of the current sheet \citep{joshi+2016ApJ,sahu+2020ApJ}.  Expanding fields  manifest  as an intensity decrease in extreme-ultraviolet emission,  i.e.  transient pre-flare coronal dimmings \citep{forbes+2000}. Hence, our  model  shows  the  possible magnetic configuration which is likely related to the dimming sites.
Investigating the intensity distribution within the pre-flare dimming in Figure~\ref{f7:pre_flare_sigmoid}(b), we speculate that red regions, indicating regions of the strongest intensity decrease, correspond to field lines that opened-up, while light blue to white regions, i.e.~areas of small to moderate intensity changes are formed as a result of field lines that expanded.

 Figure \ref{f8:mhd_sig2rope} depicts the transfer of twist from the underlying sigmoid (in orange) to the overlying sheared field lines (in red). 
 Consequently, at $t \approx 20$,  a magnetic flux rope is formed, depicted by red field lines in Figure \ref{f8:mhd_sig2rope}(c). In order to clearly identify the flux rope, we show the twist number for the field lines in panels (a)-(d) of the figure. Panel (c) shows that the flux rope has field lines with twist number close to 1 turn. The low-lying field lines become almost perpendicular to the bottom PIL, indicating that they are close to potential field (Figure \ref{f8:mhd_sig2rope}(d)).  Notably, such a transfer of twist from the sigmoidal MFLs to the overlying MFLs indicates the occurrence of magnetic reconnections which can contribute to pre-flaring activities as well as to the formation of a flux rope.
However, because of the computational constraints, we could not resolve these reconnections. Further, we notice that the negative-polarity footpoint of the flux rope undergoes a significant movement to the right.
 Observations confirm the shift of the right footpoint of the flux rope as well as magnetic reconnections along the sigmoid in the form of small-scale brightenings (cf.~Figure~\ref{f1:event_overview}(d)).
 Moreover, the magnetic reconnections also initiate in field lines comprising the outer envelope of the flux rope (Figure \ref{f8:mhd_sig2rope}(d)) which is discussed in more detail in Section \ref{subsec:reconnections}. %Furthermore, in Figure~\ref{f8:mhd_sig2rope}(e-f), \ap{to update: we overlay the bottom boundary with the panels Figure~\ref{f1:event_overview}(d) and Figure~\ref{f1:event_overview}(h) respectively. We choose a later time step that it shown in 1d.}
 Figure~\ref{f8:mhd_sig2rope}(e) highlights that the brightenings observed in the hotter 94 \AA~channel shortly after the impulsive flare onset ($\sim$22:17~UT) are co-spatial with the footpoints of the reconnecting field lines, manifesting a causal connection between the magnetic reconnections and these brightenings.  Figure \ref{f8:mhd_sig2rope}(f) shows a correspondence of the footpoints of the simulated rising flux rope to those identified in the observation of the SDO/AIA 335 \AA  ~channel
 (marked by red crosses in Figure \ref{f1:event_overview}(h)). Although  the  erupting  filament  footpoint  locations  match  quite  well with observation,  the corresponding erupting structure is not fully reproduced by the simulation.

\begin{figure}[ht!]
\gridline{
        \fig{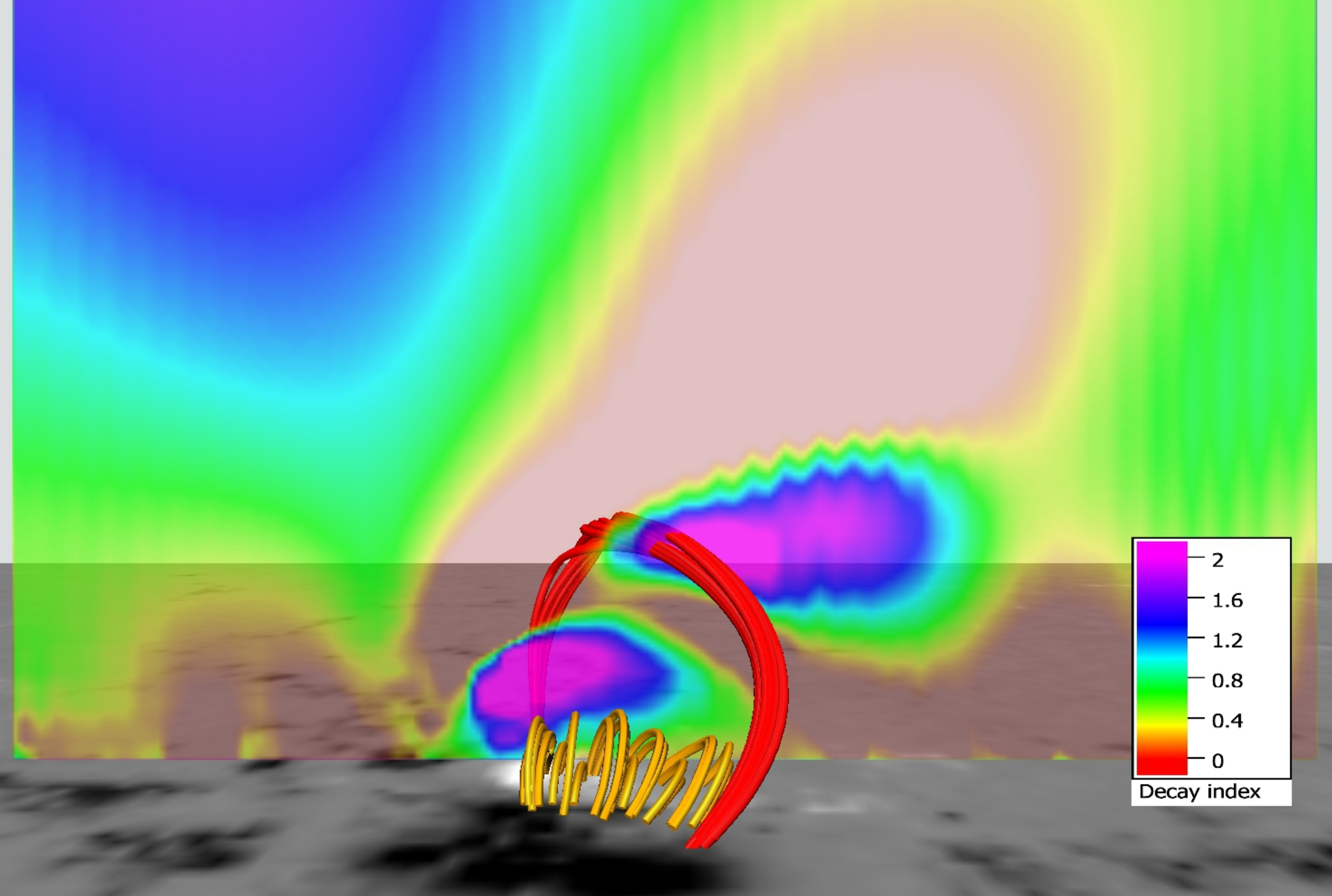}{0.48\textwidth}{(a. t=20)}
        \fig{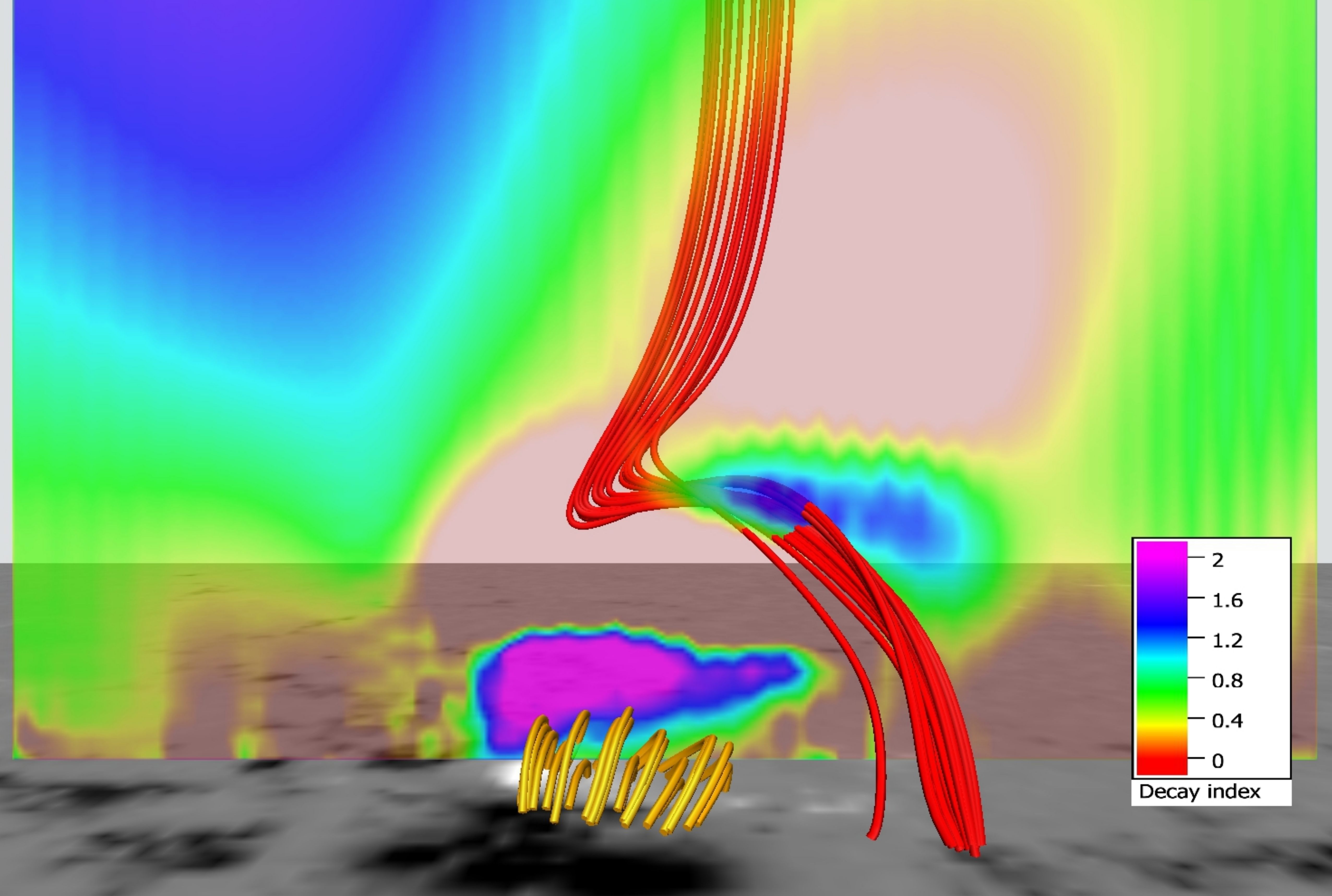}{0.48\textwidth}{(b. t=40)}
        }
\caption{Depiction of the dynamic rise of the flux rope between $t=20$ and $t=40$ as it starts reconnecting at one end. }
%The rise can be due to torus instability as the flux rope can be seen to be centered around a region with decay index $\geq$ 1.5, which is shown in the $y-z$ plane passing through the flux rope.}
\label{f9:decay_index}
\end{figure}

Noticeably, with a potential-field-like configuration of the lower-lying  orange MFLs at $t \approx 20$ in Figure \ref{f8:mhd_sig2rope}(c), the twist transfer and, hence, the magnetic reconnections between orange and red MFLs cease and the flux rope is fully developed. Subsequently, in absence of magnetic reconnections, the evolution of the rope appears to be governed by ideal MHD for the approximate time period $t ~\varepsilon ~\{20,30\}$. To explore the possibility of the torus instability \citep{kliem&torok2006prl}, in
Figure \ref{f9:decay_index}, we show snapshots of the decay index in the $y-z$ plane passing through the flux rope, which measures the decay of the external field, superposed with the flux rope. Following \citet{jiang+2016nat}, the decay index is defined as $n= - d\log(B)/d\log(h)$, where $h$ is the height and $B$ is the strength of the overlying strapping field.
Figure \ref{f9:decay_index} illustrates the rise of the flux rope between $t=20$ and $t=40$.
%{The decay index varies sharply from 0 to 2 in the vicinity of the rope which makes it inconclusive to infer the role  of torus instability in the rise and calls for a more refined calculation of decay index \citep[see,][]{duan+2019ApJ}. Alternatively, in this case, the torus instability may not play a role in the rise, because the  rope is already in a dynamic phase after the reconnections.}
Notably, at $t=20$ the decay index is 2 in the vicinity of the flux rope. However, just above the flux rope, the decay index sharply decreases to around 0, suggesting the absence of a role of the torus instability in the rise \citep{zhou+2017ApJL,duan+2019ApJ}. Consequently, in this case, the rise of the rope seems to be naturally commenced,  as the flux rope is already in a dynamic phase after the reconnections. 
%in the vicinity of the rope, satisfying the required criteria ($n> 1.5$) to make the rope torus unstable.
%As a result, the rope shows a rise (see panel b).   

\subsection{Flaring stage: reconnections at the 3D null and the X-type MFLs}
\label{subsec:reconnections}
\begin{figure}[ht!]
\gridline{
        %reconnection0000,0010,0020,0030,0040,0050
        \fig{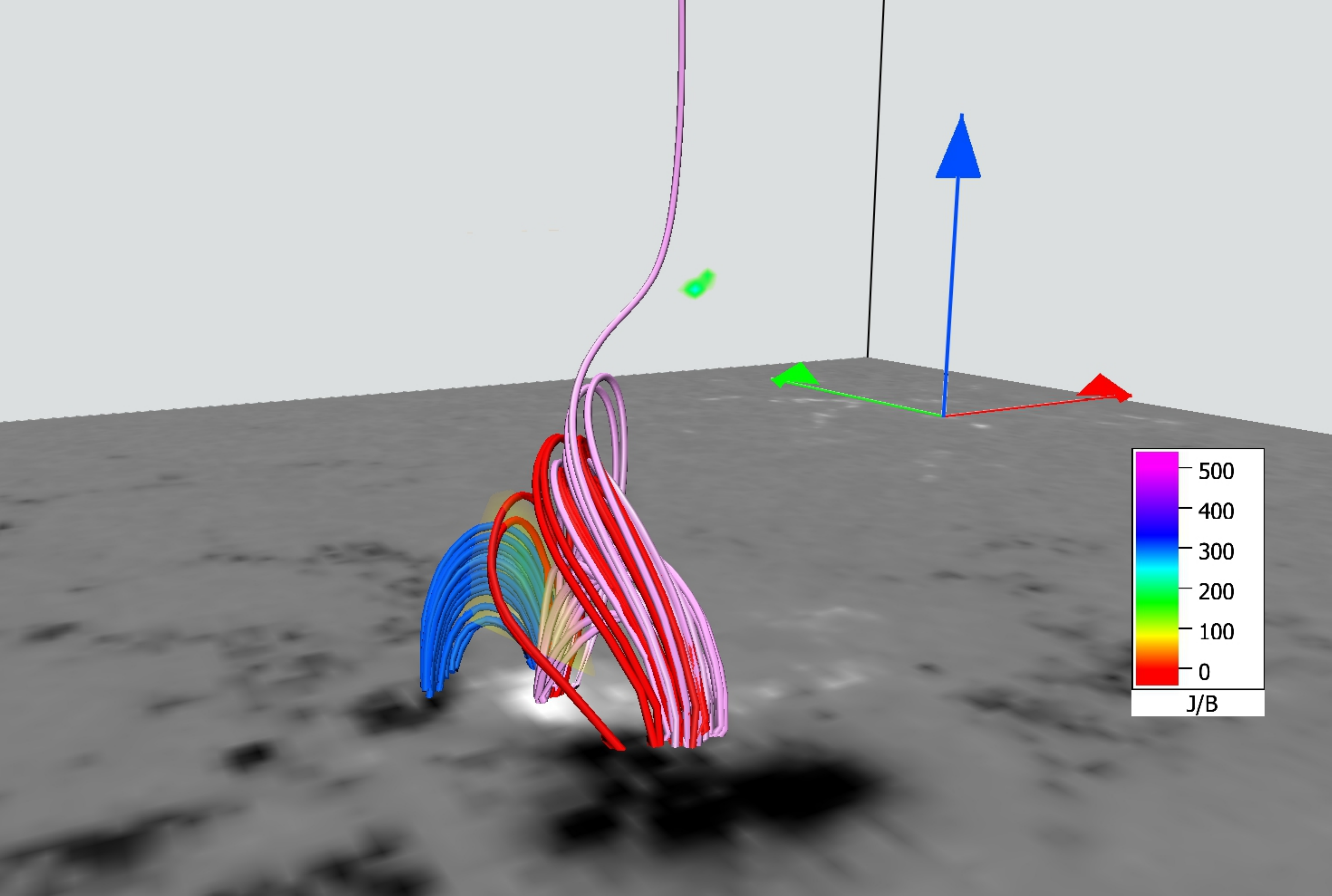}{0.48\textwidth}{(a. t=0)}
        \fig{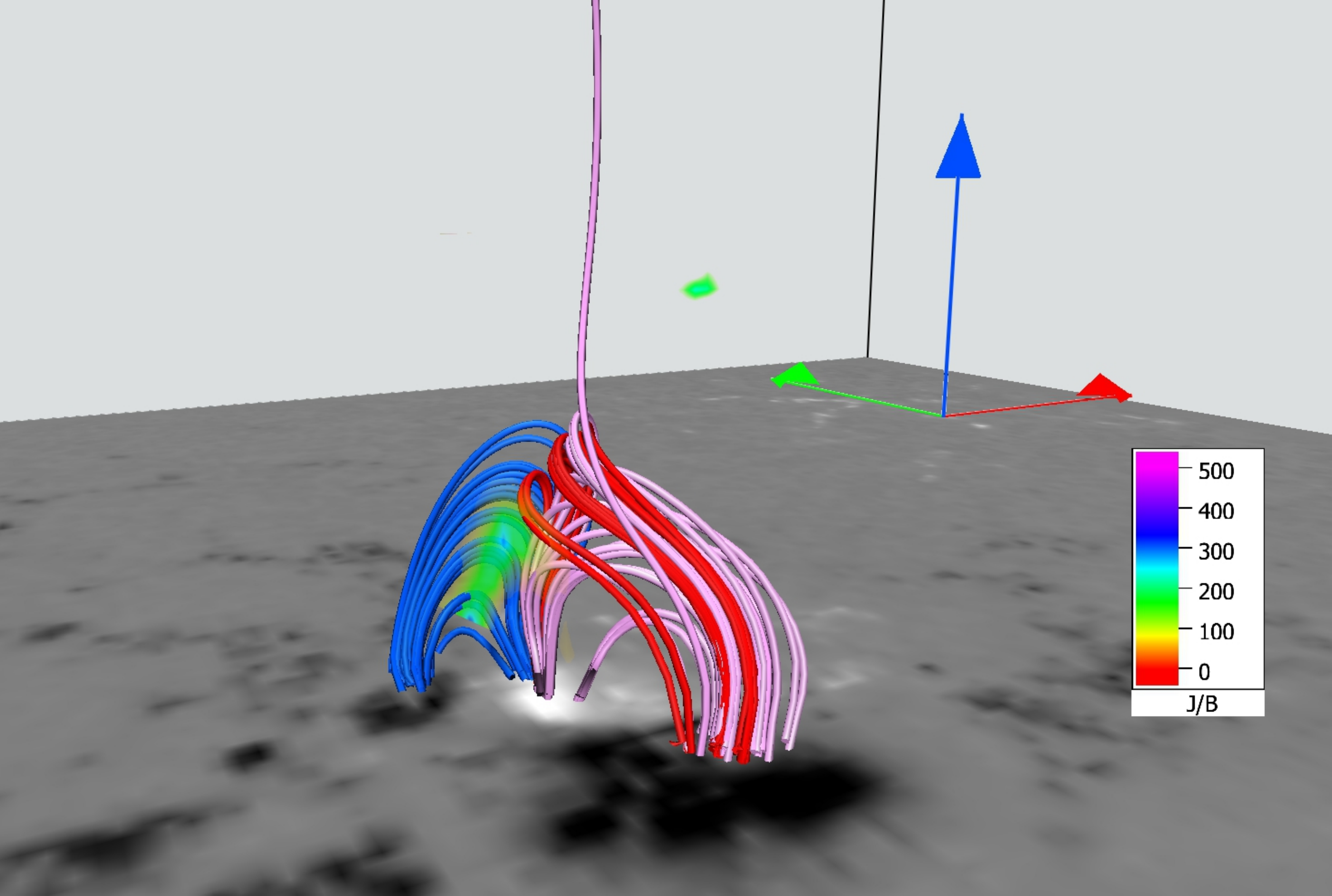}{0.48\textwidth}{(b. t=10)}
        }
\gridline{
        \fig{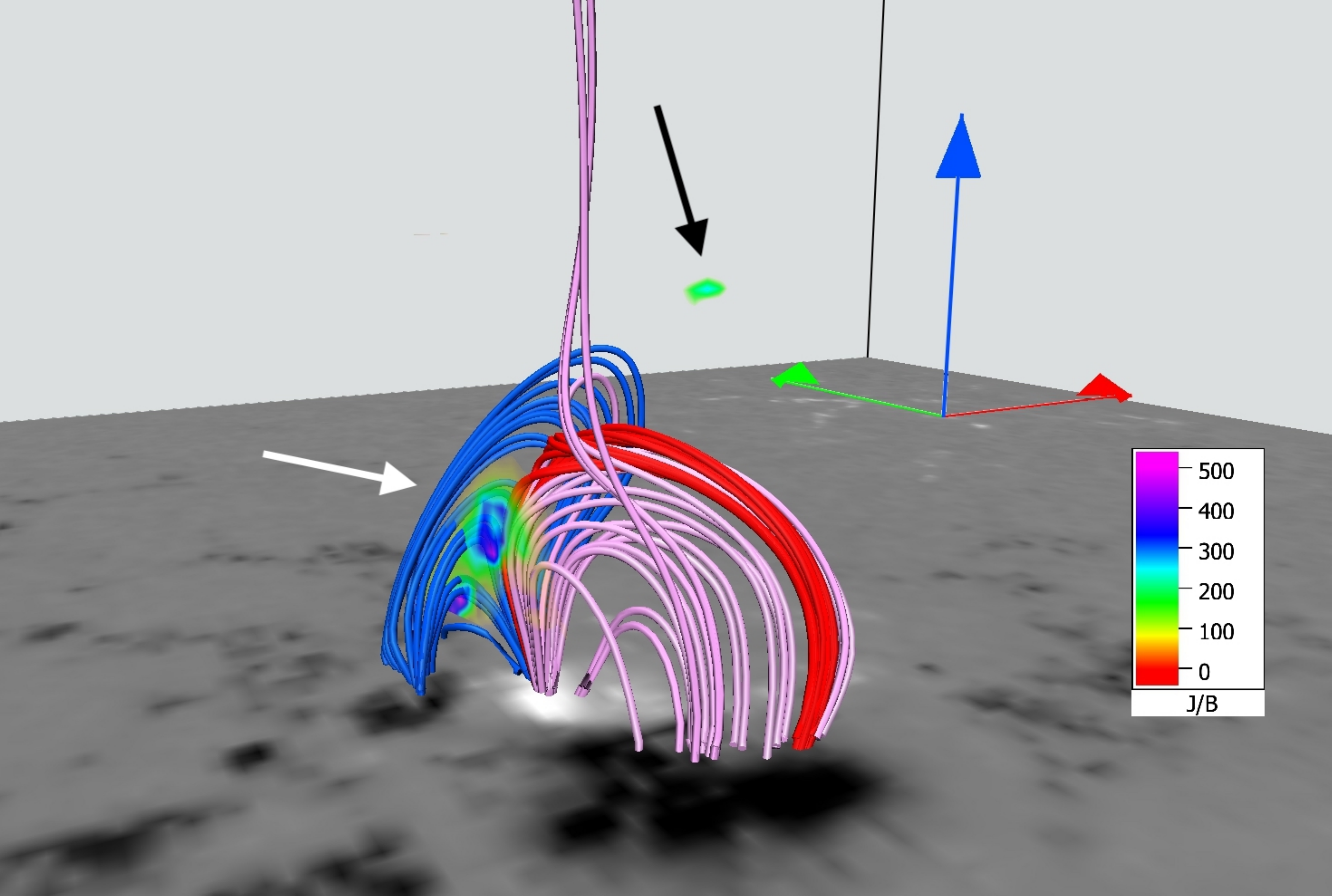}{0.48\textwidth}{(c. t=20)}
        \fig{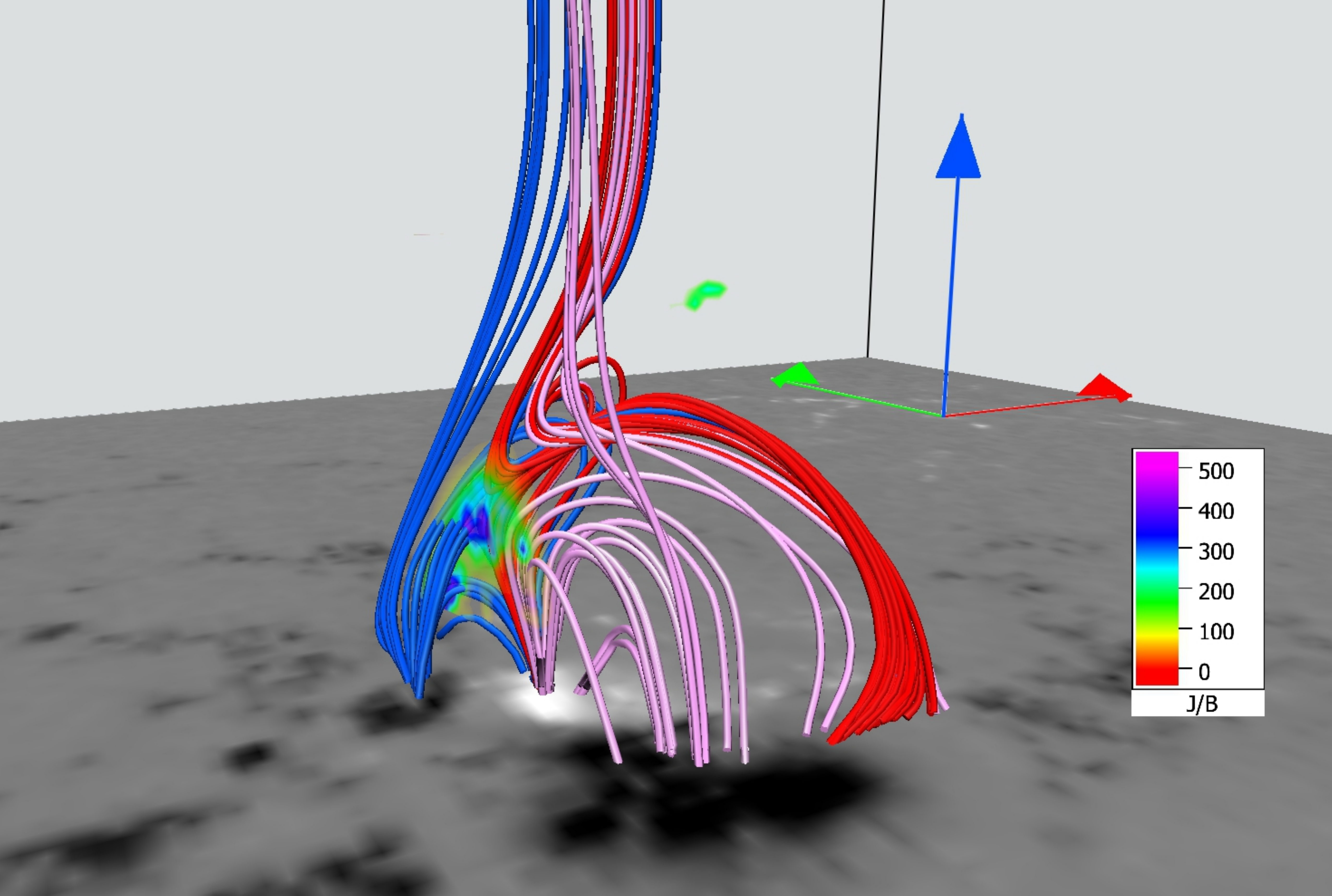}{0.48\textwidth}{(d. t=30)}
        }
\gridline{
        \fig{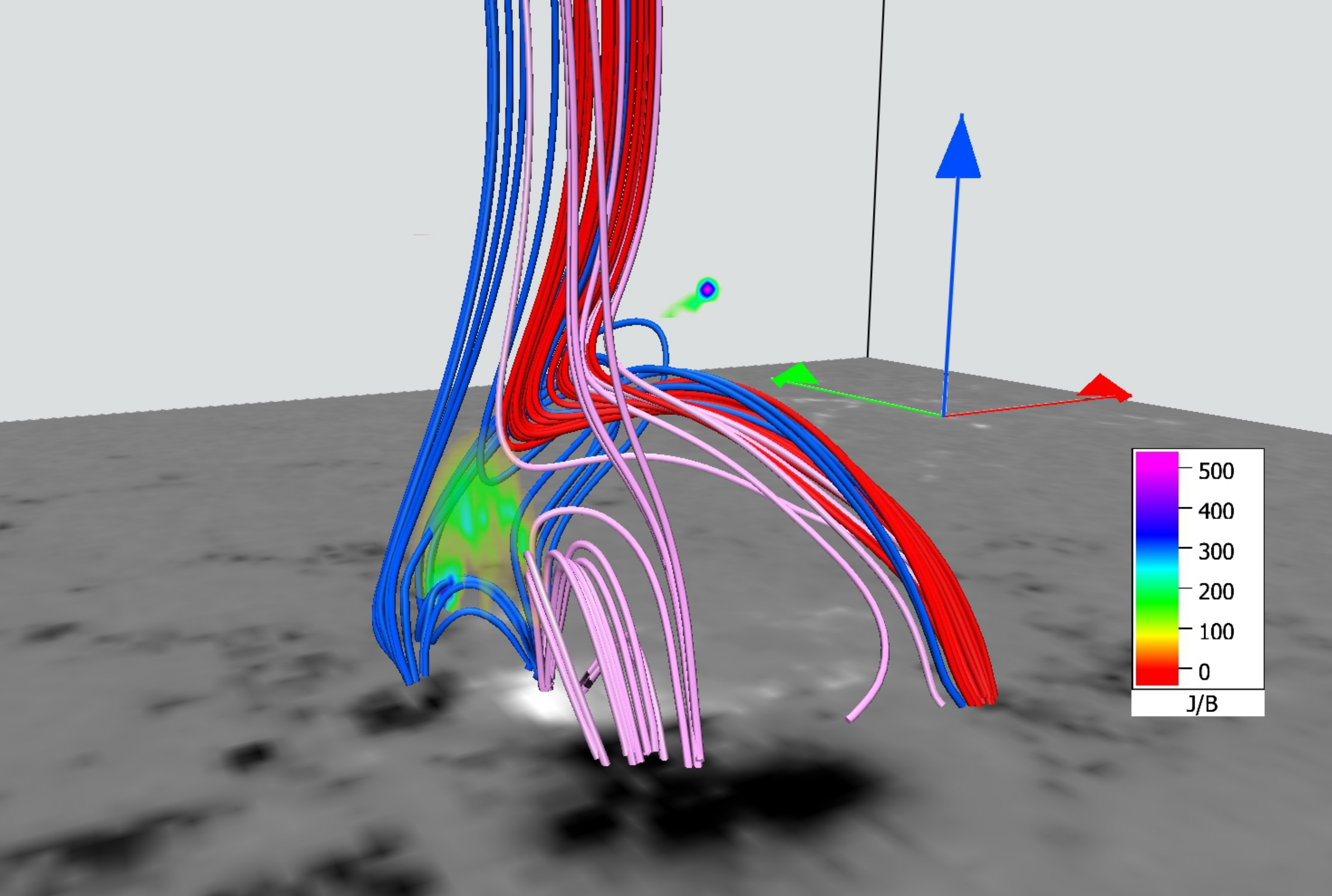}{0.48\textwidth}{(e. t=40)}
        \fig{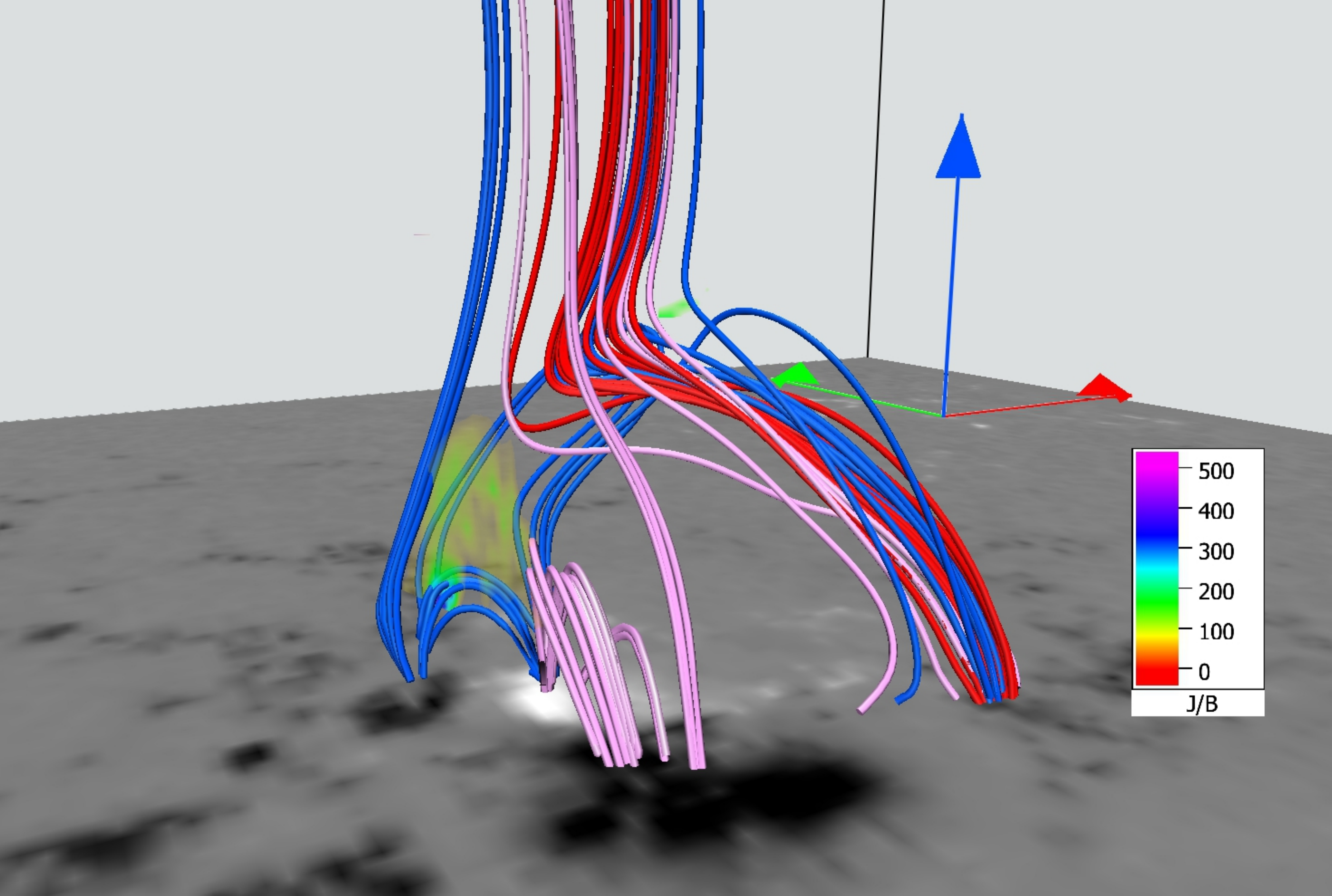}{0.48\textwidth}{(f. t=50)}
        }
\caption{Time sequence showing the formation and dissipation of a current sheet near the X-type MFLs reconnection site. Panel (a) depicts the initial field, where the outer envelope of the flux rope is seen reconnecting at the 3D null (black arrow, also see Figure \ref{f7:pre_flare_sigmoid}(b)). Panels (b)-(d) show the movement of non-parallel MFLs in the vicinity and, development of X-type geometry (white arrow) and a consequent current sheet (with high $J/B$) in that region. In panels~(e-f), simultaneous reconnections at both the 3D null and the X-type MFLs along with the dissipation of the current sheet occur.}
(An animation of this figure is available.) 
%\href{https://youtu.be/EE1HWT83-bM}{Click here for the movie})}
\label{f10:mhd_reconnections}
\end{figure}

Here we focus on the physical processes leading to magnetic reconnections which can play a key role in the flare evolution. In Figure \ref{f10:mhd_reconnections}, we show three sets of MFLs, plotted in color red (corresponding to the flux rope identified in Figure \ref{f8:mhd_sig2rope}), purple (predominately representing the outer envelope of the flux rope), and blue (nearby loops which represent the post flare arcade after the magnetic reconnections). The purple field lines start to reconnect at the pre-existing 3D null (see Figure~\ref{f10:mhd_reconnections}(a--b)). This is in agreement with the rising of overlying loops observed in SDO/AIA 335 \AA~at the start of the flare (cf.~Figure~\ref{f1:event_overview}(e)).
These magnetic reconnections are also expected to further contribute to the pre-flare activity discussed above.

\begin{figure}[ht!]
\gridline{
        \fig{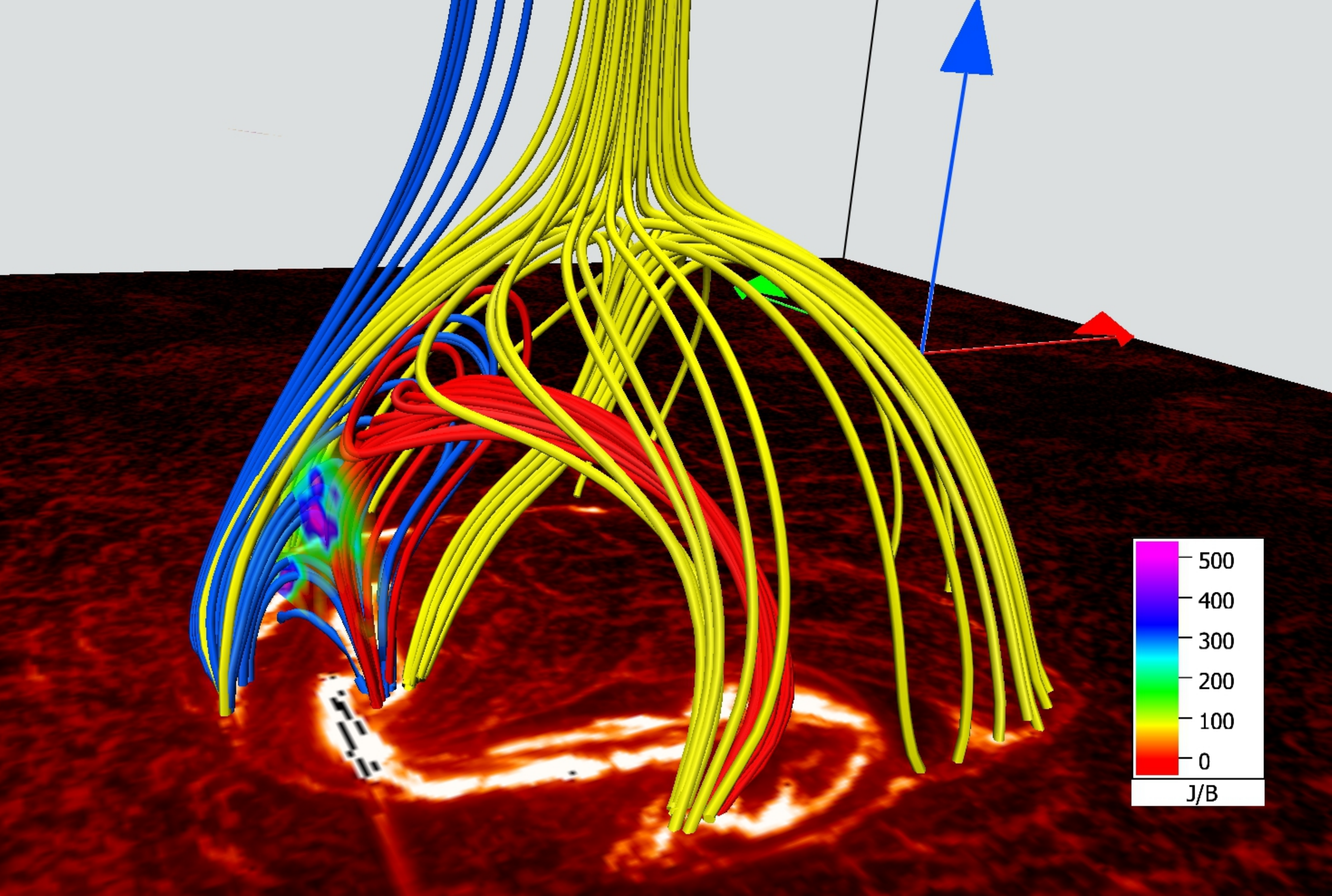}{0.48\textwidth}{(a)}
        %aia304_match_t24.pdf
        \fig{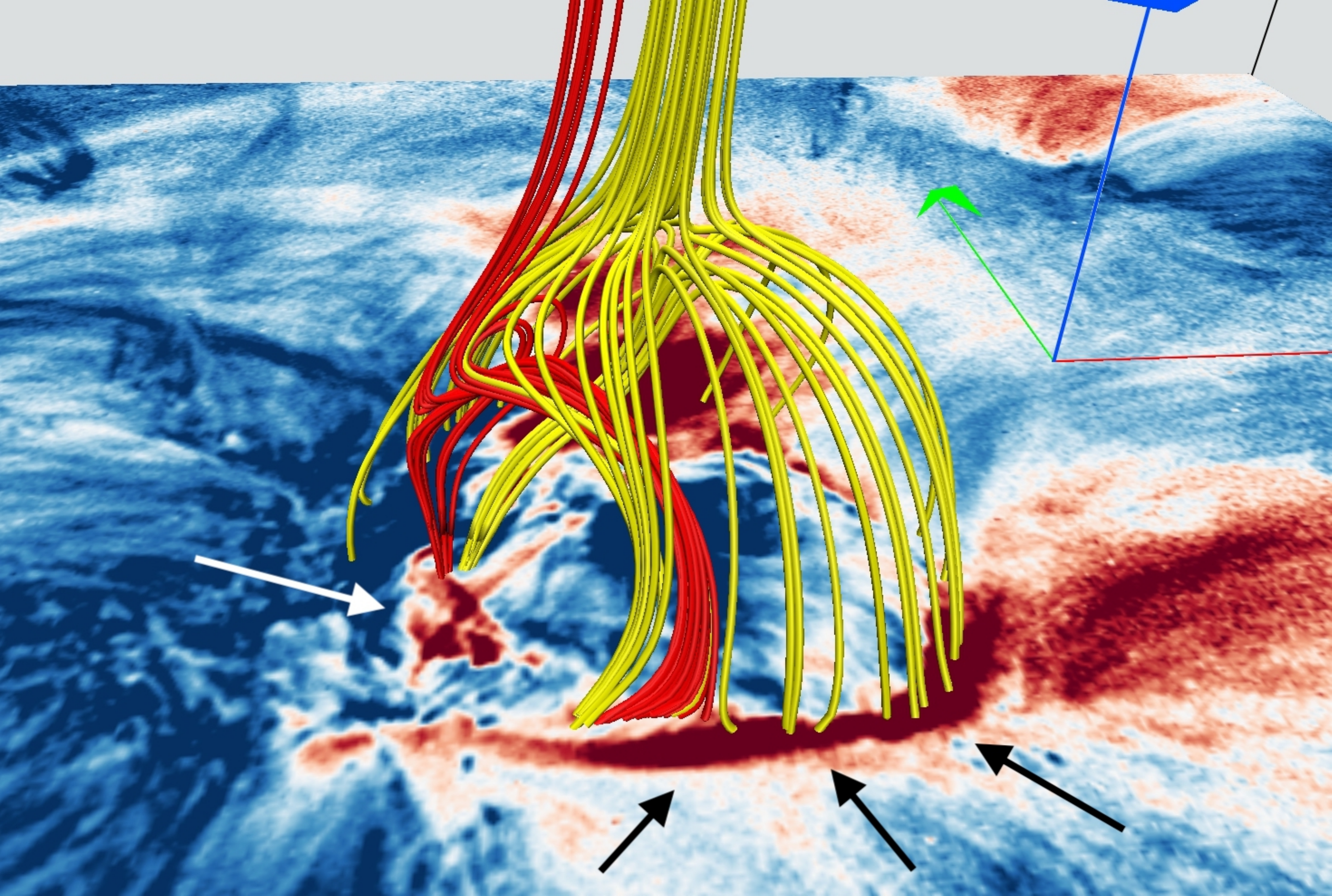}{0.48\textwidth}{(b)}
        %compare_fig2f.jpg
        }
\caption{Comparison of MFL topology  (a) at $t=25$ with the flare ribbons observed in the SDO/AIA 304 \AA~ channel shown in Figure \ref{f1:event_overview}(f) and (b) at $t=35$ with the ring-shaped dimming region shown in Figure \ref{f2:dimming_evolution}(e).  We find excellent agreement with the field lines constituting the dome of the 3D null, the circular flare ribbons and the ring-shaped dimming region (indicated by the black arrows), while the footpoints of the X-type MFLs correspond well to the parallel flare ribbons. In addition, the white arrow marks the dimming region corresponding to the left footpoint of the flux rope.}
\label{f11:ribbons_dimming}
\end{figure}

The subsequent evolution illustrates that the non-parallel field lines of the flux rope (in purple and red) and the nearby loops (in blue; cf.~Figure \ref{f10:mhd_reconnections}(c)) come in close proximity. When viewed from a vantage point (Figure \ref{f10:mhd_reconnections}(d)), the non-parallel MFLs show the near-resemblance to the
X-type geometry. Therefore, we name these MFLs as X-type field lines.  As the gradient of $\bf{B}$  steepens, a strong electric current originates in the vicinity of the X-type MFLs at $t=20$, shown by the $J/B$ probe placed on the $y-z$ plane. 
Consequently, the scales become under-resolved which onset magnetic reconnections 
%As a result, magnetic reconnection takes place as the scales become under-resolved and such magnetic reconnections, 
that repeatedly occur in time, and
are responsible for the bifurcation of the flux rope (clearly identifiable in Figure \ref{f10:mhd_reconnections}(d--f)). Importantly, from $t=20$ onwards, the evolution discerns the co-occurrence of reconnections at both sites: the X-type MFLs as well as the 3D null. Remarkably, such co-temporal reconnections at these two sites can provide a potential explanation of the simultaneously observed standard parallel ribbons and circular ribbons, as shown in Figure \ref{f11:ribbons_dimming}(a). The bottom boundary in the figure shows the cooler AIA 304 \AA ~channel after the start of the flare at \mbox{$\sim$22:17~UT}, highlighting the chromospheric flare ribbons (cf.~Figure~\ref{f1:event_overview}(f)). As the MFLs constituting the fan surface of the 3D null intersect with the chromosphere, the corresponding footpoints form a closed circle. A circular flare ribbon is then expected because the magnetic reconnections at the 3D null can accelerate charged particles which travel along the MFLs of the dome-shaped fan surface and deposit their energy in the chromosphere \citep{masson+2009apj,jiang+2013apjl,devi+2020SoPh}. In a similar way, the reconnections at the X-type MFLs can cause the ``standard" parallel flare ribbons. 
Moreover, field lines from the inner spine (which are initially closed) get transformed into the open field lines of the outer spine. The footpoints of these open field lines correspond to the ring-shaped dimming region D1 (cf. Figure~\ref{f2:dimming_evolution}(e)), tracing the circular dome as seen in Figure~\ref{f11:ribbons_dimming}(b), indicated by the black arrows. 
The close co-spatiality between the dimming region and the circular flare ribbon supports this result. Further, the white arrow in Figure~\ref{f11:ribbons_dimming}(b) marks the dimming region corresponding to the left footpoint of the flux rope.

% \begin{figure}[ht!]
% \plottwo{aia304_match_t24.pdf}{loqq_t24_dvr.jpg}
% \caption{Comparison of flare ribbon and footpoints at t=24. The scale for J/B is same as that used in Figure \ref{f:mhd_recon}.}
% \label{f:304ribbon}
% \end{figure}

% \begin{figure}[ht!]
% \plottwo{compare_fig1ga.jpg}{compare_fig2f.jpg}
% \caption{(a) Comparison of flare ribbon and footpoints as shown in Figure 1g. (b) Footpoints of the dome as compared to the dimming regions.}
% \label{f:compare}
% \end{figure}
%\ap{can add the Q-map figure here as the second panel. can we make an estimate the energy released from the dissipation of the current sheet?}\qh{ Are we talking about $\mathbf E\cdot \mathbf J$ here? Both can be derived I think. The question is whether they are physical, especially the former, which is basically the reconnection rate; I suppose?}

%\subsection{Correspondence of simulated evolution with observed coronal dimming}

\begin{figure}[ht!]
\includegraphics[width=1.0\textwidth]{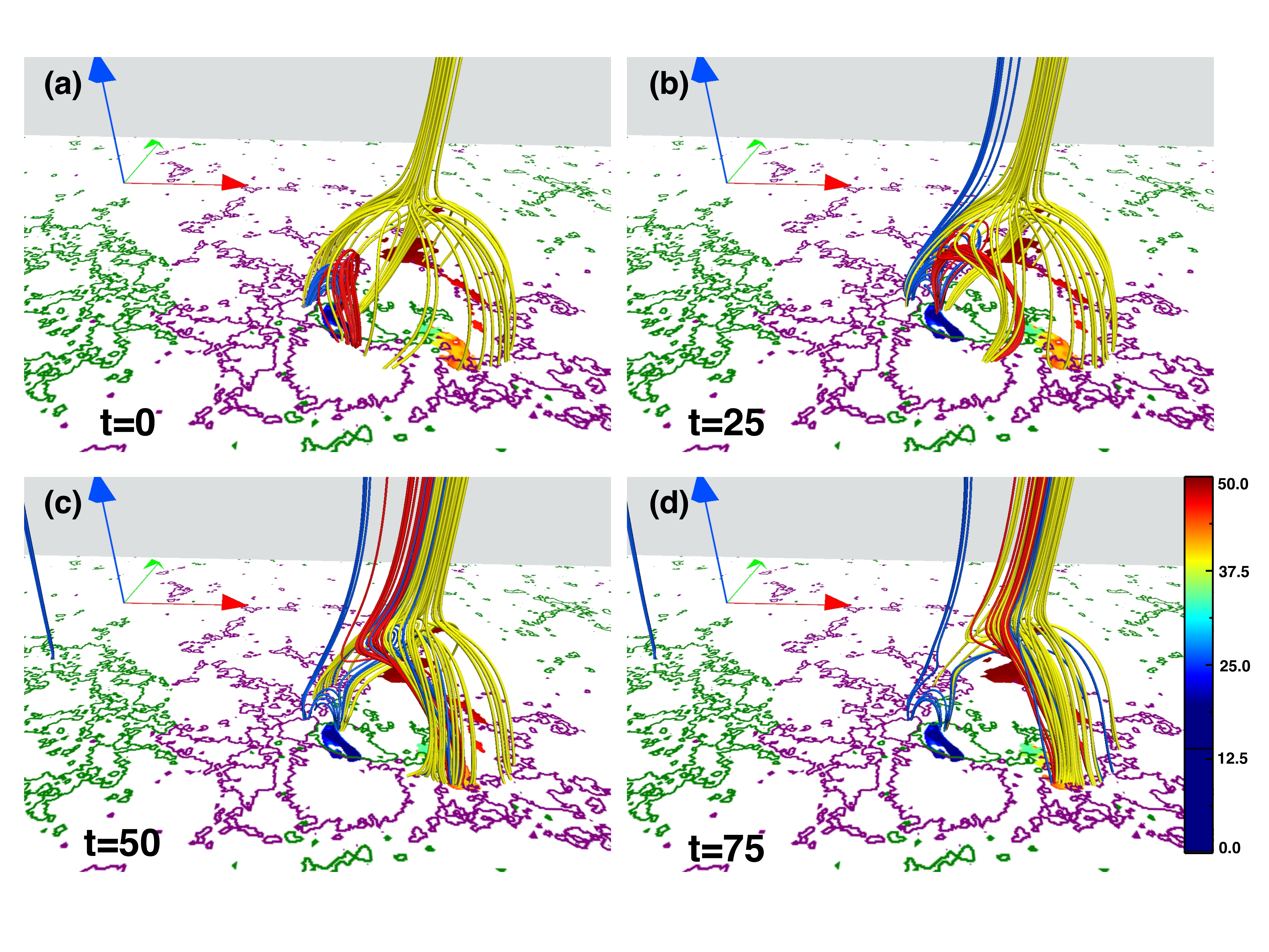}
%\gridline{
        %dimm0000,0025,0050,0075
%        \fig{r1_fig12a.jpg}{0.45\textwidth}{(a. t=0)}
%        \fig{r1_fig12b.jpg}{0.45\textwidth}{(b. t=25)}
%        }
%\gridline{
%        \fig{r1_fig12c.jpg}{0.45\textwidth}{(c. t=50)}
%        \fig{r1_fig12d.jpg}{0.45\textwidth}{(d. t=75)}
%        }
\caption{Correspondence of the magnetic field evolution and early development of the coronal dimming regions. The bottom boundary shows the contours with $B_z$ together with dimming pixels marked in color with respect to their time of first appearance (in minutes after 21:45~UT). We can observe that while in panel (a) the footpoint of the flux rope corresponds to blue pixels (pre-flare dimming), with time it moves due to slipping reconnections to an orange region marked in panel (d), where the dimming is observed at a later time.}
(An animation of this figure is available.) 
%\href{https://youtu.be/-LREu6KZPLM}{Click here for the movie.})}
\label{f12:mhd_dimming}
\end{figure}

To explain the coronal dimmings during the flare in more detail, in Figure \ref{f12:mhd_dimming}, we illustrate the evolution of the flux rope footpoints with respect to the coronal dimming timing maps (Figure \ref{f2:dimming_evolution}(f)). The bottom boundary in Figure \ref{f12:mhd_dimming} shows contours of $B_z$ (green showing positive polarity and purple showing negative polarity) and the locations of dimming pixels are  marked in color, with respect to the time of their first appearance in minutes after 21:45~UT. The blue pixels represent regions where the dimming was observed first, while red pixels represent all the sites where coronal dimmings occurred later. Notably, the evolution shows the movement of the negative polarity footpoint of the flux rope to the right due to slipping reconnections. The footpoint then approaches new dimming pixels (in orange) appearing to the right. A similar movement of the flux rope footpoint was also reported in \citet{jiang+2013apjl}.
%\ap{Can add a figure with Qmap of the bottom boundary to show slipping reconnections and refer to it here}.
At the same time, the other end of the flux rope undergoes magnetic reconnection at the X-type geometry and the field lines reconnect to the positive polarity on the far left (see Figure \ref{f12:mhd_dimming}(c)). This bifurcation of the flux rope leads to the generation of  open magnetic field lines.  The plasma loss along the open field lines from the footpoint location may result in the observed dimming in this region (marked by the white arrow in Figure~\ref{f11:ribbons_dimming}(b)). A flux rope bifurcation for this event was also reported in \citet{prasad+2017apj}. The slipping reconnections also result in the rotation of the field lines comprising the dome of the 3D null.

\subsection{Evolution of field lines in the full domain}
\begin{figure}[ht!]
        %global0000.jpg,0020,0040,0060,0080
\gridline{\fig{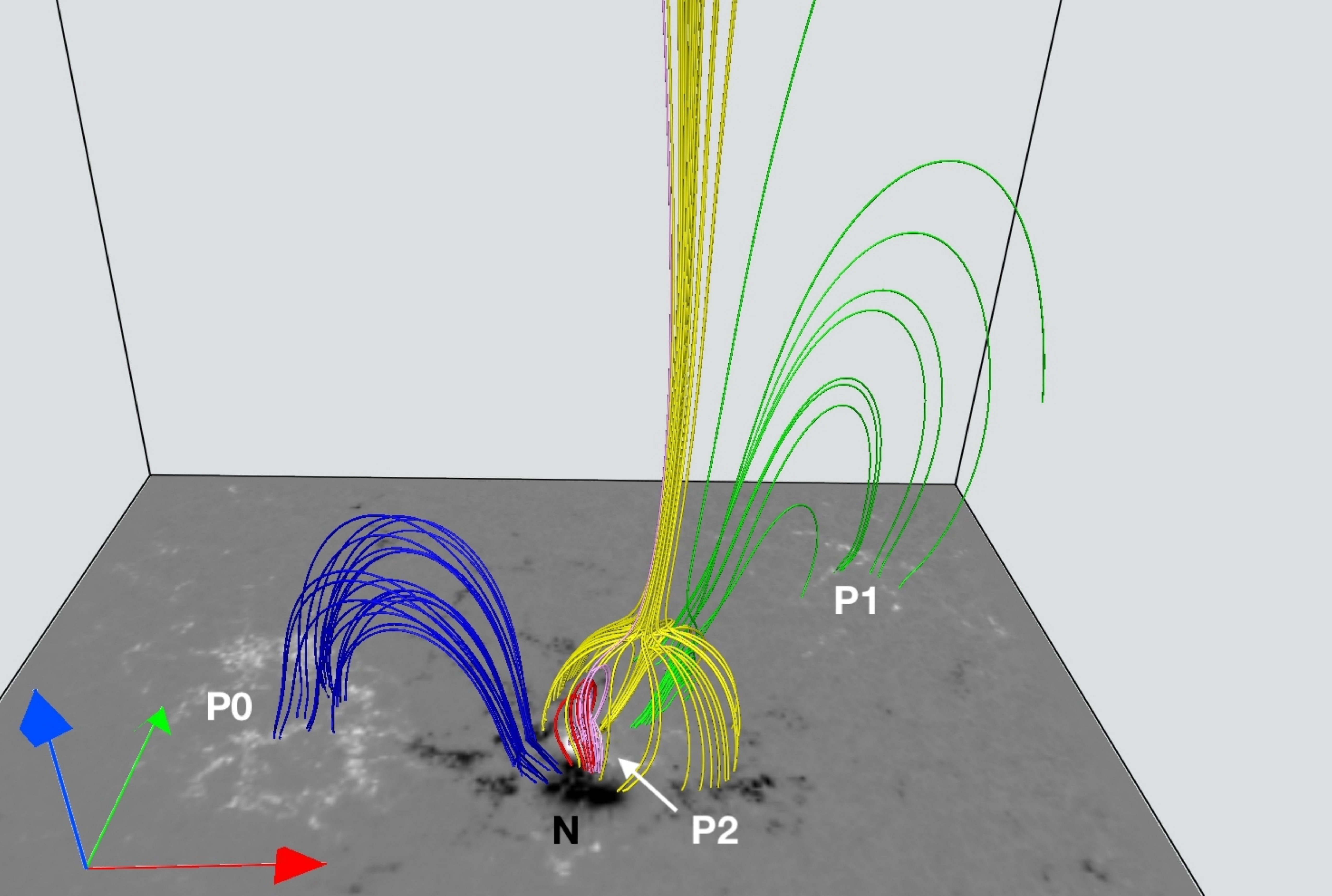}{0.48\textwidth}{(a. t=0)}
          \fig{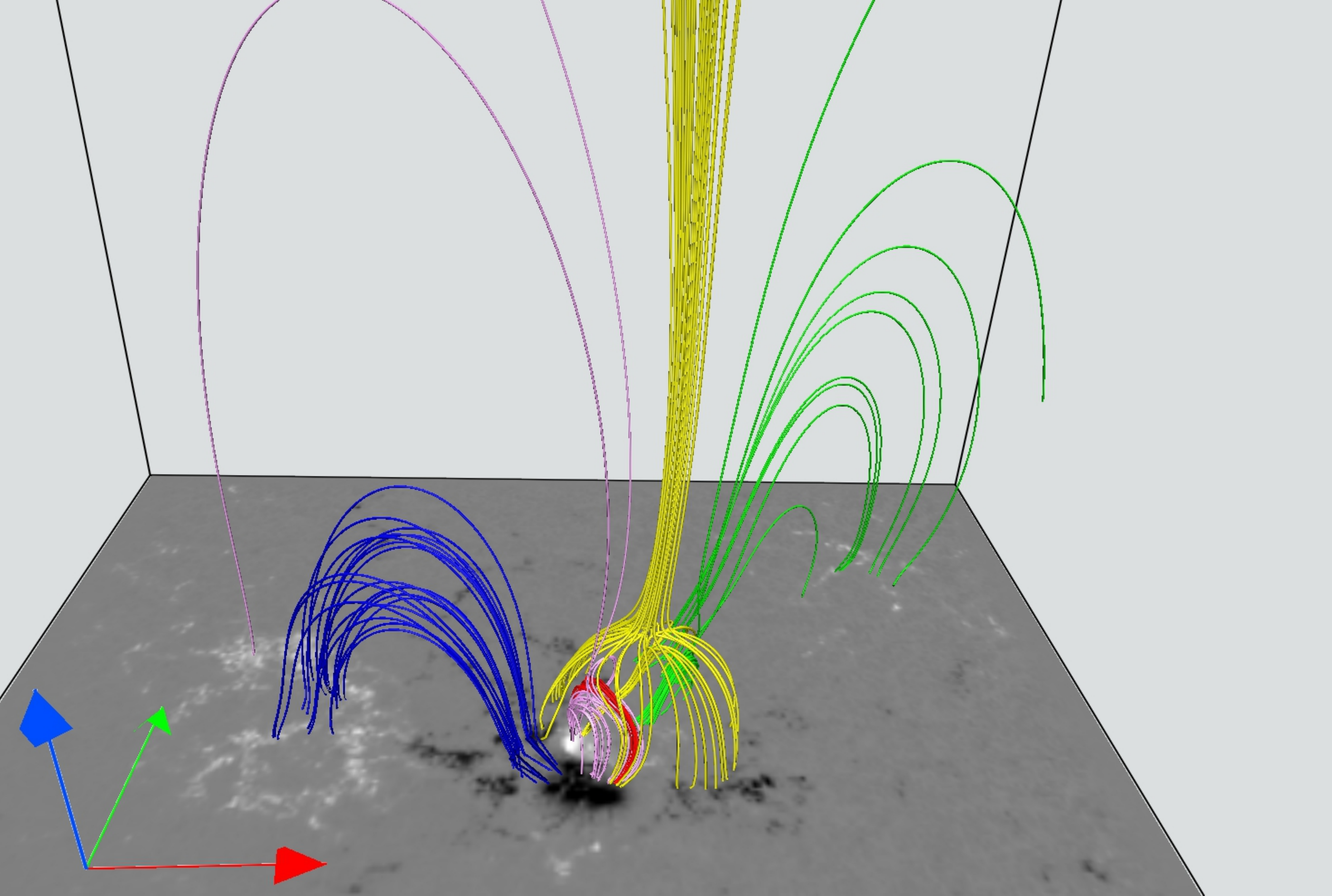}{0.48\textwidth}{(b. t=20)}}
\gridline{\fig{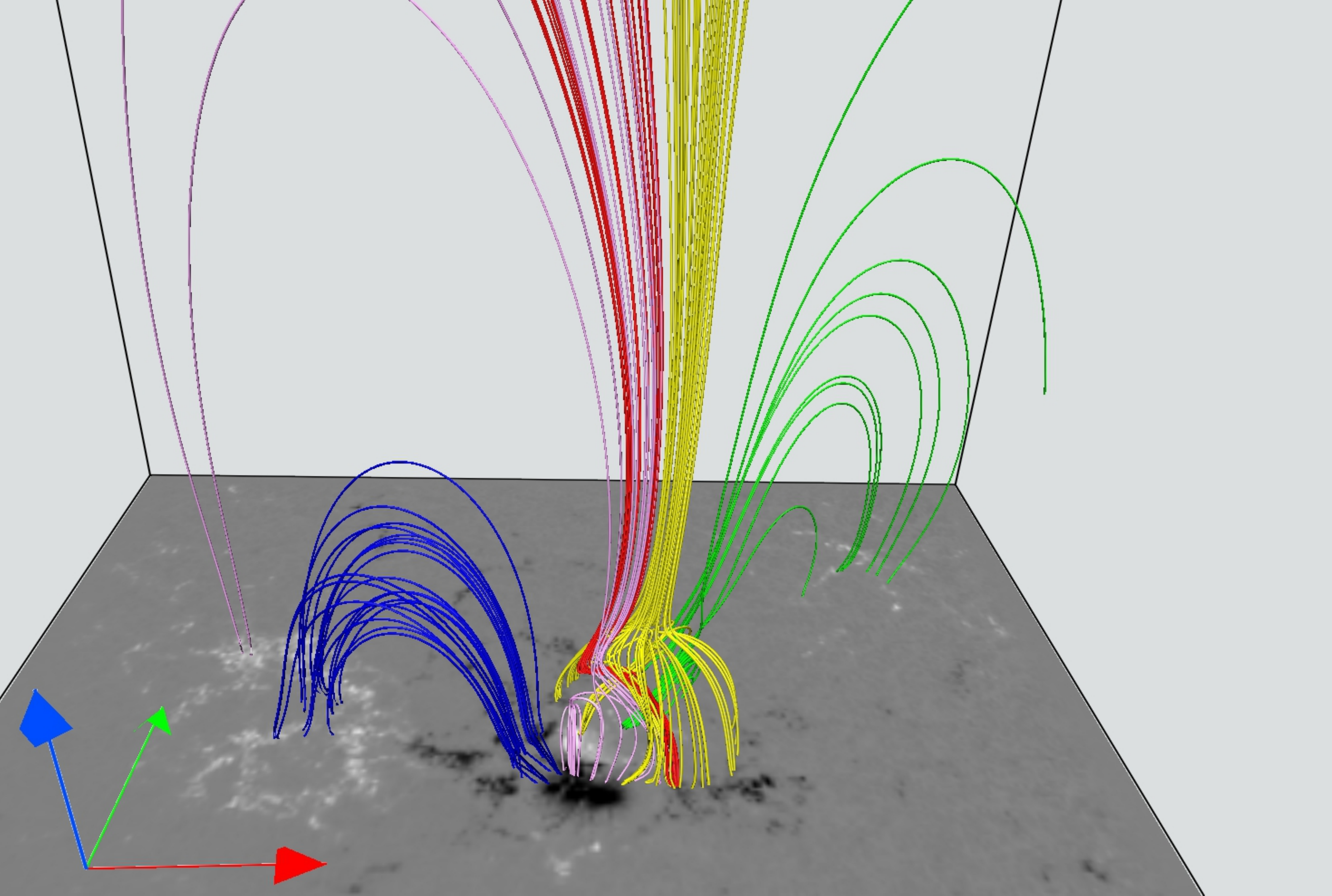}{0.48\textwidth}{(c. t=40)}
          \fig{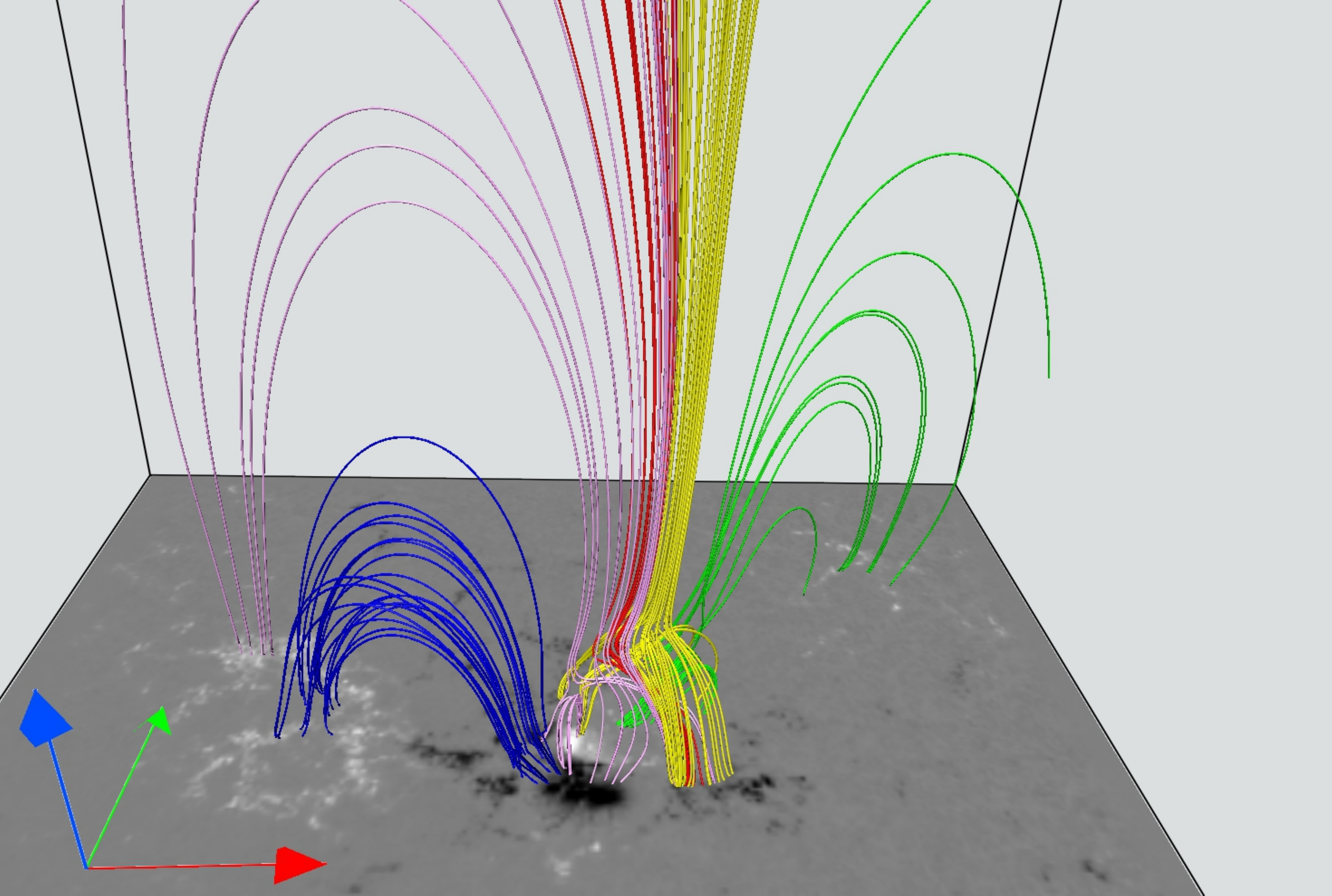}{0.48\textwidth}{(d. t=60)}}
\gridline{\fig{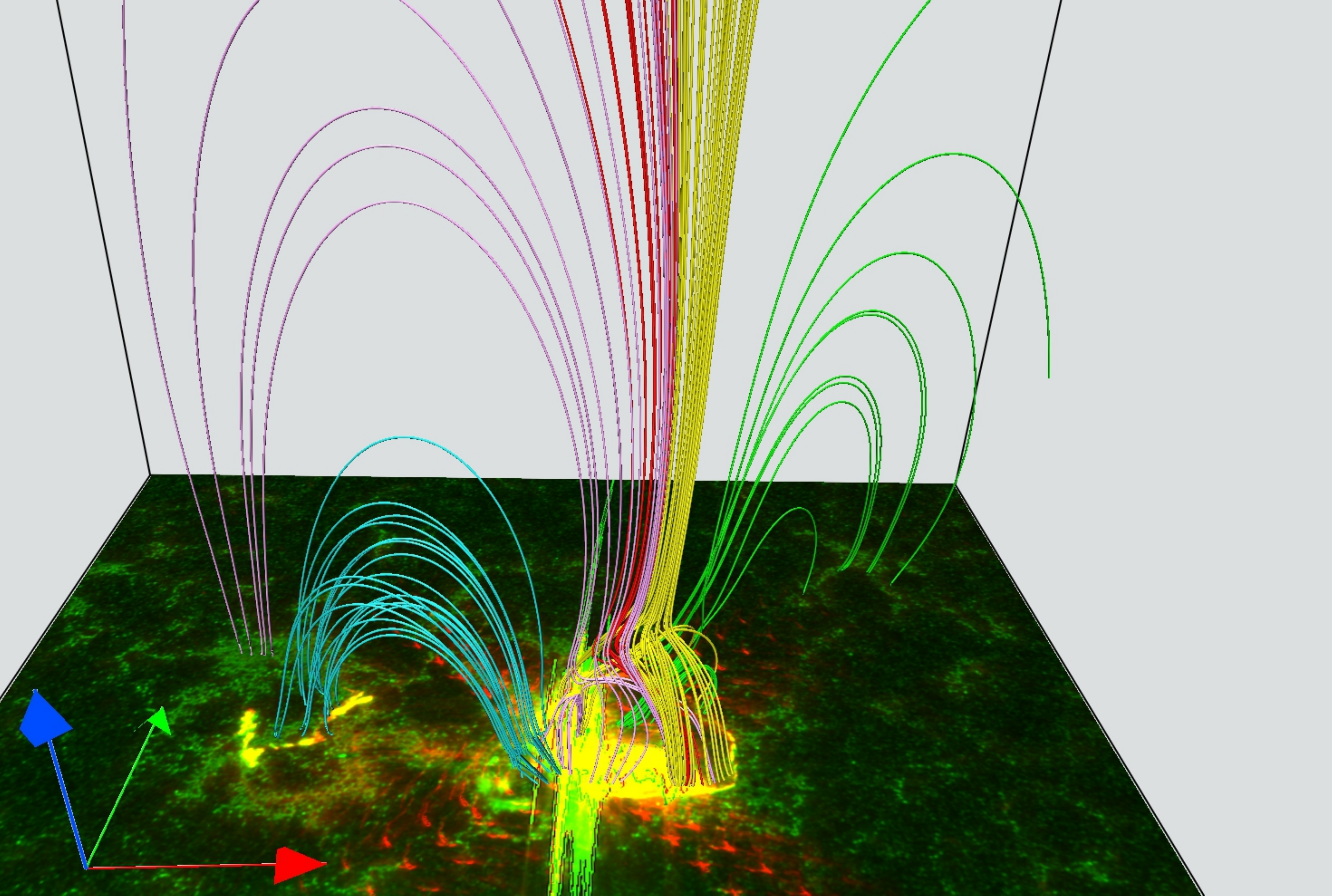}{0.48\textwidth}{(e. t=60)}
          \fig{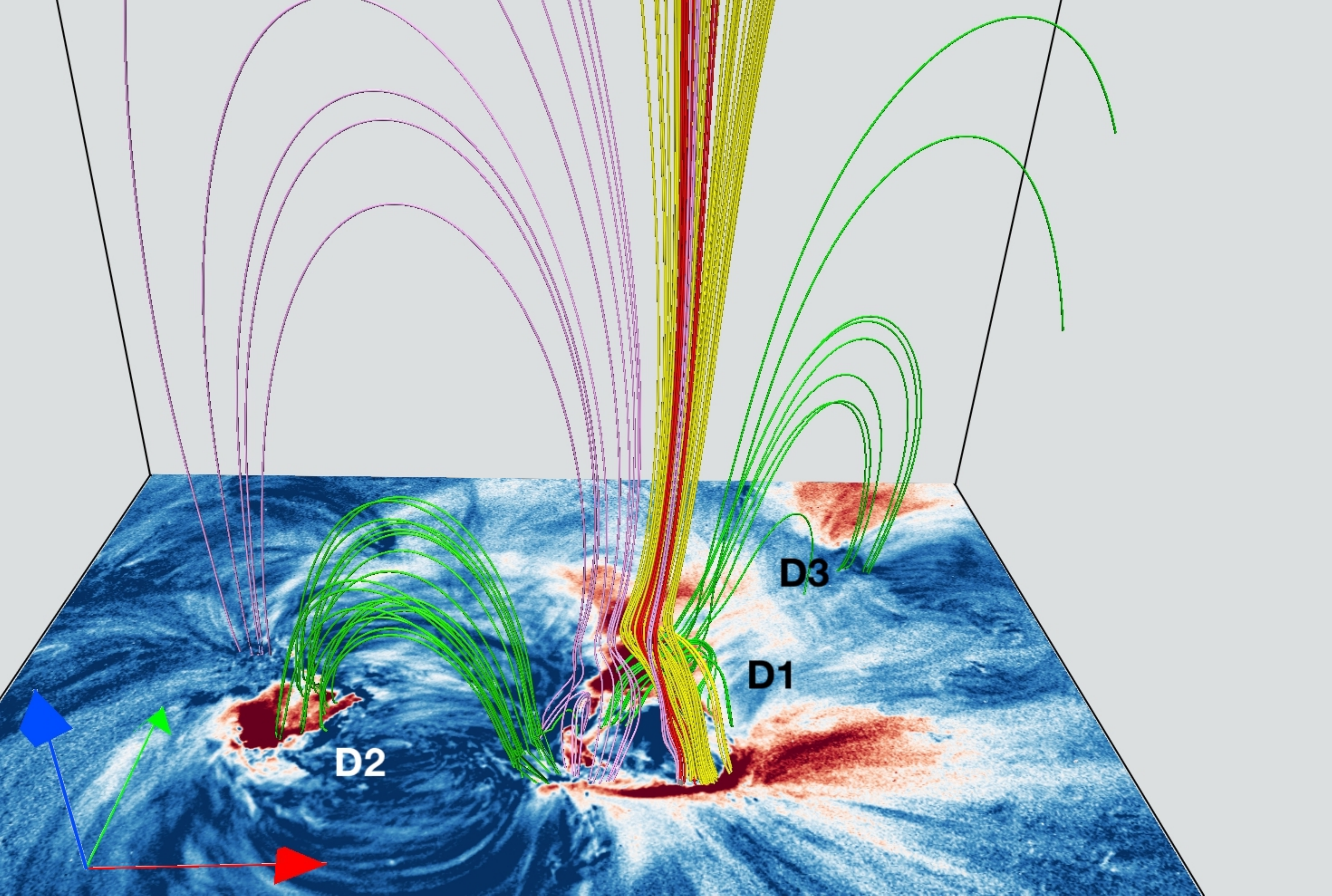}{0.48\textwidth}{(f. t=100)}}
          %20110906_2142_log_base_ratio_362_t100.jpg
\caption{Global dynamics of the field lines during the simulation highlighting the remote connectivities that form due to the reconnections.  Panels (e) and (f) use Figure \ref{f1:event_overview}~(i) and Figure \ref{f2:dimming_evolution}~(e) as bottom boundary for comparing the locations of MFLs with respect to the flare ribbons and dimming locations. The color of the blue field lines from panel (a) have been changed in panel (e) and (f) to cyan and green for better visibility.}
(An animation of this figure is available.)
%\href{https://youtu.be/EXWaam39hcg}{Click here for the movie.})}
\label{f13:global_evolution}
\end{figure}

In Figure \ref{f13:global_evolution}, we show the overall MFL dynamics in the full computational domain. In panels (a--d), the bottom boundary is overlaid with the magnetogram, alongwith the field lines previously shown in Figure \ref{f5:initial_field_overview}. 
The panels clearly illustrate a change in the connectivity of MFLs from P2 to P0 (purple field lines) which resulted from the magnetic reconnections at the X-type geometry and the 3D null. Moreover, few of the MFLs plotted in blue and green are found to be rising and, ultimately, opening up, indicating an outward expansion of coronal loops during the flare.
To establish an overall correspondence between the field line evolution and the observed flare ribbons, in Figure~ \ref{f13:global_evolution}(e), we overplot the bottom boundary with a composite image of SDO/AIA 304 and 1600 \AA~ shown in Figure~\ref{f1:event_overview}(i). It is clearly seen that the footpoints of the dome surface trace the circular flare ribbon. This strongly suggets that the magnetic reconnections at the 3D null play a key role for the development of the flare ribbons. Moreover, the change in connectivity of MFLs from P2 to P0 during magnetic reconnections may be indicative of a causal connection between the magnetic reconnections and the remote flare ribbon 1 (see Figure~\ref{f1:event_overview}(g)). Further, the connectivity of green MFLs favors a possible 
relationship between magnetic reconnections at the 3D null and the remote flare ribbon 2 (see Figure~\ref{f1:event_overview}(i)). 

In comparison, \citet{jiang+2018apj} found three episodes of reconnection occurring at different locations in the corona through which the initial sigmoidal magnetic reconnection breaks out at one of its ends, and through subsequent reconnections, gives rise to a highly twisted field having a complex magnetic topology (see Figures 7, 8 and 10 in \citet{jiang+2018apj} for details of the three stages of reconnections).
To further highlight the close connection between the post-flare MFL topology and the location of the dimming regions, in Figure \ref{f13:global_evolution}(f), we overlay the bottom boundary with the final dimming evolution plotted in Figure~\ref{f2:dimming_evolution}(e). The figure also demonstrates the similarity between the ring-shaped dimming region D1 and the dome structure of the 3D null. The present simulation, however, does not clearly identify the cause for the formation of the dimming regions marked as D2 and D3 in Figure \ref{f2:dimming_evolution}(e). This can be attributed to the absence of a significant eruption corresponding to the sudden and rapid rise of the flux rope as shown in \citet{jiang+2013apjl, jiang+2018apj}. The absence of eruption in our simulation can be ascribed to the viscous dissipation which leads to a faster depletion of the free magnetic energy required to produce the eruption. 
From the observations, these dimming regions co-locate with the remote flare ribbons. This suggests that the repeated reconnections, occurring higher-up in the corona,  can be responsible for the dimming regions.
In addition, the magnetic energy released during the simulated evolution is about $2.15 \times 10^{31}$ erg. Interestingly, the energy estimate is comparable to the ones derived from the observations in  \citet{feng+2013apj}. 
%In the simulation, we find that the field lines in pink, which form the outer envelope of the flux rope (shown in red) start reconnecting at the 3D null point from the first time step onwards, which can be interpreted as the pre-flare activity {that are} also noticeable in the dimming images (see Figure \ref{f7:pre_flare_sigmoid} (b)).
%\ap{pre-flare activity to be mentioned in the dimming images.}
%In Figure \ref{f7:pre_flare_sigmoid} (b), we see appearance of the dimming pixels in this region. 

\section{Summary and Discussion}\label{sec:summary}
In this paper, we perform an MHD simulation of magnetic field evolution during a X2.1 flare in AR 11283. The simulation is initiated by extrapolated non-force-free magnetic field, which is based on the principle of minimum dissipation rate and the photospheric vector magnetogram of the active region obtained from SDO/HMI serves as lower boundary condition. 
%For the extrapolated field, we utilize a well-established non-force-free numerical extrapolation  model based on the principle of minimum dissipation rate in which photospheric vector magnetograms of the active region obtained from SDO/HMI serves as lower boundary condition. 
Particularly, we aim to explain various observational features of the complex X2.1 flare and associated coronal dimmings that occurred on 2011 September 6, around 22:21 UT (SOL2011-09-06T22:21X2.1). SDO/AIA multi-wavelength observations show the signature of a hot sigmoid, pre-flare activities, the formation of the flux rope, and the evolution of circular as well as remote flare ribbons. SDO/AIA 211 \AA~logarithmic base-ratio images are analysed to locate the coronal dimming regions and to identify their fine structure. Notably, about 30 minutes prior to the onset of the flare, small-scale, bipolar dimmings are observed near the main flare site. Moreover, during the impulsive phase of the flare, three main dimming regions are of interest: a ring-shaped dimming region (D1) in the vicinity of the main flare site, a circular dimming region (D2) at the far east to the flare location, and a remote dimming region (D3) at the far north-west to the main flare site.  
%To optimize the computational cost, the simulation is performed on $256\times200\times200$ grids along the $x$, $y$ and, $z$ axes, respectively, resolving a physical domain of $370 \times 290 \times 290$ Mm$^3$.
The absence of any significant flux-emergence during the event allows us to use the line-tied boundary condition at the bottom boundary. The initial Lorentz force pushes the magnetofluid from its initial motion-less state and generates the evolution autonomously. The following is a summary of the main results:
\begin{itemize}
 
    \item The non-force-free extrapolation was able to sucessfully capture the presence of highly sheared/twisted field lines over the central PIL and also the magnetic topology of a 3D null point located close to the flaring region.  Importantly, these sheared field lines explain the sigmoidal brightenings observed in the SDO/AIA 94 \AA~channel (cf.~Figure~\ref{f1:event_overview}(c), \ref{f7:pre_flare_sigmoid}(a)).  %Figure 7
    These findings are in agreement with the recent observational study by \citet{sahu+2020ApJ} where a flux rope is found to be co-spatial with an HXR coronal channel, implying pre-flare brightenings caused by an activated flux rope.
    \item The distribution of the Lorentz force is shown to be concentrated near the bottom boundary, justifying the use of the NFFF description for the solar corona (cf.~Figure~\ref{f6:lorentz_force}). The Lorentz forces are also critical in generating self-consistent flows that initiate the dynamics and trigger the magnetic reconnections.
    \item The observed small-scale, bipolar pre-flare coronal dimming region is formed due to (i) the rising of the outer envelope of the flux rope and (ii) magnetic reconnection at the pre-existing 3D null point resulting in open fields (cf.~Figure~\ref{f2:dimming_evolution}(a), \ref{f7:pre_flare_sigmoid}(b)). 
    This is in agreement with predictions of such pre-flare coronal dimmings in \cite{forbes+2000} and their observation in case studies by \cite{qiu+2017apj} and \cite{zhang+2017aa}. %To our knowledge this is the first time their existence could be confirmed in data-constrained simulations and observations simultaneously.
    %\item In addition, the simulated evolution documents the onset of magnetic reconnections at the pre-existing 3D null point. The location of the foot-points of the reconnecting field lines corresponds well with the pre-flare coronal dimmings inferred from the observations.
    \item  In the simulation, we notice the transfer of twist from the low-lying sheared/twisted coronal field lines to overlying loops through magnetic reconnections, leading to the formation of a flux rope. This is in agreement with small-scale brightenings observed during the pre-flare phase in all EUV wavelengths (cf.~Figure~\ref{f1:event_overview}(d)). This result is in excellent agreement with the observational study by \citet{joshi+2017apj} in which episodic pre-flare brightenings (an evidence of small-scale magnetic reconnections) were reported during the build-up of a hot-channel flux rope. During this transfer, the right footpoint of the establishing flux rope is moving. Accumulation of filament material is observed at this newly established right flux rope footpoint (cf.~Figure~\ref{f1:event_overview}(d), \ref{f2:dimming_evolution}(b)).
    \item The role of these magnetic reconnections in flaring activities is established by co-locating the footpoints of the reconnecting field lines with the  emission in the AIA 94 \AA~channel. Moreover, the locations of the footpoints of the flux rope are found to be in good agreement with those inferred from SDO/AIA \mbox{335 \AA}~observations (cf.~Figure~\ref{f1:event_overview}(h), \ref{f8:mhd_sig2rope}(f)). 
    \item The simulation results reveal a rise of the flux rope.
    %, caused by the torus instability.
    During the rise, the non-parallel field lines constituting the rope and lower-lying coronal arcades develop an X-type geometry which lead to repeated magnetic reconnections (see Figure~\ref{f10:mhd_reconnections}).
    \item The concurrent occurrence of magnetic reconnections at the X-type geometry and the 3D null is noted to induce a bifurcation of the flux rope. These simultaneous reconnections provide a potential explanation for the co-temporal appearances of parallel as well as circular flare ribbon patterns observed in chromospheric emissions (cf.~Figure~\ref{f11:ribbons_dimming}(a)). 
    \item In addition, the footpoints of the dome surface of the 3D null are co-spatial with the ring-shaped dimming region, suggesting a causal connection between the magnetic reconnections at the 3D null and the dimming due to the transformation of field lines of the inner spine to open field lines of the outer spine (cf. black arrows in Figure~\ref{f11:ribbons_dimming}(b)).
    \item Importantly, the bifurcation of the flux rope opens up the field lines of the rope which can lead to the loss of plasma trapped inside the rope. This may explain the presence of dimming regions near the footpoints of the rope (cf. white arrow in Figure~\ref{f11:ribbons_dimming}(b)). The simulations also reveal the apparent motion of one footpoint of the flux rope due to slipping reconnections.  The motion was found to match well to those inferred from the timing maps of the dimming images (cf.~Figure~\ref{f12:mhd_dimming}).  This means that the bifurcation of the flux rope and magnetic reconnections at the 3D null are key to the dimming in the neighbourhood of the main flare site.    
\end{itemize}

We speculate that the fine structure of coronal dimmings, resulting from a different intensity distribution within the overall dimming region, %evident in logarithmic base-ratio images %(i.e. strongly decreased dimming regions in red vs. small to moderate decreased dimming regions in lightblue and white) 
is caused by two different physical mechanisms. The locations of the strongest intensity decrease (i.e.~red regions in logarithmic base-ratio images) could correspond to opened-up field lines, whereas dimming regions showing a smaller decrease in intensity (i.e.~light blue and white regions in logarithmic base-ratio images) might correspond to the expanded and stretched overlying fields. 
Our interpretation of the formation of the ring-shaped dimming region D1 (cf.~Figure~\ref{f11:ribbons_dimming}(b)) and the intensity distribution within the pre-flare dimmings (cf.~Figure~\ref{f7:pre_flare_sigmoid}(b)) where both mechanisms are at work support this view.

The origin of coronal dimming regions at the far east (D2) and north-west (D3) is not fully clear as no movement or change of connectivity of field lines in those locations was identified in the full domain (see Figure~\ref{f13:global_evolution}).
However, we note that from an observational point of view, the locations of these dimming regions match the positions of the remote flare ribbons, indicating that magnetic reconnections may play a role at these remote locations, potentially initiated higher up in the corona.
%\ap{\citet{janvier+2016} were able to obtain a morphology that matched the EUV flare ribbons of the main flare site and inferred the distribution of currents through the evolution of QSLs via a magnetofrictional method while \citet{jiang+2018apj} modelled the formation and eruption of the sigmoid as a multistage reconnection process.}

Overall, although successful in simulating parts of a particular complex flare/CME event, the combined model (extrapolation + MHD) can be advanced by using simultaneous magnetograms from two different heights, inclusion of an apt physical resistivity and accounting for any photopsheric motion. We leave these as future endeavors.
%====== ACKNOWLEDGEMENTS ======
\acknowledgments
We acknowledge the use of the visualization software VAPOR (www.vapor.ucar.edu) for generating relevant graphics. Data and images are courtesy of NASA/SDO and the HMI and AIA science teams. SDO/HMI is a joint effort of many teams and individuals to whom we are greatly indebted for providing the data. Q.H. and A.P. acknowledge partial support of NASA grant 80NSSC17K0016 and NSF award AGS-1650854. K.D. and A.M.V. acknowledge funding by the Austrian Space Applications Programme of the Austrian Research Promotion Agency FFG, BMVIT: projects ASAP-11 4900217, ASAP-14 865972, as well as the Austrian Science Fund (FWF):  projects P24092-N16, P27292-N20. A.M.V., BJ, K.D. and A.P. acknowledge the Indo-Austrian joint research project No. INT/AUSTRIA/BMWF/P-05/2017 and OeAD project No. IN 03/2017. The authors are thankful to Dr. P. K. Smolarkiewicz for his support. We are grateful to Dr. Ronald L. Moore for the discussion about simulation results. We are
also thankful to the anonymous referee for providing insightful
suggestions, which led to the overall betterment of this paper.

%====== APPENDICES ========
%\appendix
%\section{Appendix information}
%Appendices 

%==== BIBLIOGRAPHY ======
\bibliography{ms}

\begin{thebibliography}{}
\expandafter\ifx\csname natexlab\endcsname\relax\def\natexlab#1{#1}\fi

\bibitem[{{Alfv{\'e}n}(1942)}]{alfven1942nat}
{Alfv{\'e}n}, H. 1942, \nat, 150, 405

\bibitem[{{Amari} {et~al.}(2014){Amari}, {Canou}, \& {Aly}}]{amari+2014nat}
{Amari}, T., {Canou}, A., \& {Aly}, J.-J. 2014, \nat, 514, 465

\bibitem[{{Aschwanden}(2004)}]{aschwanden2004book}
{Aschwanden}, M.~J. 2004, {Physics of the Solar Corona. An Introduction}
  (Praxis Publishing Ltd)

\bibitem[{{Aulanier} {et~al.}(2005){Aulanier}, {Pariat}, \&
  {D{\'e}moulin}}]{aulanier+2005aa}
{Aulanier}, G., {Pariat}, E., \& {D{\'e}moulin}, P. 2005, \aap, 444, 961

\bibitem[{{Bhattacharyya} {et~al.}(2007){Bhattacharyya}, {Janaki}, {Dasgupta},
  \& {Zank}}]{bhattacharyya+2007soph}
{Bhattacharyya}, R., {Janaki}, M.~S., {Dasgupta}, B., \& {Zank}, G.~P. 2007,
  \solphys, 240, 63

\bibitem[{{Bhattacharyya} {et~al.}(2010){Bhattacharyya}, {Low}, \&
  {Smolarkiewicz}}]{bhattacharyya+2010phpl}
{Bhattacharyya}, R., {Low}, B.~C., \& {Smolarkiewicz}, P.~K. 2010, Physics of
  Plasmas, 17, 112901

\bibitem[{Courant {et~al.}(1967)Courant, Friedrichs, \& Lewy}]{courant1967jrd}
Courant, R., Friedrichs, K., \& Lewy, H. 1967, IBM journal of Research and
  Development, 11, 215

\bibitem[{{Dahlburg} {et~al.}(1991){Dahlburg}, {Antiochos}, \&
  {Zang}}]{dahlburg+1991apj}
{Dahlburg}, R.~B., {Antiochos}, S.~K., \& {Zang}, T.~A. 1991, \apj, 383, 420

\bibitem[{{Devi} {et~al.}(2020){Devi}, {Joshi}, {Chandra}, {Mitra}, {Veronig},
  \& {Joshi}}]{devi+2020SoPh}
{Devi}, P., {Joshi}, B., {Chandra}, R., {et~al.} 2020, \solphys, 295, 75

\bibitem[{{Dissauer} {et~al.}(2016){Dissauer}, {Temmer}, {Veronig},
  {Vanninathan}, \& {Magdaleni{\'c}}}]{dissauer+2016apj}
{Dissauer}, K., {Temmer}, M., {Veronig}, A.~M., {Vanninathan}, K., \&
  {Magdaleni{\'c}}, J. 2016, \apj, 830, 92

\bibitem[{{Dissauer} {et~al.}(2019){Dissauer}, {Veronig}, {Temmer}, \&
  {Podladchikova}}]{dissauer+2019apj}
{Dissauer}, K., {Veronig}, A.~M., {Temmer}, M., \& {Podladchikova}, T. 2019,
  \apj, 874, 123

\bibitem[{{Dissauer} {et~al.}(2018{\natexlab{a}}){Dissauer}, {Veronig},
  {Temmer}, {Podladchikova}, \& {Vanninathan}}]{dissauer+2018a_apj}
{Dissauer}, K., {Veronig}, A.~M., {Temmer}, M., {Podladchikova}, T., \&
  {Vanninathan}, K. 2018{\natexlab{a}}, \apj, 855, 137

\bibitem[{{Dissauer} {et~al.}(2018{\natexlab{b}}){Dissauer}, {Veronig},
  {Temmer}, {Podladchikova}, \& {Vanninathan}}]{dissauer+2018b_apj}
---. 2018{\natexlab{b}}, \apj, 863, 169

\bibitem[{{Duan} {et~al.}(2019){Duan}, {Jiang}, {He}, {Feng}, {Zou}, \&
  {Cui}}]{duan+2019ApJ}
{Duan}, A., {Jiang}, C., {He}, W., {et~al.} 2019, \apj, 884, 73

\bibitem[{{Duan} {et~al.}(2017){Duan}, {Jiang}, {Hu}, {Zhang}, {Gary}, {Wu}, \&
  {Cao}}]{duan+2017apj}
{Duan}, A., {Jiang}, C., {Hu}, Q., {et~al.} 2017, \apj, 842, 119

\bibitem[{{Feng} {et~al.}(2013){Feng}, {Wiegelmann}, {Su}, {Inhester}, {Li},
  {Sun}, \& {Gan}}]{feng+2013apj}
{Feng}, L., {Wiegelmann}, T., {Su}, Y., {et~al.} 2013, \apj, 765, 37

\bibitem[{{Forbes} \& {Lin}(2000)}]{forbes+2000}
{Forbes}, T.~G., \& {Lin}, J. 2000, Journal of Atmospheric and
  Solar-Terrestrial Physics, 62, 1499

\bibitem[{{Gary}(2001)}]{gary2001soph}
{Gary}, G.~A. 2001, \solphys, 203, 71

\bibitem[{{Gary}(2009)}]{gary2009soph}
---. 2009, \solphys, 257, 271

\bibitem[{{Gary} \& {Hagyard}(1990)}]{gary&hagyard1990soph}
{Gary}, G.~A., \& {Hagyard}, M.~J. 1990, \solphys, 126, 21

\bibitem[{Grinstein {et~al.}(2007)Grinstein, Margolin, \&
  Rider}]{grinstein2007book}
Grinstein, F.~F., Margolin, L.~G., \& Rider, W.~J. 2007, Implicit large eddy
  simulation: computing turbulent fluid dynamics (Cambridge university press)

\bibitem[{{Harrison} \& {Lyons}(2000)}]{harrison&lyons2000aa}
{Harrison}, R.~A., \& {Lyons}, M. 2000, \aap, 358, 1097

\bibitem[{{Hernandez-Perez} {et~al.}(2017){Hernandez-Perez}, {Thalmann},
  {Veronig}, {Su}, {G{\"o}m{\"o}ry}, \& {Dickson}}]{hernandez-perez+2017apj}
{Hernandez-Perez}, A., {Thalmann}, J.~K., {Veronig}, A.~M., {et~al.} 2017,
  \apj, 847, 124

\bibitem[{{Hu} \& {Dasgupta}(2008)}]{hu&dasgupta2008soph}
{Hu}, Q., \& {Dasgupta}, B. 2008, \solphys, 247, 87

\bibitem[{{Hu} {et~al.}(2008){Hu}, {Dasgupta}, {Choudhary}, \&
  {B{\"u}chner}}]{hu+2008apj}
{Hu}, Q., {Dasgupta}, B., {Choudhary}, D.~P., \& {B{\"u}chner}, J. 2008, \apj,
  679, 848

\bibitem[{{Hu} {et~al.}(2010){Hu}, {Dasgupta}, {Derosa}, {B{\"u}chner}, \&
  {Gary}}]{hu+2010jastp}
{Hu}, Q., {Dasgupta}, B., {Derosa}, M.~L., {B{\"u}chner}, J., \& {Gary}, G.~A.
  2010, Journal of Atmospheric and Solar-Terrestrial Physics, 72, 219

\bibitem[{{Hudson} {et~al.}(1996){Hudson}, {Acton}, \&
  {Freeland}}]{hudson+1996apj}
{Hudson}, H.~S., {Acton}, L.~W., \& {Freeland}, S.~L. 1996, \apj, 470, 629

\bibitem[{{Inoue}(2016)}]{inoue2016peps}
{Inoue}, S. 2016, Progress in Earth and Planetary Science, 3, 19

\bibitem[{{Inoue} {et~al.}(2014){Inoue}, {Hayashi}, {Magara}, {Choe}, \&
  {Park}}]{inoue+2014apj}
{Inoue}, S., {Hayashi}, K., {Magara}, T., {Choe}, G.~S., \& {Park}, Y.~D. 2014,
  \apj, 788, 182

\bibitem[{{Inoue} {et~al.}(2015){Inoue}, {Hayashi}, {Magara}, {Choe}, \&
  {Park}}]{inoue+2015apj}
---. 2015, \apj, 803, 73

\bibitem[{{Janvier} {et~al.}(2016){Janvier}, {Savcheva}, {Pariat}, {Tassev},
  {Millholland}, {Bommier}, {McCauley}, {McKillop}, \& {Dougan}}]{janvier+2016}
{Janvier}, M., {Savcheva}, A., {Pariat}, E., {et~al.} 2016, \aap, 591, A141

\bibitem[{{Jiang} \& {Feng}(2014)}]{jiang&feng2014soph}
{Jiang}, C., \& {Feng}, X. 2014, \solphys, 289, 63

\bibitem[{{Jiang} {et~al.}(2018){Jiang}, {Feng}, \& {Hu}}]{jiang+2018apj}
{Jiang}, C., {Feng}, X., \& {Hu}, Q. 2018, \apj, 866, 96

\bibitem[{{Jiang} {et~al.}(2013){Jiang}, {Feng}, {Wu}, \&
  {Hu}}]{jiang+2013apjl}
{Jiang}, C., {Feng}, X., {Wu}, S.~T., \& {Hu}, Q. 2013, \apjl, 771, L30

\bibitem[{{Jiang} {et~al.}(2014){Jiang}, {Wu}, {Feng}, \& {Hu}}]{jiang+2014apj}
{Jiang}, C., {Wu}, S.~T., {Feng}, X., \& {Hu}, Q. 2014, \apjl, 786, L16

\bibitem[{{Jiang} {et~al.}(2016){Jiang}, {Wu}, {Feng}, \& {Hu}}]{jiang+2016nat}
---. 2016, Nature Communications, 7, 11522

\bibitem[{{Joshi} {et~al.}(2016){Joshi}, {Kushwaha}, {Veronig}, \&
  {Cho}}]{joshi+2016ApJ}
{Joshi}, B., {Kushwaha}, U., {Veronig}, A.~M., \& {Cho}, K.~S. 2016, \apj, 832,
  130

\bibitem[{{Joshi} {et~al.}(2017){Joshi}, {Kushwaha}, {Veronig}, {Dhara},
  {Shanmugaraju}, \& {Moon}}]{joshi+2017apj}
{Joshi}, B., {Kushwaha}, U., {Veronig}, A.~M., {et~al.} 2017, \apj, 834, 42

\bibitem[{{Kliem} {et~al.}(2013){Kliem}, {Su}, {van Ballegooijen}, \&
  {DeLuca}}]{kliem+2013apj}
{Kliem}, B., {Su}, Y.~N., {van Ballegooijen}, A.~A., \& {DeLuca}, E.~E. 2013,
  \apj, 779, 129

\bibitem[{{Kliem} \& {T{\"o}r{\"o}k}(2006)}]{kliem&torok2006prl}
{Kliem}, B., \& {T{\"o}r{\"o}k}, T. 2006, Physical Review Letters, 96, 255002

\bibitem[{{Kumar} \& {Bhattacharyya}(2011)}]{kumar&bhattacharyya2011phpl}
{Kumar}, D., \& {Bhattacharyya}, R. 2011, Physics of Plasmas, 18, 084506

\bibitem[{{Kumar} {et~al.}(2015{\natexlab{a}}){Kumar}, {Bhattacharyya}, \&
  {Smolarkiewicz}}]{kumard+2015phpl}
{Kumar}, D., {Bhattacharyya}, R., \& {Smolarkiewicz}, P.~K. 2015{\natexlab{a}},
  Physics of Plasmas, 22, 012902

\bibitem[{{Kumar} \& {Bhattacharyya}(2016)}]{kumar&bhattacharyya2016phpl}
{Kumar}, S., \& {Bhattacharyya}, R. 2016, Physics of Plasmas, 23, 044501

\bibitem[{{Kumar} {et~al.}(2016){Kumar}, {Bhattacharyya}, {Joshi}, \&
  {Smolarkiewicz}}]{kumar+2016apj}
{Kumar}, S., {Bhattacharyya}, R., {Joshi}, B., \& {Smolarkiewicz}, P.~K. 2016,
  \apj, 830, 80

\bibitem[{{Kumar} {et~al.}(2014){Kumar}, {Bhattacharyya}, \&
  {Smolarkiewicz}}]{kumar+2014phpl}
{Kumar}, S., {Bhattacharyya}, R., \& {Smolarkiewicz}, P.~K. 2014, Physics of
  Plasmas, 21, 052904

\bibitem[{{Kumar} {et~al.}(2015{\natexlab{b}}){Kumar}, {Bhattacharyya}, \&
  {Smolarkiewicz}}]{kumar+2015phpl}
---. 2015{\natexlab{b}}, Physics of Plasmas, 22, 082903

\bibitem[{{Lau} \& {Finn}(1990)}]{lau&finn1990apj}
{Lau}, Y.-T., \& {Finn}, J.~M. 1990, \apj, 350, 672

\bibitem[{{Lemen} {et~al.}(2012){Lemen}, {Title}, {Akin}, {Boerner}, {Chou},
  {Drake}, {Duncan}, {Edwards}, {Friedlaender}, {Heyman}, {Hurlburt}, {Katz},
  {Kushner}, {Levay}, {Lindgren}, {Mathur}, {McFeaters}, {Mitchell}, {Rehse},
  {Schrijver}, {Springer}, {Stern}, {Tarbell}, {Wuelser}, {Wolfson}, {Yanari},
  {Bookbinder}, {Cheimets}, {Caldwell}, {Deluca}, {Gates}, {Golub}, {Park},
  {Podgorski}, {Bush}, {Scherrer}, {Gummin}, {Smith}, {Auker}, {Jerram},
  {Pool}, {Soufli}, {Windt}, {Beardsley}, {Clapp}, {Lang}, \&
  {Waltham}}]{lemen+2012soph}
{Lemen}, J.~R., {Title}, A.~M., {Akin}, D.~J., {et~al.} 2012, \solphys, 275, 17

\bibitem[{{Liu} {et~al.}(2020){Liu}, {Prasad}, {Lee}, \& {Wang}}]{liu+2020ApJ}
{Liu}, C., {Prasad}, A., {Lee}, J., \& {Wang}, H. 2020, \apj, 899, 34

\bibitem[{{Liu} {et~al.}(2016){Liu}, {Kliem}, {Titov}, {Chen}, {Wang}, {Wang},
  {Liu}, {Xu}, \& {Wiegelmann}}]{liu+2016ApJ}
{Liu}, R., {Kliem}, B., {Titov}, V.~S., {et~al.} 2016, \apj, 818, 148

\bibitem[{{Margolin} {et~al.}(2006){Margolin}, {Rider}, \&
  {Grinstein}}]{margolin+2006jtb}
{Margolin}, L.~G., {Rider}, W.~J., \& {Grinstein}, F.~F. 2006, Journal of
  Turbulence, 7, 15

\bibitem[{{Masson} {et~al.}(2009){Masson}, {Pariat}, {Aulanier}, \&
  {Schrijver}}]{masson+2009apj}
{Masson}, S., {Pariat}, E., {Aulanier}, G., \& {Schrijver}, C.~J. 2009, \apj,
  700, 559

\bibitem[{{Mitra} {et~al.}(2018){Mitra}, {Joshi}, {Prasad}, {Veronig}, \&
  {Bhattacharyya}}]{mitra+2018apj}
{Mitra}, P.~K., {Joshi}, B., {Prasad}, A., {Veronig}, A.~M., \&
  {Bhattacharyya}, R. 2018, \apj, 869, 69

\bibitem[{{Nayak} {et~al.}(2019){Nayak}, {Bhattacharyya}, {Prasad}, {Hu},
  {Kumar}, \& {Joshi}}]{nayak+2019apj}
{Nayak}, S.~S., {Bhattacharyya}, R., {Prasad}, A., {et~al.} 2019, \apj, 875, 10

\bibitem[{{Parker}(1972)}]{parker1972apj}
{Parker}, E.~N. 1972, \apj, 174, 499

\bibitem[{{Parker}(1988)}]{parker1988apj}
---. 1988, \apj, 330, 474

\bibitem[{{Parker}(1994)}]{parker1994book}
---. 1994, Spontaneous current sheets in magnetic fields : with applications to
  stellar x-rays.~ International Series in Astronomy and Astrophysics, Vol.~1.~
  New York : Oxford University Press, 1994., 1

\bibitem[{{Pesnell} {et~al.}(2012){Pesnell}, {Thompson}, \&
  {Chamberlin}}]{pesnell+2012soph}
{Pesnell}, W.~D., {Thompson}, B.~J., \& {Chamberlin}, P.~C. 2012, \solphys,
  275, 3

\bibitem[{{Petrie}(2012)}]{petrie:2012apj}
{Petrie}, G.~J.~D. 2012, \apj, 759, 50

\bibitem[{{Prasad} \& {Bhattacharyya}(2016)}]{prasad&bhattacharyya2016phpl}
{Prasad}, A., \& {Bhattacharyya}, R. 2016, Physics of Plasmas, 23, 114504

\bibitem[{{Prasad} {et~al.}(2018){Prasad}, {Bhattacharyya}, {Hu}, {Kumar}, \&
  {Nayak}}]{prasad+2018apj}
{Prasad}, A., {Bhattacharyya}, R., {Hu}, Q., {Kumar}, S., \& {Nayak}, S.~S.
  2018, \apj, 860, 96

\bibitem[{{Prasad} {et~al.}(2017){Prasad}, {Bhattacharyya}, \&
  {Kumar}}]{prasad+2017apj}
{Prasad}, A., {Bhattacharyya}, R., \& {Kumar}, S. 2017, \apj, 840, 37

\bibitem[{{Priest}(2014)}]{priest2014book}
{Priest}, E. 2014, {Magnetohydrodynamics of the Sun} (Cambridge University
  Press)

\bibitem[{{Prusa} {et~al.}(2008){Prusa}, {Smolarkiewicz}, \&
  {Wyszogrodzki}}]{prusa2008cf}
{Prusa}, J.~M., {Smolarkiewicz}, P.~K., \& {Wyszogrodzki}, A.~A. 2008,
  Computers \& Fluids, 37, 1193

\bibitem[{{Qiu} \& {Cheng}(2017)}]{qiu+2017apj}
{Qiu}, J., \& {Cheng}, J. 2017, \apjl, 838, L6

\bibitem[{{Romano} {et~al.}(2015){Romano}, {Zuccarello}, {Guglielmino},
  {Berrilli}, {Bruno}, {Carbone}, {Consolini}, {de Lauretis}, {Del Moro},
  {Elmhamdi}, {Ermolli}, {Fineschi}, {Francia}, {Kordi}, {Landi
  Degl'Innocenti}, {Laurenza}, {Lepreti}, {Marcucci}, {Pallocchia},
  {Pietropaolo}, {Romoli}, {Vecchio}, {Vellante}, \&
  {Villante}}]{roman0+2015aa}
{Romano}, P., {Zuccarello}, F., {Guglielmino}, S.~L., {et~al.} 2015, \aap, 582,
  A55

\bibitem[{{Ruderman} \& {Roberts}(2002)}]{ruderman&roberts2002apj}
{Ruderman}, M.~S., \& {Roberts}, B. 2002, \apj, 577, 475

\bibitem[{{Sahu} {et~al.}(2020){Sahu}, {Joshi}, {Mitra}, {Veronig}, \&
  {Yurchyshyn}}]{sahu+2020ApJ}
{Sahu}, S., {Joshi}, B., {Mitra}, P.~K., {Veronig}, A.~M., \& {Yurchyshyn}, V.
  2020, \apj, 897, 157

\bibitem[{{Savcheva} {et~al.}(2016){Savcheva}, {Pariat}, {McKillop},
  {McCauley}, {Hanson}, {Su}, \& {DeLuca}}]{savcheva+2016apj}
{Savcheva}, A., {Pariat}, E., {McKillop}, S., {et~al.} 2016, \apj, 817, 43

\bibitem[{{Savcheva} {et~al.}(2015){Savcheva}, {Pariat}, {McKillop},
  {McCauley}, {Hanson}, {Su}, {Werner}, \& {DeLuca}}]{savcheva+2015apj}
---. 2015, \apj, 810, 96

\bibitem[{{Schou} {et~al.}(2012){Schou}, {Scherrer}, {Bush}, {Wachter},
  {Couvidat}, {Rabello-Soares}, {Bogart}, {Hoeksema}, {Liu}, {Duvall}, {Akin},
  {Allard}, {Miles}, {Rairden}, {Shine}, {Tarbell}, {Title}, {Wolfson},
  {Elmore}, {Norton}, \& {Tomczyk}}]{schou+2012soph}
{Schou}, J., {Scherrer}, P.~H., {Bush}, R.~I., {et~al.} 2012, \solphys, 275,
  229

\bibitem[{{Shibata} \& {Magara}(2011)}]{shibata&magara2011lrsp}
{Shibata}, K., \& {Magara}, T. 2011, Living Reviews in Solar Physics, 8, 6

\bibitem[{{Smolarkiewicz}(2006)}]{smolarkiewicz2006ijnmf}
{Smolarkiewicz}, P.~K. 2006, International Journal for Numerical Methods in
  Fluids, 50, 1123

\bibitem[{{Smolarkiewicz} \&
  {Charbonneau}(2013)}]{smolarkiewicz&charbonneau2013jcoph}
{Smolarkiewicz}, P.~K., \& {Charbonneau}, P. 2013, Journal of Computational
  Physics, 236, 608

\bibitem[{{Sterling} \& {Hudson}(1997)}]{sterling&hudson1997apj}
{Sterling}, A.~C., \& {Hudson}, H.~S. 1997, \apjl, 491, L55

\bibitem[{{Thompson} {et~al.}(2000){Thompson}, {Cliver}, {Nitta},
  {Delann{\'e}e}, \& {Delaboudini{\`e}re}}]{thompson+2000georl}
{Thompson}, B.~J., {Cliver}, E.~W., {Nitta}, N., {Delann{\'e}e}, C., \&
  {Delaboudini{\`e}re}, J.~P. 2000, \grl, 27, 1431

\bibitem[{{Vanninathan} {et~al.}(2018){Vanninathan}, {Veronig}, {Dissauer}, \&
  {Temmer}}]{vanninathan+2018apj}
{Vanninathan}, K., {Veronig}, A.~M., {Dissauer}, K., \& {Temmer}, M. 2018,
  \apj, 857, 62

\bibitem[{{Veronig} {et~al.}(2019){Veronig}, {G{\"o}m{\"o}ry}, {Dissauer},
  {Temmer}, \& {Vanninathan}}]{veronig+2019apj}
{Veronig}, A.~M., {G{\"o}m{\"o}ry}, P., {Dissauer}, K., {Temmer}, M., \&
  {Vanninathan}, K. 2019, \apj, 879, 85

\bibitem[{{Wang} \& {Liu}(2012)}]{wang&liu2012apj}
{Wang}, H., \& {Liu}, C. 2012, \apj, 760, 101

\bibitem[{{Wiegelmann}(2008)}]{wiegelmann2008jgra}
{Wiegelmann}, T. 2008, Journal of Geophysical Research (Space Physics), 113,
  A03S02

\bibitem[{{Wiegelmann} {et~al.}(2006){Wiegelmann}, {Inhester}, \&
  {Sakurai}}]{wiegelmann+2006soph}
{Wiegelmann}, T., {Inhester}, B., \& {Sakurai}, T. 2006, \solphys, 233, 215

\bibitem[{{Wiegelmann} \& {Sakurai}(2012)}]{wiegelmann&sakurai2012lrsp}
{Wiegelmann}, T., \& {Sakurai}, T. 2012, Living Reviews in Solar Physics, 9,
  arXiv:1208.4693

\bibitem[{{Yalim} {et~al.}(2020){Yalim}, {Prasad}, {Pogorelov}, {Zank}, \&
  {Hu}}]{yalim+2020ApJ}
{Yalim}, M.~S., {Prasad}, A., {Pogorelov}, N.~V., {Zank}, G.~P., \& {Hu}, Q.
  2020, \apjl, 899, L4

\bibitem[{{Yang} {et~al.}(2014){Yang}, {Chen}, {Hsieh}, {Wu}, {He}, \&
  {Tsai}}]{yang:2014apj}
{Yang}, Y.-H., {Chen}, P.~F., {Hsieh}, M.-S., {et~al.} 2014, \apj, 786, 72

\bibitem[{{Zarro} {et~al.}(1999){Zarro}, {Sterling}, {Thompson}, {Hudson}, \&
  {Nitta}}]{zarro+1999apj}
{Zarro}, D.~M., {Sterling}, A.~C., {Thompson}, B.~J., {Hudson}, H.~S., \&
  {Nitta}, N. 1999, \apjl, 520, L139

\bibitem[{{Zhang} {et~al.}(2017){Zhang}, {Su}, \& {Ji}}]{zhang+2017aa}
{Zhang}, Q.~M., {Su}, Y.~N., \& {Ji}, H.~S. 2017, \aap, 598, A3

\bibitem[{{Zhou} {et~al.}(2017){Zhou}, {Zhang}, {Wang}, \&
  {Wheatland}}]{zhou+2017ApJL}
{Zhou}, G.~P., {Zhang}, J., {Wang}, J.~X., \& {Wheatland}, M.~S. 2017, \apjl,
  851, L1

\end{thebibliography}
%=== LIST OF CHANGES ===
\listofchanges
%==== END OF FILE =====
\end{document}